\documentclass[11pt]{article}

\usepackage{pdflscape}
\usepackage{placeins}
\usepackage{iftex}
\ifPDFTeX
  \usepackage[T1]{fontenc}
  \usepackage[utf8]{inputenc}
  \usepackage{lmodern}
\else
  \usepackage{fontspec}
\fi

\usepackage[main=english,ngerman]{babel}
\usepackage[a4paper,margin=1in]{geometry}
\usepackage{authblk}
\usepackage{microtype}
\usepackage{setspace}
\usepackage{amsmath}
\usepackage{graphicx}
\usepackage{booktabs}
\usepackage{longtable}
\usepackage{tabularx}
\usepackage{array}
\usepackage{enumitem}
\usepackage{csquotes}
\usepackage{xcolor}
\usepackage{float}
\usepackage{caption}
\usepackage{tcolorbox}
\usepackage{hyperref}
\usepackage[nameinlink,noabbrev]{cleveref}
\usepackage[backend=biber,style=authoryear,maxcitenames=2,maxbibnames=99,natbib=true]{biblatex}

\hypersetup{
  colorlinks=true,
  linkcolor=blue!50!black,
  citecolor=blue!50!black,
  urlcolor=blue!50!black,
  pdftitle={Scaling Up Semi-Structured and In-depth Interviews},
  pdfauthor={Alexander Wuttke, Max Lang, Christopher Klamm, Quirin Würschinger, Frauke Kreuter}
}

\title{\textbf{AI Conversational Interviewing: Scaling Up Semi-Structured and In-depth Interviews}\\[0.35em]}

\author[1]{Alexander Wuttke\thanks{Both authors contributed equally.}\textsuperscript{,}\thanks{Corresponding Author: \href{mailto:A.Wuttke@lmu.de}{A.Wuttke@lmu.de}}}
\author[2]{Max Melchior Lang\thanks{Both authors contributed equally.}}
\author[3,4]{Christopher Klamm}
\author[1]{Quirin Würschinger}
\author[1,5]{Frauke Kreuter}

\affil[1]{LMU Munich}
\affil[2]{University of Oxford}
\affil[3]{University of Mannheim}
\affil[4]{University of Cologne}
\affil[5]{University of Maryland}

\begin{document}
\maketitle

\begin{abstract}
Public opinion research has long faced a trade-off between depth and scale: standardized surveys enable large-scale measurement but restrict respondents to researcher-defined categories, obscuring the diversity of unexpected considerations that underlie public sentiment. More conversational interviews provide richer insights through open-ended probing, but their reliance on trained human interviewers has kept them difficult to scale. This study introduces AI Conversational Interviewing as a method for collecting open-ended public opinion data at scale, pursuing three objectives: to demonstrate the analytical value of conversational text data for questions beyond the reach of closed-ended items; to assess the method's practical viability through participants' own evaluations; and to inform implementation by experimentally comparing voice-based, chat-based, and free-choice interview modes. We conducted a study combining an AI-led interview with a standardized survey on migration policy among 571 respondents recruited via Prolific and Payback Panel. The findings establish AI Conversational Interviewing as a viable and valuable addition to the social-science toolkit. The conversational transcripts surface considerations and reasoning that a comprehensive standardized battery does not capture such as markedly different mental models of migration among subgroups with similar attitudes levels. Among respondents who completed the interview, evaluations of the AI interview were at or above those of the standardized survey across modes, although completion itself varied by condition. By releasing open data and open-source pipeline materials, the study contributes to a growing literature on harnessing artificial intelligence to expand the methods of public opinion measurement.

\end{abstract}

\section{Introduction}

Standardized surveys are the primary instrument of public opinion research. They rely on a fixed set of identical questions and predefined response options to measure attitudes, intentions, and behaviours, typically with the goal of producing accurate inferences about broader populations \parencite{conrad_new_2008,fowler_standardized_1998}. Especially in web-based administration, standardization delivers major benefits: surveys are scalable, cost-efficient, and are, if carefully designed, comparable across respondents, countries, and time points. They also allow sample sizes large enough for subgroup analysis, experiments, and estimates with pre-specified levels of precision \parencite{berinsky_measuring_2017,groves_survey_2011,konig_conceptualizing_2022}.

Conversational interviews, rooted in the qualitative tradition with semi-structured or in-depth interviews as prominent subtypes, offer the opposite bundle of strengths by centering on respondents' subjective realities \parencite{flick_doing_2022,rubin_qualitative_2005,seidman_interviewing_2019}. Conversational interviewing allows respondents to speak freely while interviewers can probe, clarify, and adapt to respondents' answers. This flexibility allows respondents to articulate views in their own terms and enables researchers to reconstruct how individuals interpret the world, which considerations they draw upon, and how those considerations connect within broader belief systems. But conversational interviews are resource intensive. They require skilled human interviewers and therefore have historically been limited to comparatively small samples \parencite{berg_toward_2025}. 

\begin{figure}
    \centering
    \includegraphics[width=1\linewidth]{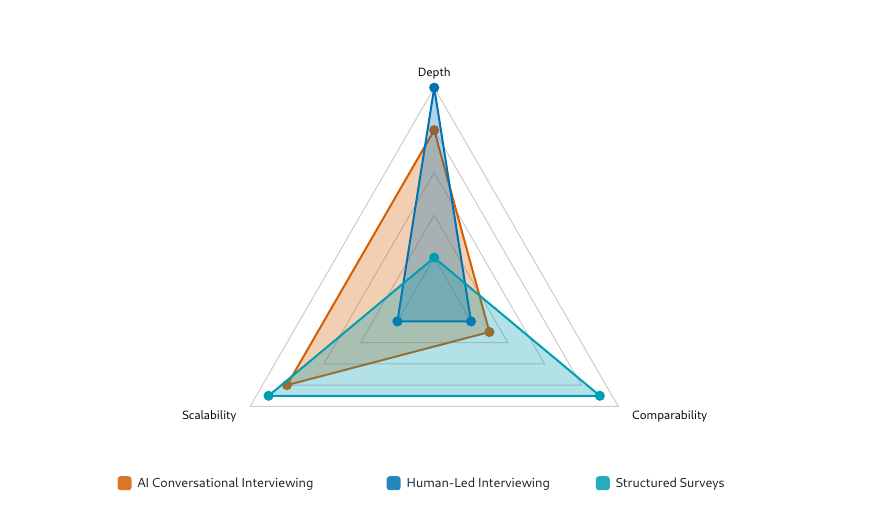}
    \caption{Comparative Strengths: Attitude Measurement Approaches}
    \label{fig:radar}
\end{figure}

Public opinion researchers accordingly face a familiar trade-off (Figure \ref{fig:radar}): they must often choose between the scale of standardized surveys and the depth of conversational interviews.

\subsection{Making conversational interviews scalable}

Recent advances in natural language processing create a plausible path beyond this dilemma. Large Language Models possess conversational capacities that make them potential adaptive interviewers. By relying on AI agents, researchers can conduct semi-structured interviews at sample sizes that would previously have been infeasible. Based on an interview agenda and explicit ground rules for interviewer behaviour, we employ large language models to act as adaptive interviewers in conversational interviews carried out via text or voice chat modes. We refer to this approach as \emph{AI Conversational Interviewing}.

Notably, this approach differs from a growing literature that seeks to increase the flexibility or accuracy of standardized surveys, for example, by clarifying question meaning, adding conversational probes within otherwise fixed interviews, or translating open-ended responses into structured outputs \parencite{barari_ai-assisted_2025,mittereder_interviewerrespondent_2018,schober_does_1997,shih_rationalizer_2026,wei_leveraging_2024,west_can_2018,xiao_tell_2020}. These are important and promising developments, but they remain within the standardized-survey paradigm. While their goal is to improve the measurement properties of structured questionnaires, our approach uses LLMs to generate unstructured textual data by making a different method of attitude measurement, i.e. conversational interviewing, scalable.

\subsection{Avenues of inquiry enabled by AI Conversational Interviewing}

Making conversational interviews scalable does not merely increase the value of surveys but enables a qualitatively distinct form of inquiry that addresses what standardized surveys are structurally unable to capture. Standardized surveys are indispensable for measuring the direction and intensity of attitudes: they establish whether respondents support or oppose a policy, how strongly they do so, and how these positions vary across groups and over time. 

Yet, public opinion consists of more than stances on predefined scales. Belief systems also vary in content, sophistication, and internal structure \parencite{converse_nature_2006,hernandez_democracy_2019,howe_attitude_2017}. Citizens differ in which considerations are available to them, which they treat as central or peripheral, how they connect ideas to one another, whether they can justify their views, and whether their attitudes are coherent, ambivalent, or internally contradictory. These dimensions are not incidental properties of an attitude, but rather reflect whether a person's view is strong or superficial, principled or cue-driven and, consequently, whether an attitude is likely to anchor related beliefs and guide behaviour, or to give way under scrutiny \parencite{howe_attitude_2017}.

One reason why closed-ended instruments are poorly suited to capture these features is that researchers need to specify all relevant considerations in advance. Even a carefully designed survey battery can only measure responses the researcher anticipated and built into the questionnaire. In this sense, standardized surveys embody the streetlight problem \parencite{kaplan_conduct_2017}, meaning that they reflect what the researcher expected, leaving in the dark the unanticipated arguments, memories, and forms of reasoning that respondents themselves would have introduced. The problem is compounded by a further limitation: by presenting predefined response options as the default vocabulary, surveys may not only fail to observe what is on people's minds but actively shape it \parencite{fictitious}, priming frames that had not previously figured in respondents' awareness and thereby distorting the very object surveys seek to measure \parencite{Poulsen2026VisualNetwork, schwarz, Schwarz1999SelfReports}. This limitation is especially consequential when citizens organize political issues through frameworks that diverge from expert categories, when they rely on personal experience or local knowledge or even on nonsensical reasoning, all of which may be theoretically important. 

AI Conversational Interviewing addresses this limitation by making the respondent's subjective worldview the starting point of measurement \parencite{flick_doing_2022,rubin_qualitative_2005}. Because respondents can answer in their own terms and the interviewer can follow up with neutral probes across the full course of an exchange, the resulting data shows which arguments, beliefs and emotions people have available on their minds and can express on their own. Moreover, AI Conversational Interviews reveal not only what respondents like but how they think: which considerations arise spontaneously, which remain absent until prompted, how respondents move from one idea to another, where they provide reasons or examples, and where their accounts become vague, unstable, or internally inconsistent. The analytical promise of AI Conversational Interviewing therefore lies not simply in adding expressive depth to surveys, but in the ability to capture people's reasoning process when reflecting on an issue. The generated data allows researchers to translate people's utterances into analytically useful categories that inform about people's underlying belief systems. And because such interviews can be conducted at survey-like scale, they enable comparisons of people's mental models across social groups, partisan camps, countries, and experimental conditions.

To make these advantages concrete, consider what a researcher can learn about migration attitudes when conversational data supplements a standardized survey. Suppose three respondents record identical anti-immigration positions on a Likert-type scale — valuable comparative information that locates them on a shared dimension. Conversational data can reveal that the same scale position rests on distinct mental models in way that a researcher might not have anticipated and thus could not capture even with extensive survey batteries. One respondent may organize her position primarily around housing shortages in her city, grounded in recent difficulties finding an apartment. The second holds a structurally more complex belief system: he endorses immigration restriction as a policy position while describing immigrant colleagues with genuine warmth, resolving the tension through a distinction between ``sustainable'' and ``excessive'' immigration that is his own conceptual invention. A third may offer few available considerations and reveal, under probing, that her scale position reflects a norm absorbed from her family rather than a view she has actively worked through. Three identical survey scores; three qualitatively distinct mental models, differing in their salient considerations, internal complexity, and reasoning process.

AI Conversational Interviewing is not intended as a replacement for existing approaches. Standardized surveys continue to have unique advantages, such as the ability to record discrete factual information swiftly and with high comparability across time. Nor is it intended to replace qualitative interviewing conducted by human researchers with their unique empathic qualities. AI Conversational Interviewing is best understood as a complementary addition to the toolkit of public opinion scholars that makes accessible a class of research questions that existing approaches cannot address well on their own.

\subsection{Contributions of this study}

This study contributes to an emerging interdisciplinary literature on AI-led interviewing in computer science \parencite{handa_introducing_2025, degen_probing_2025, wong_ai_2025}, survey methodology \parencite{lang_telephone_2025, vonDerHeyde2025WhoCounts}, linguistics \parencite{kim_llm-as--interviewer_2025, ma_too_2025}, the social sciences \parencite{cavusoglu_deveci_experimental_2026, cuevas_collecting_2025}, and human-computer interaction \parencite{anugraha_sparkme_2026, liu_mimitalk_2025, ,hwang_scale_2025,jacobsen_chatbots_2025}. Previous work has shown the feasibility of LLM-based interviewing in smaller-scale settings, demonstrating that AI interviewers closely follow the rules for good conversational interviewing and achieve interview performance similar to student interviewers    \parencite{budig_towards_2025, guven_comparing_2025, wuttke_ai_2025, wong_ai_2025}. Data collected by LLM interviewers are not as rich compared to expert interviewers \parencite{cuevas_collecting_2025}, but AI agents nonetheless elicit responses that experts rate as relevant and informational \parencite{anugraha_sparkme_2026}. Interview data from AI interviews have proven useful to study unexplored topics \parencite{handa_introducing_2025, geiecke_conversations_2024} and passed validity tests such as predicting individual  behavior six month into the future \parencite{chopra_conducting_2023}. So, recent research has shown that frontier LLMs possess the capabilities of conducting conversational interviews and collect textual belief system data that is suitable for research purposes. 

Building on these contributions, our study pursues three interconnected objectives. First, we demonstrate the analytical value of conversational text data for public opinion research by conducting an illustrative text-as-data analysis of citizens' views on migration policy. By pairing conversational interview data with a comprehensive standardized survey battery drawn from the migration-related items included in the German Longitudinal Election Study (GLES), we show that the considerations, arguments, and cognitive frames respondents articulate are substantially more diverse than any closed-ended instrument could capture and that conversational data provides unique insights into people's mental maps that is inaccessible to structured surveys. Second, we assess the practical viability of AI Conversational Interviewing at scale by examining participants' subjective evaluations of the method and comparing these with their evaluations of the standardized survey component. Third, having established the method's value and viability, we provide experimental evidence to guide implementation choices by comparing voice-based interviews, chat-based interviews, and a condition in which respondents could choose their preferred modality. By providing open data and open-source pipeline materials\footnote{\url{ANONYMIZED}}, this study aims to lower barriers to adoption and facilitate replication and extension by other researchers.

\section{Methods}

\subsection{Study design, recruitment, and measures}

The study combined an AI-led Conversational Interview on migration and migration policy in Germany with a standardized survey and post-interview evaluations. The study was approved by the IRB of \textsc{anonymized}. 

Participants received detailed information about study content, procedures, and data privacy at the outset.  Participants were informed about the upcoming interaction with a large language model and received technical instructions. The study only began after participants provided informed consent. 

Data were collected in Germany from Prolific between 25 September 2025 and 31 October 2025 and from Payback Panel between 16 October 2025 and 27 October 2025. Recruitment was initially planned exclusively through Prolific. Because Prolific recruited respondents at a slower rate than anticipated, we recruited additional participants through Payback Panel to reach the target sample size within the projected fieldwork period. Participants from both survey vendors received the same survey instrument. The analysis pools both sources.

Data collection consisted of three subsequent components: the AI conversational interview on migration politics, standardized survey questions on migration politics, and standardized survey questions on the interview experience. Data collection was conducted via Qualtrics. The customized interface for AI Conversational Interviewing (see below for details) was embedded within the Qualtrics environment so that respondents could move directly from the conversational component to the closed-ended questionnaire without leaving the Qualtrics environment. 

We selected migration policy as the substantive topic of data collection given its prominence in contemporary public opinion discourse. The standardized survey battery includes all migration-related items from the German Longitudinal Election Study (see Appendix \Cref{sec:appendix-questionnaire} for full survey instrument), with the aim of replicating the full range of items available to researchers working with secondary survey data on this topic \parencite{glesZA10100}. 

For the conversational component, we implemented a semi-structured interview guide specifying core questions while retaining flexibility for context-specific follow-up probes (see Appendix \Cref{sec:appendix-prompts} for interview guide). This approach is well-suited for large-N interview projects because it ensures common topical coverage while preserving flexibility for individualized probing \parencite{helfferich_qualitat_2011}. Conversational interviews began by reminding respondents that there were no right or wrong answers, reducing evaluative pressure \parencite{rubin_qualitative_2005}. Interviews then moved to a broad grand-tour opening question about migration \parencite{spradley_ethnographic_2016}. 

To study the role of interview mode, respondents were randomly assigned to one of three conditions: a voice condition using spoken input and output, a chat condition using typed interaction, and a choice condition in which respondents selected their preferred mode. 

\subsection{Open science}

Our evaluation of AI Conversational Interviewing was pre-registered. We registered the hypotheses, analysis strategy, and falsifiable success criteria on the Open Science Framework\footnote{\url{https://osf.io/y5cjt/overview?view_only=6f8714dd90dc456795390874a8ec6f9f}}.  In Appendix \Cref{sec:appendix-deviations}, we document where and why the manuscript deviates from the pre-analysis plan.

\subsection{Prompting strategy}
Building on insights from the LLM prompting literature \parencite{he_does_2024,kim_detail_2025,meilan_using_2026}, the prompts define the interviewer's role and task, the substantive topic of the interview and conversational norms (see Appendix \Cref{sec:appendix-prompts}). Specifically, the interviewer was instructed to conduct a semi-structured interview in German on migration and migration policy, to remain neutral and refrain from taking a position, to practise active listening, and to use follow-up questions that allowed elaboration without steering respondents toward a preferred answer. In doing so, the prompts reflected best-practice guidance from the qualitative literature on conversational interviewing such as openness, non-directiveness, and the balancing of structure with flexibility \parencite{newcomer_conducting_2015,helfferich_qualitat_2011,rubin_qualitative_2005,seidman_interviewing_2019}.

The prompts also included two experimentally varied prompt variants (see Appendix \Cref{sec:appendix-prompts}): a short version (approximately 2{,}000 tokens) and an informed version (approximately 3{,}000 tokens), which is not reported in this study.

For voice and chat interviews, we implemented separate LLM agents with minor adaptions in the prompts for each mode. 


\begin{landscape}
\begin{figure}
\centering
\includegraphics[width=1.6\textwidth]{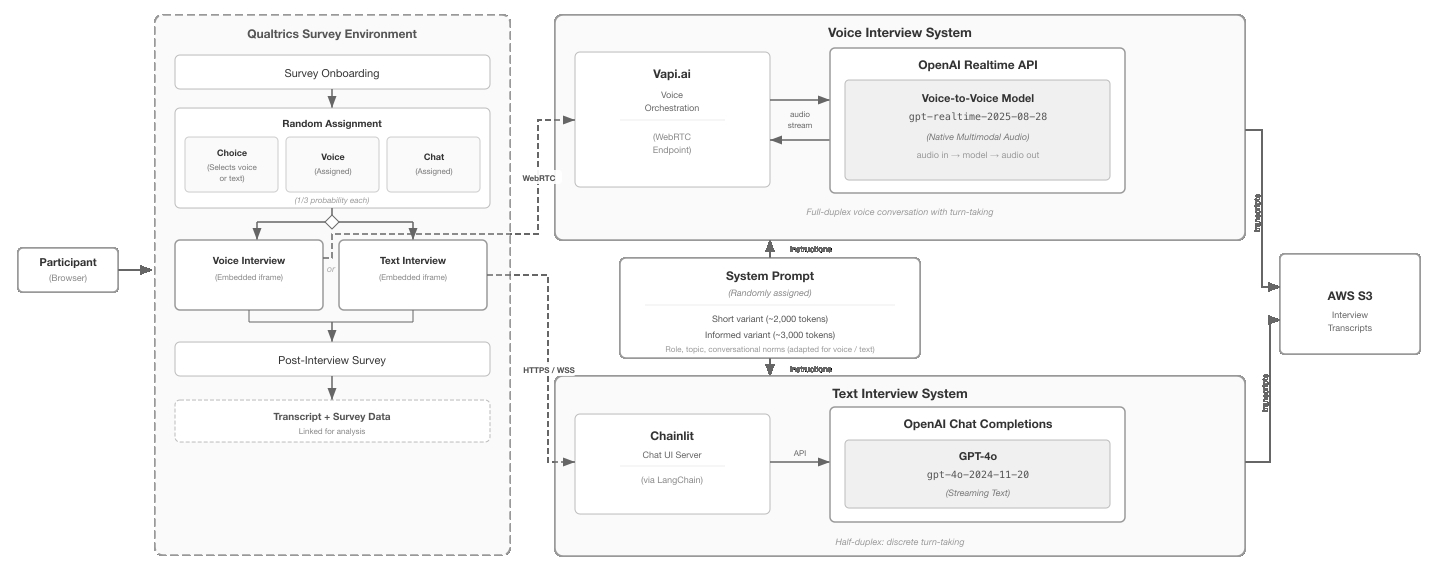}
\caption{Overall system architecture and participant flow.}
\end{figure}
\end{landscape}

\subsection{Text interviewer}
At the start of each chat session, respondents were asked to enter the numeric respondent ID displayed in Qualtrics above the chat window. This identifier was used to link interview records to the corresponding survey responses. Upon entering a numeric ID, participants were directed to the LLM interviewer, which initiated the conversational interview.  The conversation proceeded as a sequence of alternating respondent and interviewer turns. Interviewer messages were displayed incrementally as they were generated, producing a common chat experience. The text interviewer was implemented in Python using Chainlit as the browser-based chat interface and the OpenAI Chat Completions API via LangChain, with GPT-4o (\texttt{gpt-4o-2024-11-20}, temperature: 0.8) as the underlying model. 

\subsection{Voice interviewer}

As a pre-screen, respondents were asked to enter a numeric respondent ID displayed in Qualtrics above the voice module.  Once a numeric ID had been entered, respondents proceeded to the voice interview interface.

\begin{figure}[H]
  \centering
  \includegraphics[scale=0.5]{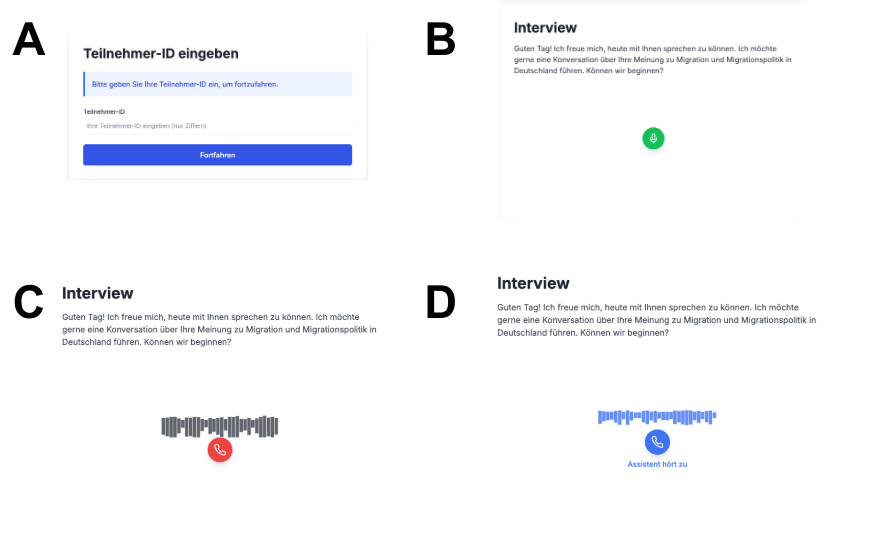}
  \caption{Voice Interface in different states. }
  \label{fig:architecture}
   \caption*{\footnotesize Note: A = Entering the ID; B = Before starting the interview; C = The interviewer speaking; D = The respondent speaking.}

\end{figure}

The voice user interface was intentionally simple and comprised three states  (Figure \ref{fig:architecture}). After entering the respondent ID, introductory text was displayed alongside a green microphone button which participants clicked to initiate the conversation. This allowed participants to prepare, find a quiet space, or familiarise themselves with the interface. Once the green button was clicked, the interviewer initiated the conversation via audio output and the interface displayed a soundwave together with a red telephone button that allowed participants to end the call. When the respondent was actively speaking, the sound wave and end-call button changed colour and a brief on-screen message indicated that the system was listening.  This visual cue was introduced after pilot testing showed that respondents were sometimes unsure whether the system could hear them.

The voice interview system used Vapi.ai for orchestration and OpenAI’s GPT Realtime model (\texttt{gpt-realtime-2025-08-28}, temperature: 0.7). Data from the voice interviewer were stored on an AWS S3 bucket under the researchers' control. The stored records consisted of a speaker-labelled transcript, timestamps, total conversation duration, and respondent ID. We did not store audio recordings.

In contrast to the text condition, the voice interviewer operated as a full-duplex speech interface with dynamic turn-taking, enabling either party to interrupt the other such that respondents could begin speaking while the system was still responding \parencite{veluri_beyond_2024}.  So, the voice and text conditions varied in several respects. The mode contrasts reported throughout this manuscript should thus not be understood as not as estimates of a pure modality effect but schould be read as comparisons of voice and text systems as deployed.

\subsection{Pre-Tests}

Before the main field period, we conducted multiple pre-tests with members of the research team and Prolific respondents to assess the technical viability of the study and refine the implementation. These pre-tests prompted several modifications prior to the main field period.  Most notably, one pre-test indicated that some respondents in the voice condition felt that the interviewer interrupted them too early while they were still thinking through their answer. We therefore adjusted the turn-taking configuration of the voice system, with the result that most respondents in the main data collection reported a reasonable balance in response timing (see  Appendix~ \Cref{sec:appendix-voice-waiting}). Another notable insight from our pre-tests was the model dependency of our prompts. When OpenAI released GPT-5 shortly before our planned field period, pre-tests revealed a marked deterioration in interviewer performance when switching to the new model. The system failed to keep the conversation focused and generated too many probing questions, resulting in excessively long interviews. Because we had a validated prompting strategy for GPT-4o, we decided to conduct the main data collection with that model. The pre-tests also included systematic stress testing (``red teaming''). In these stress tests, research assistants and members of the research team impersonated respondents with atypical behaviour  (e.g. giving nonsensical responses, sabotaging the interview etc.) and then documented model behavior based on a checklist of potentially undesirable behavior to inform prompt refinement. Data from the pre-tests is not included in the main analysis reported below. 

\section{Results}

Evaluating the viability and value of AI-assisted interviewing requires attention to three distinct criteria. First, given the novel technical implementation, the system must function reliably across a heterogeneous range of devices and user configurations. Second, respondent experience constitutes a critical dimension of viability, yet it remains empirically unclear how participants respond to this unfamiliar interview format. Third, while prior research suggests that LLM-based interviewers can adhere to conversational interviewing protocols \parencite{budig_towards_2025, guven_comparing_2025, wuttke_ai_2025, wong_ai_2025}, the analytical suitability of the resulting data for substantive social research has rarely been assessed. We  examine each criterion in turn.

\subsection{Participant flow}

\begin{figure}[H]
    \centering
    \includegraphics[width=1\linewidth]{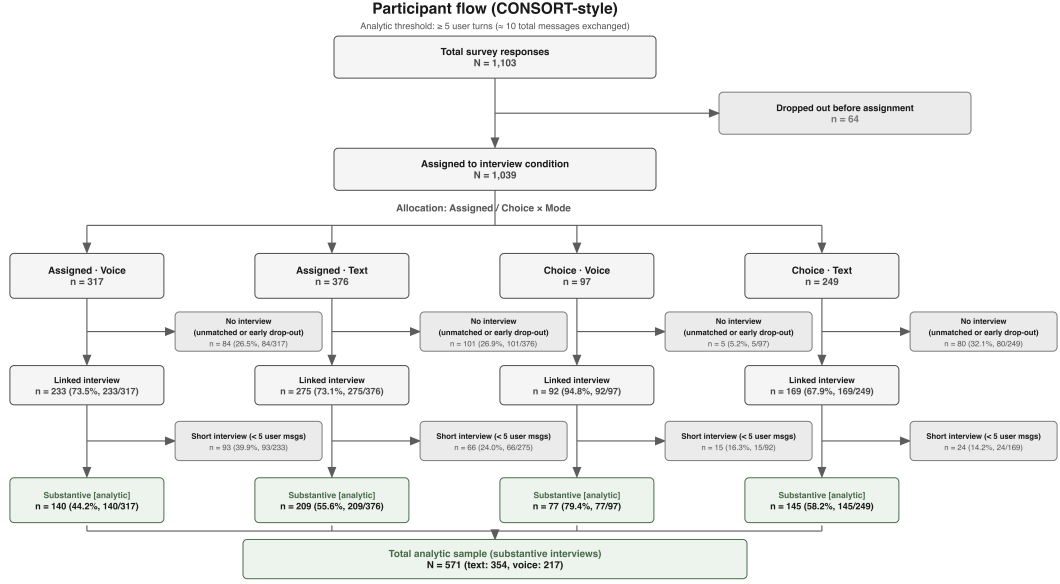}
    \caption{Participant flow}
    \label{fig:consort}
\end{figure}

The \textsc{consort}-style flow diagram \ref{fig:consort} details sample size and participant attrition across each stage of the study. A total of 1,039 respondents were randomly assigned to the voice, text, or choice conditions, with no statistically significant imbalances across experimental groups  ($\chi^2(2) = 3.67,\; p = 0.159$).  Two sources of attrition substantially reduced the analytic sample. First, survey and interview records could not be linked for approximately one quarter of respondents, owing to a reliance on manual identifier entry rather than automated linkage. These cases were excluded from all subsequent analyses. This limitation reflects a correctable implementation shortcoming; future studies should employ background ID-linkage procedures, such as URL parameter passing, that require no participant input. Second, a non-trivial proportion of respondents did not complete the full interview, with dropout concentrated in the forced-voice condition. Incomplete sessions likely reflect a combination of technical barriers and elevated task burden associated with mandatory voice input. Consistent with this interpretation, respondents assigned to the choice condition overwhelmingly selected the chat modality when given the option. Taken together, both sources of attrition point to addressable imperfections of our implementation rather than fundamental constraints of the approach. Adopting a conservative inclusion criterion — retaining only cases with complete interview data and successfully linked records — yields a final analytic sample of N=571. 

\subsection{Evaluation of the technical setup}

Technical accessibility constitutes a foundational prerequisite for successful interview completion, particularly because even respondents experienced with online survey panels may be unfamiliar with browser-based conversational interfaces — let alone voice interactions mediated by microphone and speaker input. Overall, 80.3\% of respondents reported no difficulties initiating the interview and 82.8\% of respondents reported no difficulties ending the interview, with only minor differences across modes (see Appendix \Cref{sec:appendix-technical-problems} for full tabulations). Significant mode differences emerged for giving answers (text: 79.4\%, voice: 55.6\%) and receiving questions (text: 83.1\%, voice: 72.4\%), where respondents in voice modes reported problems more frequently. Note that this analysis includes the full dataset of all respondents, including respondents who could not be linked successfully or who discontinued the interview early.

Although the majority of respondents reported neither minor nor major problems, we conducted a review of verbatim open-ended responses from respondents who indicated problems. Two recurring themes emerged. The first concerned implementation and usability issues: respondents cited confusion about the unique identifier entry procedure at the outset of the interview, as well as interface rendering failures on certain devices. The second concerned interviewer behavior and was concentrated in the voice condition, where respondents reported that AI response latency was either too short or too long, disrupting the conversational flow. Together, these findings identify concrete targets for improvement, particularly regarding interface robustness and participant onboarding, while confirming that the system functioned without meaningful difficulty for the large majority of respondents.

Another crucial aspect is that the interviewer does not engage in harmful or inappropriate behavior. A concern that is especially salient given that LLM outputs are not fully constrained by the researchers' prompting instructions and thus carry some residual behavioral risk. To address this, we implemented a monitoring mechanism enabling respondents to flag behavior they perceived as offensive or inappropriate. Consistent with our pre-registered success criterion, fewer than 1\% of respondents reported any such incidents. Those who did were offered direct contact with the principal investigator (an opportunity none of the respondents used). These findings suggest that, when appropriate safeguards and prompting protocols are in place, the risk of harmful interviewer behavior can be kept at negligible levels.

\subsection{Respondent evaluation of interviewer behavior}

We now turn to investigating how respondents evaluated AI Conversational Interviewing. We pre-registered the success criterion that respondents would evaluate the AI interviewer positively, operationalized as mean ratings above the scale midpoint on relevant attributes. The data support this expectation. Respondents rated the AI interviewer as polite, motivating, clear, and impartial, with all four attributes exceeding the scale midpoint. Ratings for compassion, while comparatively lower, likewise remained above the midpoint. Notably, these positive evaluations emerged despite full respondent awareness of interacting with an artificial agent, suggesting that the system was capable of projecting qualities associated with effective interviewing practice in semi-structured settings \parencite{newcomer_conducting_2015,helfferich_qualitat_2011}.

\begin{figure}[H]
  \centering
  \includegraphics[width=\textwidth]{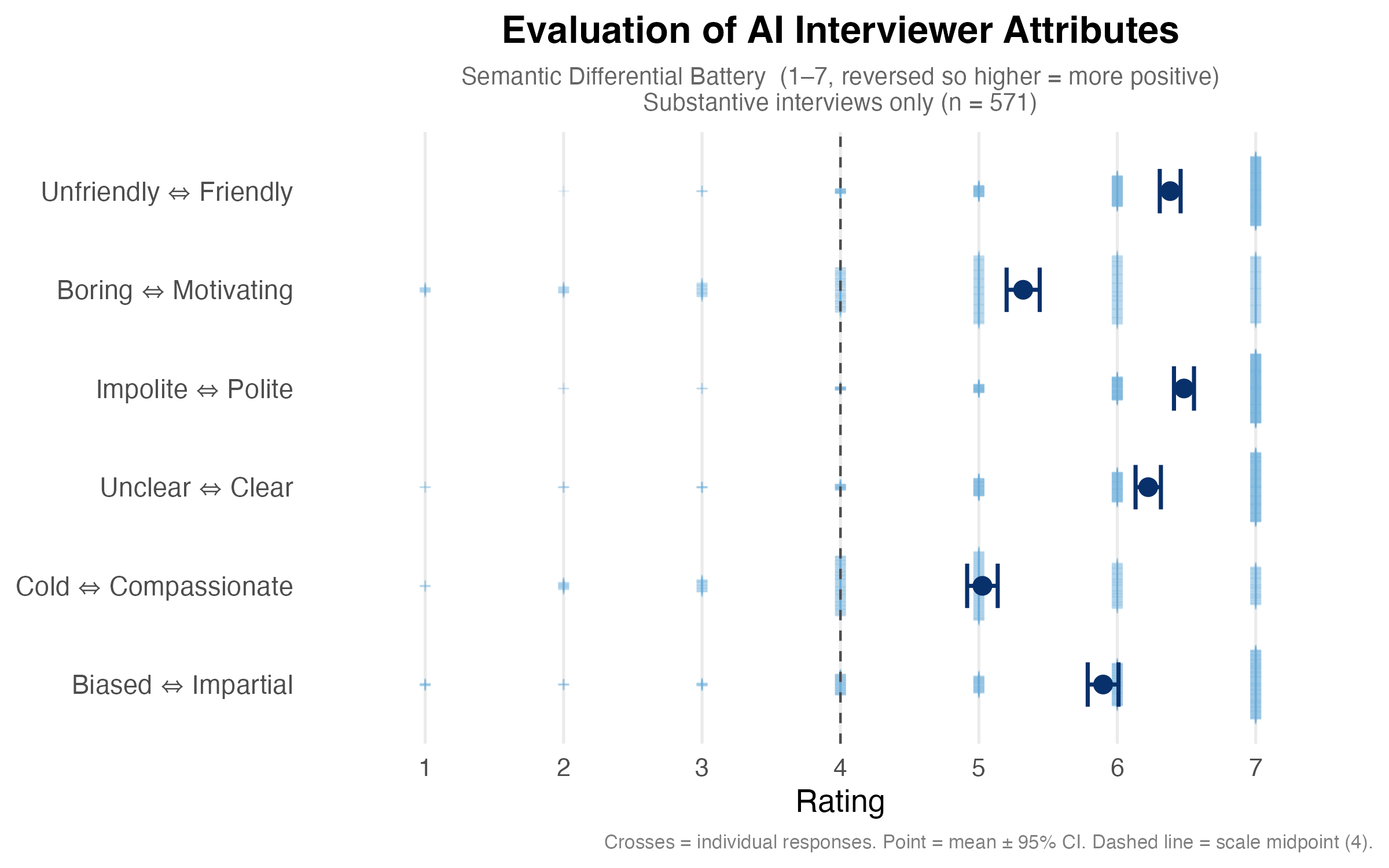}
  \caption{Evaluation of the AI conversational interviewer.}
  \label{fig:interviewer-evaluation}
\end{figure}

\begin{figure}[H]
  \centering
\includegraphics[width = \textwidth]{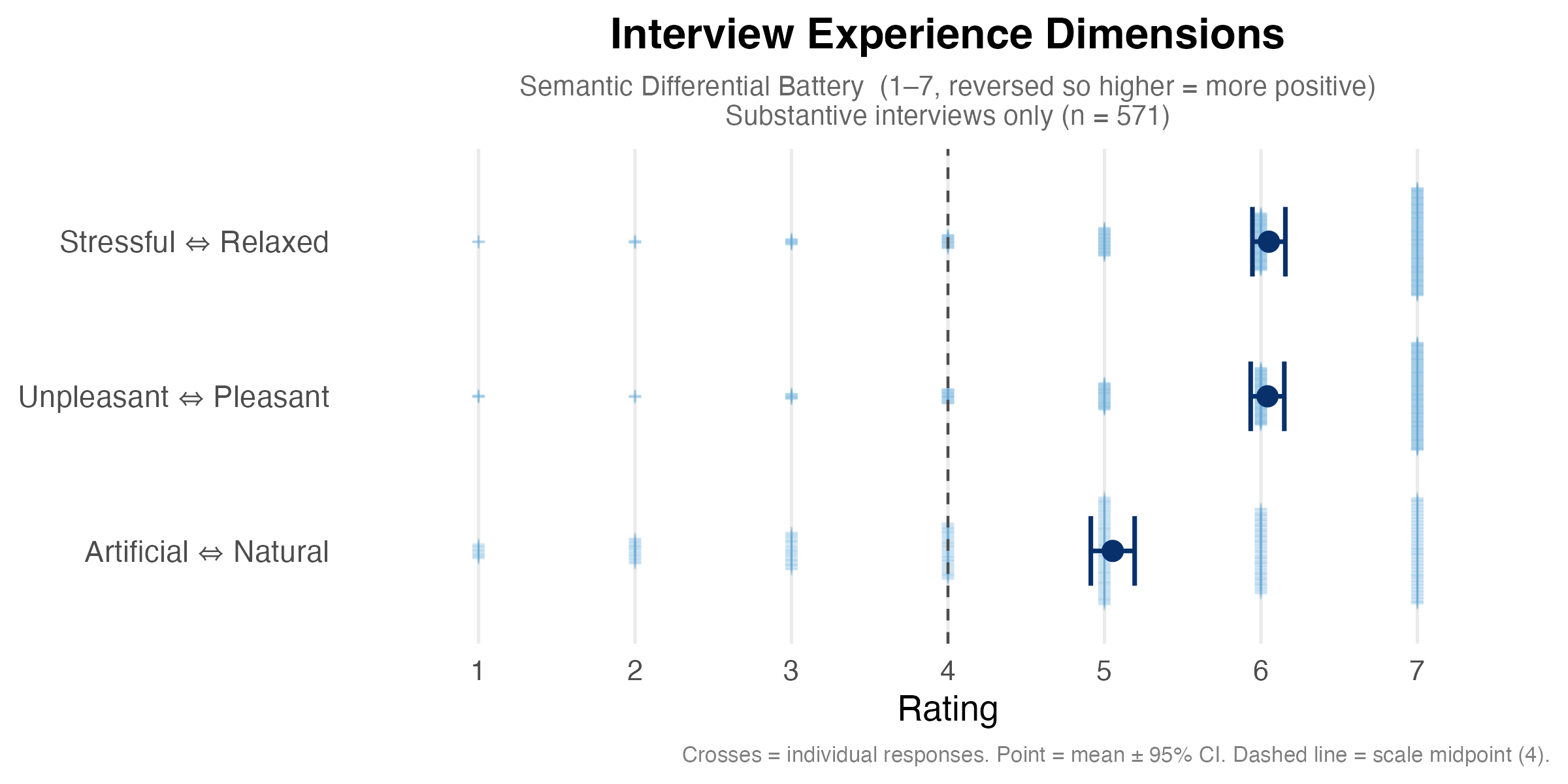}
  \caption{Evaluation of AI Conversational Interviewing.}
  \label{fig:ai-experience}
\end{figure}

Figure \ref{fig:ai-experience} presents respondent evaluations of the interview process. On average, participants experienced the AI-mediated conversation as relaxed, pleasant, and natural, again pointing to an overall positive interview experience. 

\subsection{Comparing evaluations of AI Conversational Interviewing with standardized surveys}

\begin{figure}[H]
  \centering
  \includegraphics[width=\textwidth]{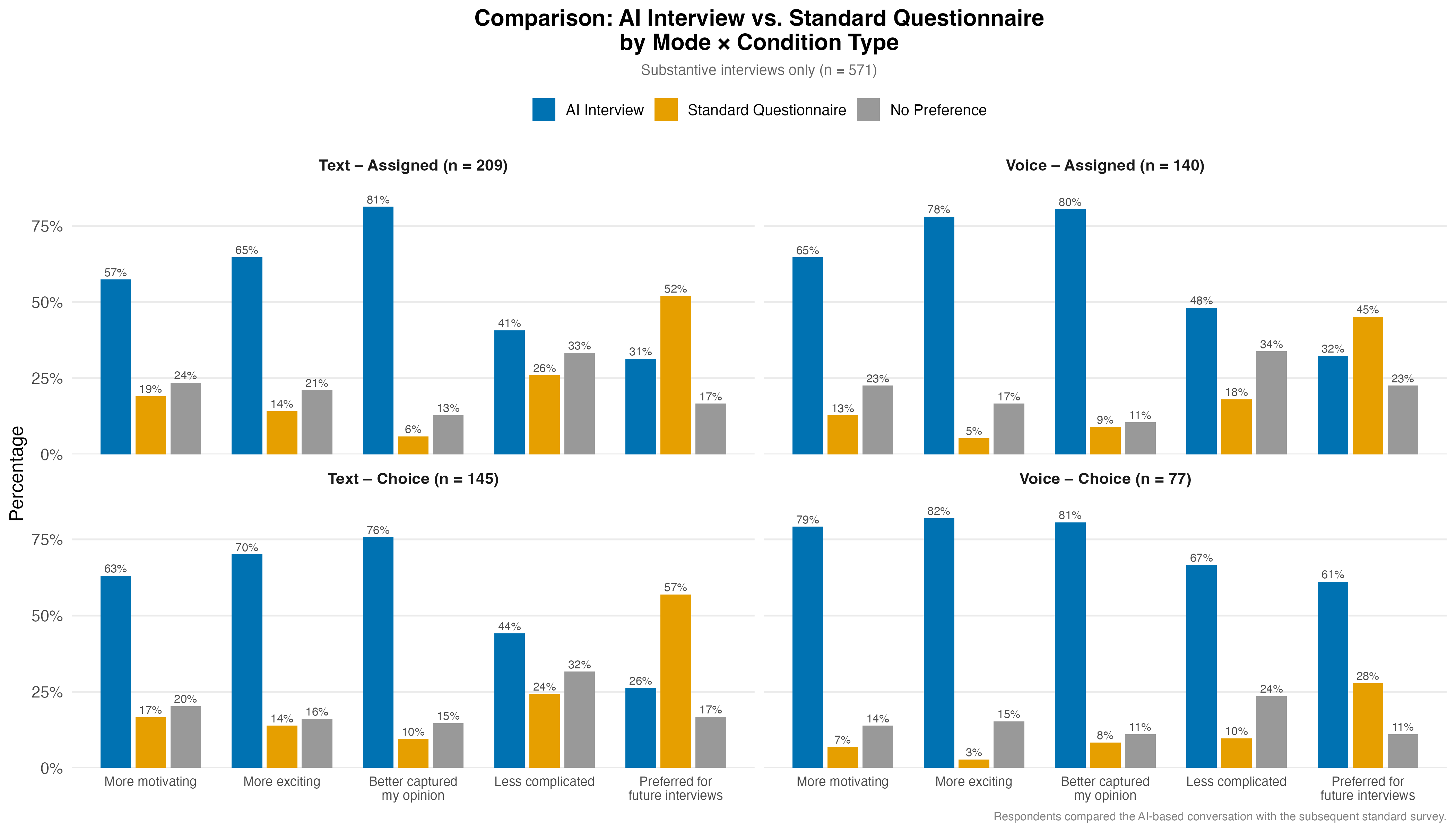}
  \caption{Comparison of AI Conversational Interviewing and the standardized survey.}
  \label{fig:comparison}
\end{figure}

How does this experience compare with participating in a standardized survey? Our pre-registered success criterion specified that evaluations of the AI interview should not fall substantially below evaluations of the subsequent standardized survey. Across both text and voice modalities, respondents rated the AI interview as more motivating and engaging than the standardized survey (Figure \ref{fig:comparison}).  Consistent with one of the central theoretical advantages attributed to AI Conversational Interviewing, the large majority of respondents also reported that the conversational format captured their individual views much better.  Across most evaluative dimensions and both modalities, the conversational interview met the pre-registered success criterion, with ratings at or above those observed for the standardized survey. We thus conclude that respondents who completed an interview accepted and in several respects appreciated the format — while keeping in mind that the reported differential completion across conditions is itself a substantive finding about the method's current viability, which we address below. 

\begin{figure}[H]
  \centering
  \includegraphics[width=\textwidth]{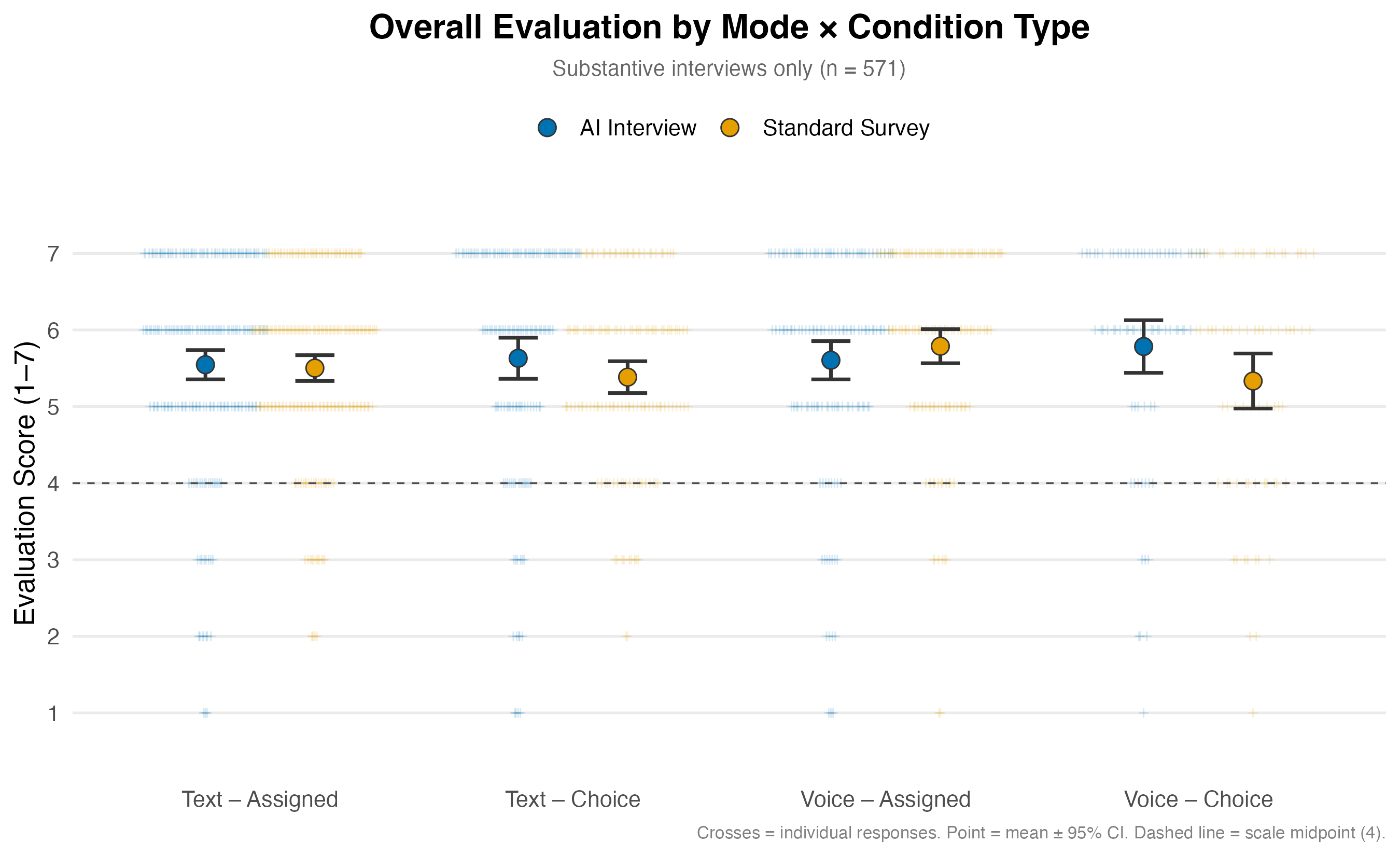}
  \caption{Overall evaluation across interview modes.}
  \label{fig:overall-evaluation}
\end{figure}

There is one exception that merits closer attention as it stands in partial tension with all the other previously reported  favorable ratings of the conversational interview: many respondents nonetheless expressed an overall preference for the standardized survey format for the next round of data collection  despite the conversational interviewing receiving ratings  consistently on par with the standardized survey on most recorded attributes   (Figure \ref{fig:overall-evaluation}). Interestingly, the preference for the standardized surveys was most pronounced among respondents who completed their interviews via chat. In contrast, respondents who voluntarily chose the voice format report a strong preference for the AI Conversational Interview. We return to this issue in the next section.

\subsection{Interview duration and content}

Turning from self-reported evaluations to the textual data generated by the conversational interviews, Table~\ref{tab:descriptives} presents descriptive summary statistics disaggregated by interview mode. On average, interviews lasted approximately 10 minutes and yielded several hundred words per respondent. However, considerable variation around these averages indicates that the interview experience differed substantially across individuals. Notably, the text condition exhibits a small number of unusually long sessions, likely reflecting respondents who paused during the task or required additional time to compose written responses. 
Although self-reported satisfaction was broadly comparable across modalities, voice and text interviews exhibit markedly different response characteristics. Voice interviews generated approximately twice as many words per respondent as text interviews (mean = 608.8 vs. 299.8), at a pace nearly 2.5 times faster (52.5 vs. 21.1 words per minute). Voice session durations cluster tightly around 9–10 minutes (mean = 8.3, median = 9.6, SD = 5.2), whereas text sessions were somewhat longer on average and roughly twice as variable (mean = 11.5, median = 9.9, SD = 9.9). Linguistically, voice responses feature shorter sentences (17.9 vs. 29.3 words) and more accessible language (Flesch Reading Ease: 61.2 vs. 40.4; Wiener Sachtextformel: 7.2 vs. 11.0), a pattern consistent with the well-documented distinction between spoken and written registers \parencite{gavras_innovating_2022, atkeson_nonresponse_2014}. We found that voice responses are more paratactic and conversational, whereas typed responses exhibit greater syntactic density and lexical complexity. The substantially wider dispersion in readability scores for text respondents (Flesch SD: 9.2 vs. 40.4; Wiener SD: 1.8 vs. 5.9) further points to greater stylistic heterogeneity in the text condition. In sum, the conversation modes generate qualitatively distinct textual data, while text responses induce deeper elaboration and greater cognitive effort, voice interviews generate substantially more verbose responses overall. 

\begin{table}[H]
  \centering
  \caption{Descriptive statistics by interview modality. Word-based metrics use the analysis cohort ($N = 568$, $n_\text{voice} = 214$, $n_\text{text} = 354$). Duration and Words/min use the R-equivalent cohort (no word-count filter, 99th-percentile trim; $n_\text{voice} = 282$, $n_\text{text} = 435$).}
  \label{tab:descriptives}
  \small
  \begin{tabular}{l r r r r r r}
    \toprule
    & \multicolumn{3}{c}{Voice} & \multicolumn{3}{c}{Text} \\
    \cmidrule(lr){2-4} \cmidrule(lr){5-7}
    Measure & $M$ & $\mathit{Mdn}$ & $\mathit{SD}$ & $M$ & $\mathit{Mdn}$ & $\mathit{SD}$ \\
    \midrule
    Words                       & 608.8  & 555.5  & 377.6  & 299.8  & 222.0  & 931.2 \\
    Characters                  & 3902.9 & 3453.0 & 2396.7 & 2036.6 & 1508.0 & 6161.4 \\
    Avg.\ words / sentence      & 17.9   & 17.3   & 5.9    & 29.3   & 19.0   & 33.9 \\
    Flesch Reading Ease         & 61.2   & 61.3   & 9.2    & 40.4   & 49.4   & 34.8 \\
    Wiener Sachtextformel       & 7.2    & 7.3    & 1.8    & 11.0   & 9.7    & 5.9 \\
    \midrule
    Duration (min)$^{\dagger}$  & 8.3    & 9.6    & 5.2    & 11.5   & 9.9    & 9.9 \\
    Words / min$^{\dagger}$     & 52.5   & 52.2   & 22.8   & 21.1   & 18.0   & 36.8 \\
    \bottomrule
  \end{tabular}
  \par\smallskip
\end{table}

These patterns also bear on the finding, noted above, that many respondents in the text mode expressed a preference for standardized surveys over conversational interviews despite broadly positive evaluations of the AI interview. Two complementary explanations emerge from the data. First, this preference plausibly reflects the substantially higher cognitive effort associated with typed responding. The factor effort in the text mode is corroborated by an exploratory analysis showing that response length in the chat condition declines progressively after approximately the tenth interaction turn (much more than in the voice condition), indicating relevant fatigue in the chat interviews (see Appendix \Cref{sec:appendix-turn-statistic}). Second, experimental assignments seems an important factor for the preference pattern in the voice condition. Because a substantial share of respondents preferred chat when given the option (see Figure \ref{fig:consort}), participants involuntarily assigned to the voice condition were more likely to express a preference for standardized surveys. By contrast, respondents who voluntarily selected the voice mode when given the choice — and who therefore presumably felt comfortable and competent with voice-based interaction on their device — overwhelmingly preferred conversational over standardized interviews. We interpret these patterns as underscoring the importance of giving users a choice. While voice mode seems less tiring for respondents, many (but not all) respondents prefer the more accessible and familiar text mode despite requiring more effort. 

\subsection{Demonstrating the analytical value of conversational data}
This section illustrates how the complementary strengths of standardized surveys and conversational data can be leveraged within an integrated analytical framework. Standardized survey measures are well-suited to capturing attitudinal positions that, under well-understood measurement assumptions, are comparable across groups and over time. To illustrate, our battery of standardized items on migration policy includes the widely used GLES semantic differential, in which respondents place themselves on an 11-point scale anchored at ``facilitating immigration'' (1) and ``restricting immigration'' (11). Results indicate that \textit{AfD} supporters hold the most restrictive attitudes (M = 10.3), followed by \textit{CDU/CSU} (M = 7.5), \textit{FDP} (M = 6.6), and \textit{SPD} supporters (M = 5.2), while supporters of \textit{Die Linke} report the most permissive positions (M = 2.5; see Appendix \Cref{sec:appendix-standardized-survey}).

Standardized surveys are, however, not well suited to capturing the mental models and unprompted considerations respondents associate with a given topic, or to revealing the interpretive frameworks through which they make sense of it. We therefore employ topic analysis to uncover the considerations that spontaneously come to mind when respondents think about migration, and framing analysis to examine the reasoning processes that structure their responses.

\begin{figure}[H]
  \centering
  \includegraphics[width=1\textwidth]{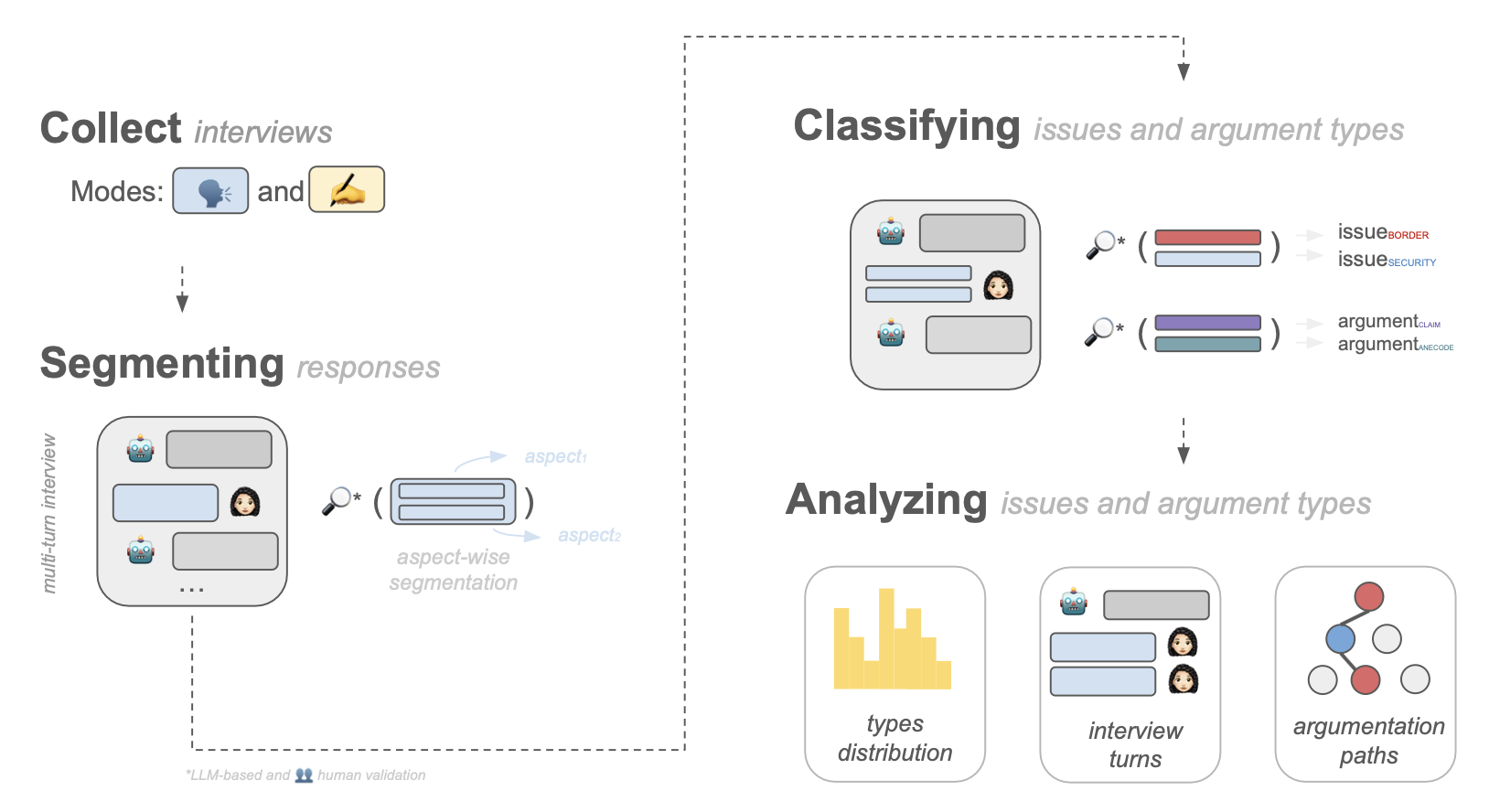}
    \caption{Analytical framework for extracting issue repertoires and argument structures from conversational interviews.}
  \label{fig:framework}
\end{figure}

Specifically, we conducted a targeted issue analysis of the conversational transcript corpus, focusing on migration-related content. Transcripts were segmented into $16{,}034$ \textit{information units} (IUs; $M = 27.2$ per interview, $\mathit{Mdn} = 23.0$), each classified along two dimensions: issue topic and argument type (Figure \ref{fig:framework}). 

Both analyses are stratified by voter group, operationalized using respondents' party identification as recorded in the standardized survey component. This choice itself illustrates an advantage of the integrated design: the two data types not only yield complementary substantive findings when analyzed separately, but can also be combined directly within a single analytical framework.

The classification draws on two taxonomies developed for this study: 12 topical areas (e.g., housing, labor market, humanitarian concerns) and 10 rhetorical frames (e.g., evidence: anecdotal; evidence: testimonial; reasoning: causal; assertion: claim), both described in detail in Appendix \Cref{sec:appendix-text-analysis}.

To classify each information unit along both dimensions, we employed a human-validated LLM annotation procedure. In a first step, an LLM-based pass generated candidate segmentations and provisional labels. In a second step, trained researchers validated these outputs against a structured codebook. To assess the reliability of the annotation procedure, we calculated intercoder agreement across both the segmentation and classification stages. Results indicate high agreement across all coding tasks (see Appendix \Cref{sec:appendix-text-analysis}).

\subsection{Topic analysis}
\begin{figure}[H]
  \centering
  \includegraphics[width=1\textwidth]{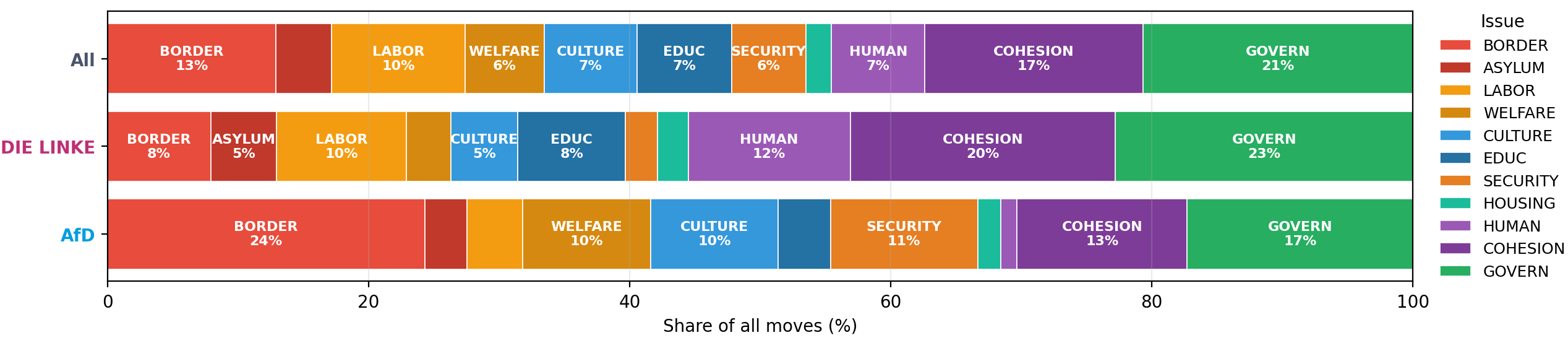}
  \caption{Issue topic distribution by party identification across \emph{all} moves in the collapsed sequences (consecutive duplicates and the residual \textsc{other} category excluded). Each bar sums to 100\%; segment labels indicate the within-party share of each issue.}
  \label{fig:party-issue-distribution}
\end{figure}

Figure \ref{fig:party-issue-distribution} reveals substantial variation in the topical considerations that come to mind when different voter groups discuss migration (see Appendix \Cref{sec:appendix-text-analysis-voters} for further evidence). While the standardized survey data document differences in the \textit{level} of immigration attitudes across partisan groups, the analysis of the textual data shows that supporters also differ markedly in \textit{how} they conceptualize the issue  (see Appendix \Cref{sec:appendix-text-analysis-voters} for results on all party groups). Descriptively, \textit{Die~Linke} supporters are considerably more likely to frame migration in terms of the labor market, social cohesion, and human rights, whereas \textit{AfD} supporters disproportionately invoke border security. Only 1\% of \textit{AfD} segments address humanitarian considerations, compared to 12\% among \textit{Die~Linke} supporters.\footnote{We present these contrasts as exploratory illustrations of the method's analytical possibilities rather than as confirmatory population estimates.} The distinction is thus not only one of more or less favorable attitudes toward immigration, but one of qualitatively different mental models of the issue. 
 
Notably, this divergence extends to party supporters whose standardized immigration attitude scores are comparatively close. \textit{SPD} supporters, who score only marginally more favorable toward immigration than \textit{FDP} supporters (5.2 vs. 6.5 on the 11-point scale), nevertheless differ substantially in topical emphasis: societal cohesion accounts for 19\% of \textit{SPD} segments, compared to just 9\% among \textit{FDP} supporters.

These differences in underlying mental models are further illustrated by the sequence in which migration-related topics are raised (see Appendix \Cref{sec:appendix-sequence-analysis}). \textit{FDP} supporters typically open their elaborations with labor market considerations, whereas \textit{Die~Linke} supporters lead with humanitarian concerns before transitioning to issues of social cohesion. Which consideration first come to  mind is typically understood as an indicator of saliency and accessibility, which is closely related to attitude importance \parencite{krosnick1989attitude}. Hence, the potential of textual opinion data to analyze the sequence in which respondents express considerations is yet another analytical avenue for mapping individual belief systems. 

\subsection{Frame analysis}

Beyond topical content, we examine the rhetorical structure of respondents' reasoning through argument type analysis \parencite{soltani-romberg-2023-general, wachsmuth-etal-2024-argument}, treating systematic variation in rhetorical signatures as a window into how different voter groups think about and justify their positions on migration.

Figure \ref{fig:argtype-distribution} presents the distribution of argument types across partisan groups (see Appendix \Cref{sec:appendix-text-analysis} for details on the taxonomy of argument types). A notable commonality across all groups is the predominance of assertion-based reasoning: respondents overwhelmingly express judgments and posit claims instead of recounting anecdotes, citing evidence or testimonials. The two assertion types — \textsc{claim} and \textsc{eval} — account for roughly half of all argumentative moves in each group (51–58\%), while references to evidence remain consistently rare across parties (\textsc{anecd}: 9–11\%; \textsc{testim}: $<$3\%; \textsc{stat}: $<$2\%).

Some differences do emerge across parties. \textit{AfD} supporters record the highest share of evaluative assertions (\textsc{eval} 33\%) while exhibiting the weakest reliance on causal reasoning  (\textsc{cause} 12\%). \textit{Die~Linke} supporters show the inverse profile: a comparatively low claim share  (\textsc{claim}: 23\%) alongside the highest causal reasoning share in the sample (\textsc{Cause}: 21\%), consistent with an argumentative style oriented toward explanation rather than prescription.

Taken together, these analyses demonstrate how conversational interview illuminates what is on people's minds, what considerations structure their thinking and which rhetorical frameworks they use to reason about and justify political attitudes.

\begin{figure}[H]
  \centering
  \includegraphics[width=1\textwidth]{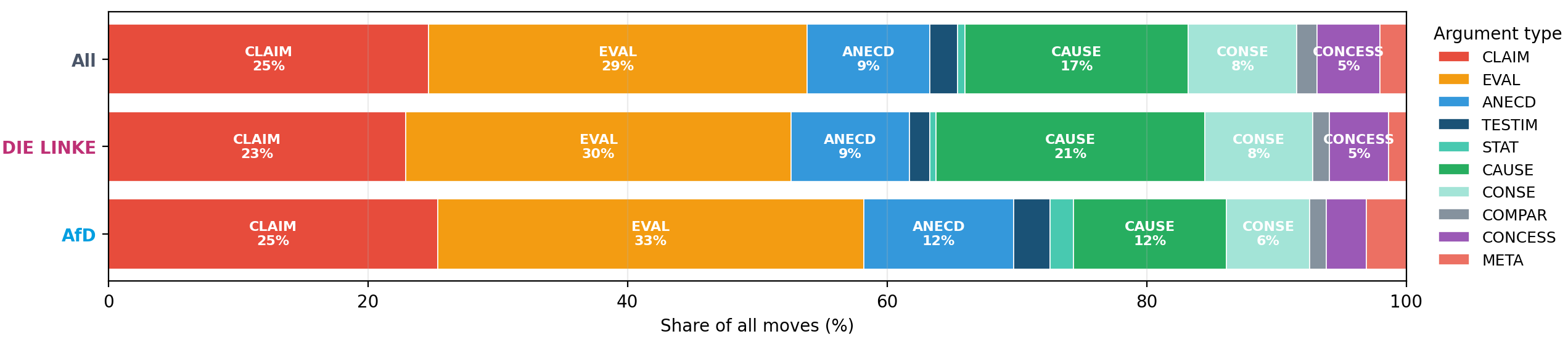}
  \caption{Argument-type distribution by party identification}
  \label{fig:argtype-distribution}
\end{figure}

\section{Discussion}

This study asks whether large language models can conduct semi-structured interviews at the scale of standardized survey research, and what such interviews add to the public-opinion toolkit. Three findings anchor our conclusions. First, the collected textual public opinion data surface considerations, arguments, and reasoning structures that a comprehensive standardized battery leaves uncaptured, enabling new avenues for research. Second, respondents judge AI-led interviews to be broadly comparable to a conventional survey, and more engaging on several dimensions. Third, respondents evaluated AI Conversational Interviewing positively in both text and voice modes but each mode elicits distinct response styles which results in qualitatively different data. We discuss each in turn.

\subsection*{Acceptance and viability}

The method was received well by respondents, satisfying our pre-registered success criteria. Respondents rated the AI interviewer above the scale midpoint on all attributes, including friendliness, clarity, impartiality, and motivation. These positive evaluations held despite respondents' full awareness that they were interacting with an artificial agent, undercutting the concern that AI-mediated interviews would be dismissed as impersonal or illegitimate.

\subsection*{Mode effects}

On subjective evaluations, voice, text, and choice conditions produced broadly similar results but mode effects affected respondents' preference for future survey modes. Despite  favorable evaluations of AI Conversational Interviewing many respondents still expressed a preference for standardized survey after having experienced both data collection modes, particularly if they participated in the effortful text-based interview. This finding emphasizes the importance of respondent burden in navigating future implementation of AI Conversational Interviewing. The choice condition showed that many respondents prefer chat when given control, while those who voluntarily selected voice were after the interview especially favorable toward conversational interviewing. Voice therefore appears promising for respondents who are comfortable with it, but risky as the only mode. Text chat, on the other hand, has a lower technical barrier but seems particularly prone to generate fatigue  over the course of longer interviews, manifesting in ever shorter messages.

The mode of conversation also influenced what data is generated. Voice interviews are full-duplex interactions that generated substantially longer responses at a faster pace and in a more conversational style. Text interviews produced shorter, more effortful, and often more syntactically dense responses. 

One caveat to these findings is that the mode effects reported here bundle multiple differences (input, output, full/half duplex, LLM model, interface) and our design cannot clearly identify which of these differences drives the effects. Nonetheless, we interpret these findings to show that one mode is not clearly better than the other. Instead, conversation mode has multi-dimensional and cross-cutting effects on respondent experience and  data quality.

\subsection*{The analytical value of conversational data}

The conversational data revealed that respondents organized their thinking about migration around governance failure, societal cohesion, and border control more often than around well-researched topics like cultural-threat perceptions. Moreover, the conversational data show that similar scale positions can rest on different mental models. Respondents did not merely vary in how restrictive or permissive they were toward immigration, but rather in how they make sense of an issue. Partisans with similar attitude-scale scores approach the topic in different ways, differing in which considerations were salient, in the sequence in which issues arose, and in the rhetorical structure of their justifications.

\subsection*{From Proof of Concept to Operational Use}

Taken together, these findings suggest that AI Conversational Interviewing can serve as a viable instrument for use in substantive social inquiry. Yet, for use in established probability-based survey programs such as the ANES important but solvable issues would need to be addressed in order to mitigate the substantial attrition observed in this study. These improvements include data linkage procedures that require no user input, greater technical robustness across heterogeneous devices, and an intuitive interface for less technically sophisticated users.

\subsection*{Limitations of this Study}

First, the study draws on respondents from online access panels. These respondents are experienced with digital survey participation and may be more comfortable with novel online interfaces than the broader population. Replication in probability-based samples and among respondents with lower digital familiarity is needed before AI Conversational Interviewing can be recommended for the full range of population surveys.   
Second, the study focuses on one salient political issue in one national context. Migration in Germany is a topic on which many respondents have available considerations. It remains open whether similarly rich data can be collected on low-salience, technical, or highly sensitive topics. 
Third, the analytic sample was reduced by practical implementation problems which potentially introduced biases in the sample. 
Fourth, the fixed ordering of components means that the standardized survey was completed in a primed state; future designs should randomize component order to estimate both the size of this priming effect and order-free format evaluations.

\section{Outlook}

\subsection*{Integrating conversational and standardized measurement}

AI Conversational Interviewing is most valuable when researchers want to understand how respondents mentally structure an issue, which considerations they spontaneously invoke, and how they justify or connect their views. It is less likely to be the most efficient tool when the goal is to measure discrete facts, estimate marginal distributions, or track a small number of well-established attitudes over time. Standardized surveys remain superior for many such purposes.  We see AI Conversational Interviewing as a complement to standardized surveys rather than a competitor, and the two are most powerful in combination, as in the present study: surveys measure where respondents stand on shared scales, while conversational interviews reveal what those positions mean in the eyes of the respondent. 

Our own implementation placed the conversational and standardized components side by side as distinct elements, but other designs are possible. In sequential designs, AI interviews can inform questionnaire development by mapping the space of considerations before researchers construct closed-ended batteries. In adaptive designs, prior survey responses can inform individualized interview probes, allowing researchers to explore why respondents selected particular response options. In reverse, live classification of interview data via LLMs can inform a subsequent standardized survey question \parencite{Poulsen2026VisualNetwork}. Instead of administering full interviews and surveys one after the other, hybrid approaches could mix both types on an item-level basis such as a standardized item followed by a short conversational probe. Each of these designs would not dissolve the distinction between surveys and interviews, but would allow researchers to exploit the strengths of each.

\subsection*{A research agenda for implementation}

AI Conversational Interviewing opens a research agenda rather than closing a methodological debate.  Decades of survey-methodological work, much of it in the pages of Public Opinion Quarterly,  established how design choices shape data quality, response rates, and respondent satisfaction in standardized surveys \parencite{schober_does_1997, poq1, poq2, poq3, poq4}.  AI Conversational Interviewing now requires a comparable program of methodological research to determine best practices \parencite{ivey2026makesgoodresponseempirical}, potentially building on existing knowledge for conducting qualitative interviews \parencite{flick_doing_2022,rubin_qualitative_2005,seidman_interviewing_2019}. Open questions include the optimal length of interviews, the number and phrasing of follow-up probes. 

A related second strand of open questions concerns technical issues, including back-end implementation  (e.g., single vs multi-agent systems, model choice, prompting strategies, turn-taking detection in voice agents) and front-end implementation (e.g., interface design). 

A third area of open questions relates to the effects of the artificial character of the interviewer on the interview (e.g. model biases, interviewer effects).

A fourth strand concerns open questions of how to analyze textual public opinion data. Existing text-as-data approaches offer a promising starting point \parencite{Grimmer2022TextAsData} but the emergent area of computational grounded theory \parencite{qual1, qual2, doi:qual3, qual4, qual5, qual6, qual7} may be particularly productive for inductive exploration of the novel considerations, arguments and beliefs people express in conversational interviews. The goal might not be to automate interpretation entirely, but to combine scalable classification with close reading, transparent codebooks, and validation procedures.

In this sense, the conversational capabilities of LLMs seem to open a phase of rapid innovation reminiscent of earlier inflection points in twentieth-century survey research, from probability sampling to web administration.

\subsection*{Open science, privacy, and ethics}

This research agenda should take reproducibility, privacy, and ethics seriously from the outset \parencite{DunleavyMonteath2026}. Our pipeline relied on proprietary closed models, which currently outperform open alternatives; open models confer important advantages for reproducibility, data privacy and transparency \parencite{spirling2023open}, and have already reached acceptable quality for chat-based interviewing, with voice-based open systems following later. Where proprietary models remain in use, data-residency arrangements that keep transcripts within researcher-controlled infrastructure should become standard. With respect to software pipelines for data collection, practitioners can rely on emerging commercial vendors, but researcher-led open-source frameworks such as OASIS\footnote{\url{https://oasis-surveys.github.io/}} are especially promising, because they avoid infrastructure dependency, may reduce costs, and permit the oversight, verification, and modification that scholarly use demands.

Because unstructured conversational data can be sensitive and personal even when researchers do not anticipate it, the ethical burden exceeds that of closed-ended surveys. Consent procedures should disclose not only that respondents are interacting with an AI system, but also what data are stored, whether transcripts are reviewed by humans, whether external vendors process them, within which jurisdiction, and how sensitive disclosures will be handled. Another challenge is how to enable data reuse and replication when privacy concerns impede the sharing of raw text data \parencite{campbell2023open}. Yet, recent research has demonstrated various avenues for enabling reuse and replication even for qualitative and sensitive textual data \parencite{Kapiszewski_Karcher_2021, qualdata, qualshare, walsh2025responsible, campbell2023open, osti_10319965}.

\subsection*{A period of innovation and experimentation}

The present study establishes that the fundamental feasibility conditions of AI Conversational Interviewing are met; the open questions concern scope, integration with established approaches, and the design choices that determine interview quality. An exciting and productive period of methodological development lies ahead.

\newpage

\printbibliography[title={References}]

@book{Grimmer2022TextAsData,
  author    = {Justin Grimmer and Margaret E. Roberts and Brandon M. Stewart},
  title     = {Text as Data: A New Framework for Machine Learning and the Social Sciences},
  publisher = {Princeton University Press},
  address   = {Princeton, NJ},
  year      = {2022},
  isbn      = {9780691207544},
  url       = {https://press.princeton.edu/books/hardcover/9780691207544/text-as-data}
}

@article{xiao_tell_2020,
	title = {Tell {Me} {About} {Yourself}: {Using} an {AI}-{Powered} {Chatbot} to {Conduct} {Conversational} {Surveys} with {Open}-ended {Questions}},
	volume = {27},
	issn = {1073-0516, 1557-7325},
	shorttitle = {Tell {Me} {About} {Yourself}},
	url = {https://dl.acm.org/doi/10.1145/3381804},
	doi = {10.1145/3381804},
	language = {en},
	number = {3},
	urldate = {2023-08-09},
	journal = {ACM Transactions on Computer-Human Interaction},
	author = {Xiao, Ziang and Zhou, Michelle X. and Liao, Q. Vera and Mark, Gloria and Chi, Changyan and Chen, Wenxi and Yang, Huahai},
	month = jun,
	year = {2020},
	pages = {1--37},
}

@article{Poulsen2026VisualNetwork,
  author       = {Victor Poulsen and Peter Steiglechner and Henrik Olsson and Mirta Galesic},
  title        = {Visual Network Tool: Individual Belief Networks from LLM-Guided Interviews and a Visual Canvas},
  year         = {2026},
  journal      = {PsyArXiv},
  publisher    = {Center for Open Science},
  doi          = {10.31234/osf.io/ypfz6},
  url          = {https://osf.io/preprints/psyarxiv/ypfz6},
  note         = {Preprint}
}

@article{schwarz,
    author = {Schwarz, Norbert},
    title = {WHAT RESPONDENTS LEARN FROM SCALES: THE INFORMATIVE FUNCTIONS OF RESPONSE ALTERNATIVES},
    journal = {International Journal of Public Opinion Research},
    volume = {2},
    number = {3},
    pages = {274-285},
    year = {1990},
    month = {10},
    issn = {0954-2892},
    doi = {10.1093/ijpor/2.3.274},
    url = {https://doi.org/10.1093/ijpor/2.3.274},
    eprint = {https://academic.oup.com/ijpor/article-pdf/2/3/274/1895195/2-3-274.pdf},
}

@inproceedings{vonDerHeyde2025WhoCounts,
  author    = {von der Heyde, Leah},
  title     = {Who Counts? The Potentials and Pitfalls of Using {LLMs} in Survey Research},
  booktitle = {Proceedings of the First Workshop on Bridging {NLP} and Public Opinion Research ({NLPOR} @ {COLM} 2025)},
  year      = {2025},
  url       = {https://openreview.net/forum?id=ww2KqnPLdK},
  note      = {Published as a conference paper at COLM 2025}
}

@article{fictitious,
    author = {Bishop, George F. and Tuchfarber, Alfred J. and Oldendick, Robert W.},
    title = {Opinions on Fictitious Issues: The Pressure to Answer Survey Questions},
    journal = {Public Opinion Quarterly},
    volume = {50},
    number = {2},
    pages = {240-250},
    year = {1986},
    month = {01},
    issn = {0033-362X},
    doi = {10.1086/268978},
    url = {https://doi.org/10.1086/268978},
    eprint = {https://academic.oup.com/poq/article-pdf/50/2/240/5388518/50-2-240.pdf},
}

@article{Schwarz1999SelfReports,
  author    = {Norbert Schwarz},
  title     = {Self-reports: How the Questions Shape the Answers},
  journal   = {American Psychologist},
  year      = {1999},
  volume     = {54},
  number     = {2},
  pages      = {93--105},
  doi        = {10.1037/0003-066X.54.2.93},
  publisher  = {American Psychological Association}
}

@book{DunleavyMonteath2026,
  author    = {Dunleavy, Patrick and Monteath, Timothy},
  title     = {Doing Open Social Science: A Guide for Researchers},
  publisher = {LSE Press},
  address   = {London},
  year      = {2026},
  doi       = {10.31389/lsepress.dos},
  url       = {https://doi.org/10.31389/lsepress.dos}
}

@article{campbell2023open,
  title={Open-science guidance for qualitative research: An empirically validated approach for de-identifying sensitive narrative data},
  author={Campbell, Rebecca and Javorka, McKenzie and Engleton, Jasmine and Fishwick, Kathryn and Gregory, Katie and Goodman-Williams, Rachael},
  journal={Advances in Methods and Practices in Psychological Science},
  volume={6},
  number={4},
  pages={25152459231205832},
  year={2023},
  publisher={Sage Publications Sage CA: Los Angeles, CA}
}

@article{qualdata,
author = {James M. DuBois  and Jessica Mozersky  and Meredith Parsons  and Heidi A. Walsh  and Annie Friedrich  and Amy Pienta },
title = {Exchanging words: Engaging the challenges of sharing qualitative research data},
journal = {Proceedings of the National Academy of Sciences},
volume = {120},
number = {43},
pages = {e2206981120},
year = {2023},
doi = {10.1073/pnas.2206981120},
URL = {https://www.pnas.org/doi/abs/10.1073/pnas.2206981120},
eprint = {https://www.pnas.org/doi/pdf/10.1073/pnas.2206981120}}

@article{walsh2025responsible,
  title={Responsible sharing of qualitative research data: Insights from a pioneering project in the United States},
  author={Walsh, Heidi A and Parsons, Meredith V and Mozersky, Jessica and Gupta, Aditi and Lai, Albert M and Friedrich, Annie B and DuBois, James M},
  journal={International journal of qualitative methods},
  volume={24},
  pages={16094069251329607},
  year={2025},
  publisher={SAGE Publications Sage CA: Los Angeles, CA}
}

@article{qualshare,
    author = {Gupta, Aditi and Lai, Albert and Mozersky, Jessica and Ma, Xiaoteng and Walsh, Heidi and DuBois, James M},
    title = {Enabling qualitative research data sharing using a natural language processing pipeline for deidentification: moving beyond HIPAA Safe Harbor identifiers},
    journal = {JAMIA Open},
    volume = {4},
    number = {3},
    pages = {ooab069},
    year = {2021},
    month = {07},
    issn = {2574-2531},
    doi = {10.1093/jamiaopen/ooab069},
    url = {https://doi.org/10.1093/jamiaopen/ooab069},
    eprint = {https://academic.oup.com/jamiaopen/article-pdf/4/3/ooab069/42112220/ooab069.pdf},
}

@article{osti_10319965, year = 2021, title = {How Data Curation Enables Epistemically Responsible Reuse of Qualitative Data}, url = {https://par.nsf.gov/biblio/10319965}, DOI = {10.46743/2160-3715/2021.5012}, abstractNote = {Data sharing and reuse are becoming the norm in quantitative research. At the same time, significant skepticism still accompanies the sharing and reuse of qualitative research data on both ethical and epistemological grounds. Nevertheless, there is growing interest in the reuse of qualitative data, as demonstrated by the range of contributions in this special issue. In this research note, we address epistemological critiques of reusing qualitative data and argue that careful curation of data can enable what we term “epistemologically responsible reuse” of qualitative data. We begin by briefly defining qualitative data and summarizing common epistemological objections to their shareability or usefulness for secondary analysis. We then introduce the concept of curation as enabling epistemologically responsible reuse and a potential way to address such objections. We discuss three recent trends that we believe are enhancing curatorial practices and thus expand the opportunities for responsible reuse: improvements in data management practices among researchers, the development of collaborative curation practices at repositories focused on qualitative data and technological advances that support sharing rich qualitative data. Using three examples of successful reuse of qualitative data, we illustrate the potential of these three trends to further improve the availability of reusable data projects.}, journal = {The Qualitative Report}, author = {Karcher, Sebastian and Kirilova, Dessislava and Pagé, Christiane and Weber, Nic} }

@article{Kapiszewski_Karcher_2021, title={Transparency in Practice in Qualitative Research}, volume={54}, DOI={10.1017/S1049096520000955}, number={2}, journal={PS: Political Science \& Politics}, author={Kapiszewski, Diana and Karcher, Sebastian}, year={2021}, pages={285--291}}

@article{spirling2023open,
  title={Why open-source generative AI models are an ethical way forward for science},
  author={Spirling, Arthur},
  journal={Nature},
  volume={616},
  number={7957},
  pages={413--413},
  year={2023},
  publisher={Nature}
}

@misc{glesZA10100,
author = "GLES",
title = "GLES Cross-Section 2025, Post-Election",
year = "2025",
howpublished = "(ZA10100; Version 2.0.0) [Data set]. GESIS, Cologne. https://doi.org/10.4232/5.ZA10100.2.0.0",
doi = "10.4232/5.ZA10100.2.0.0",
}

@article{gavras_innovating_2022,
	title = {Innovating the {Collection} of {Open}-{Ended} {Answers}: {The} {Linguistic} and {Content} {Characteristics} of {Written} and {Oral} {Answers} to {Political} {Attitude} {Questions}},
	volume = {185},
	issn = {0964-1998, 1467-985X},
	shorttitle = {Innovating the {Collection} of {Open}-{Ended} {Answers}},
	url = {https://academic.oup.com/jrsssa/article/185/3/872/7068903},
	doi = {10.1111/rssa.12807},
	language = {en},
	number = {3},
	urldate = {2023-08-09},
	journal = {Journal of the Royal Statistical Society Series A: Statistics in Society},
	author = {Gavras, Konstantin and Höhne, Jan Karem and Blom, Annelies G. and Schoen, Harald},
	month = jul,
	year = {2022},
	pages = {872--890},
}

@book{spradley_ethnographic_2016,
	address = {Long Grove, Illinois},
	title = {The ethnographic interview},
	isbn = {978-1-4786-3207-8},
	language = {eng},
	publisher = {Waveland Press, Inc},
	author = {Spradley, James P.},
	year = {2016},
}

@misc{chopra_conducting_2023,
	address = {Rochester, NY},
	type = {{SSRN} {Scholarly} {Paper}},
	title = {Conducting {Qualitative} {Interviews} with {AI}},
	url = {https://papers.ssrn.com/abstract=4572954},
	language = {en},
	urldate = {2023-09-23},
	author = {Chopra, Felix and Haaland, Ingar},
	month = sep,
	year = {2023},
}

@incollection{newcomer_conducting_2015,
	edition = {1},
	title = {Conducting {Semi}‐{Structured} {Interviews}},
	isbn = {978-1-118-89360-9 978-1-119-17138-6},
	url = {https://onlinelibrary.wiley.com/doi/10.1002/9781119171386.ch19},
	doi = {10.1002/9781119171386.ch19},
	language = {en},
	urldate = {2024-05-02},
	booktitle = {Handbook of {Practical} {Program} {Evaluation}},
	publisher = {Wiley},
	author = {Adams, William C.},
	editor = {Newcomer, Kathryn E. and Hatry, Harry P. and Wholey, Joseph S.},
	month = aug,
	year = {2015},
	pages = {492--505},
}

@article{atkeson_nonresponse_2014,
	title = {Nonresponse and {Mode} {Effects} in {Self}- and {Interviewer}-{Administered} {Surveys}},
	volume = {22},
	copyright = {https://www.cambridge.org/core/terms},
	issn = {1047-1987, 1476-4989},
	url = {https://www.cambridge.org/core/product/identifier/S1047198700013942/type/journal_article},
	doi = {10.1093/pan/mpt049},
	language = {en},
	number = {3},
	urldate = {2024-07-30},
	journal = {Political Analysis},
	author = {Atkeson, Lonna Rae and Adams, Alex N. and Alvarez, R. Michael},
	year = {2014},
	pages = {304--320},
}

@book{helfferich_qualitat_2011,
	address = {Wiesbaden},
	edition = {4. Auflage},
	series = {{SpringerLink} {Bücher}},
	title = {Die {Qualität} qualitativer {Daten}: {Manual} für die {Durchführung} qualitativer {Interviews}},
	isbn = {978-3-531-17382-5 978-3-531-92076-4},
	shorttitle = {Die {Qualität} qualitativer {Daten}},
	doi = {10.1007/978-3-531-92076-4},
	language = {ger},
	publisher = {VS Verlag für Sozialwissenschaften},
	author = {Helfferich, Cornelia},
	year = {2011},
}

@book{rubin_qualitative_2005,
	address = {2455 Teller Road, Thousand Oaks California 91320 United States},
	title = {Qualitative {Interviewing} (2nd ed.): {The} {Art} of {Hearing} {Data}},
	isbn = {978-0-7619-2075-5 978-1-4522-2665-1},
	shorttitle = {Qualitative {Interviewing} (2nd ed.)},
	url = {https://methods.sagepub.com/book/qualitative-interviewing},
	doi = {10.4135/9781452226651},
	urldate = {2025-12-09},
	publisher = {SAGE Publications, Inc.},
	author = {Rubin, Herbert and Rubin, Irene},
	year = {2005},
}

@book{seidman_interviewing_2019,
	address = {New York London},
	edition = {Fifth edition},
	title = {Interviewing as qualitative research: a guide for researchers in education and the social sciences},
	isbn = {978-0-8077-6187-8 978-0-8077-6148-9},
	shorttitle = {Interviewing as qualitative research},
	language = {eng},
	publisher = {Teachers College Press},
	author = {Seidman, Irving},
	year = {2019},
}

@book{flick_doing_2022,
	address = {Los Angeles London New Delhi Singapore Washington DC Melbourne},
	title = {Doing interview research: the essential how to guide},
	isbn = {978-1-5264-6405-7 978-1-5264-6406-4},
	shorttitle = {Doing interview research},
	language = {eng},
	publisher = {SAGE},
	author = {Flick, Uwe},
	year = {2022},
}

@article{berinsky_measuring_2017,
	title = {Measuring {Public} {Opinion} with {Surveys}},
	volume = {20},
	issn = {1094-2939, 1545-1577},
	url = {https://www.annualreviews.org/doi/10.1146/annurev-polisci-101513-113724},
	doi = {10.1146/annurev-polisci-101513-113724},
	language = {en},
	number = {1},
	urldate = {2025-12-11},
	journal = {Annual Review of Political Science},
	author = {Berinsky, Adam J.},
	month = may,
	year = {2017},
	pages = {309--329},
}

@article{cuevas_collecting_2025,
	title = {Collecting {Qualitative} {Data} at {Scale} with {Large} {Language} {Models}: {A} {Case} {Study}},
	volume = {9},
	issn = {2573-0142},
	shorttitle = {Collecting {Qualitative} {Data} at {Scale} with {Large} {Language} {Models}},
	url = {https://dl.acm.org/doi/10.1145/3710947},
	doi = {10.1145/3710947},
	language = {en},
	number = {2},
	urldate = {2026-01-21},
	journal = {Proceedings of the ACM on Human-Computer Interaction},
	author = {Cuevas, Alejandro and Scurrell, Jennifer V. and Brown, Eva M. and Entenmann, Jason and Daepp, Madeleine I. G.},
	month = may,
	year = {2025},
	pages = {1--27},
}

@misc{wong_ai_2025,
	title = {The {AI} {Interviewer}: {Exploring} the {Use} of {Conversational} {AI}-{Enabled} {Chatbots} in {Qualitative} {Data} {Collection}},
	shorttitle = {The {AI} {Interviewer}},
	url = {https://www.ssrn.com/abstract=5194078},
	doi = {10.2139/ssrn.5194078},
	urldate = {2026-01-21},
	publisher = {SSRN},
	author = {Wong, Liang Ze and Juraimi, Siti Amelia and Tan, Yin Zhien and Loh, Siyuan Brandon and Chong, Mary F-F. and Bhattacharya, Prasanta and Pink, Aimee E.},
	year = {2025},
}

@misc{geiecke_conversations_2024,
	title = {Conversations at {Scale}: {Robust} {AI}-led {Interviews} with a {Simple} {Open}-{Source} {Platform}},
	shorttitle = {Conversations at {Scale}},
	url = {https://www.ssrn.com/abstract=4974382},
	doi = {10.2139/ssrn.4974382},
	urldate = {2026-01-21},
	publisher = {SSRN},
	author = {Geiecke, Friedrich and Jaravel, Xavier},
	year = {2024},
}

@misc{barari_ai-assisted_2025,
	title = {{AI}-{Assisted} {Conversational} {Interviewing}: {Effects} on {Data} {Quality} and {Respondent} {Experience}},
	copyright = {Creative Commons Attribution 4.0 International},
	shorttitle = {{AI}-{Assisted} {Conversational} {Interviewing}},
	url = {https://arxiv.org/abs/2504.13908},
	doi = {10.48550/ARXIV.2504.13908},
	urldate = {2026-02-05},
	publisher = {arXiv},
	author = {Barari, Soubhik and Angbazo, Jarret and Wang, Natalie and Christian, Leah M. and Dean, Elizabeth and Slowinski, Zoe and Sepulvado, Brandon},
	year = {2025},
	note = {Version Number: 3},
}

@book{fowler_standardized_1998,
	address = {Newbury Park, Calif.},
	edition = {Nachdr.},
	series = {Applied social research methods series},
	title = {Standardized survey interviewing: minimizing interviewer-related error},
	isbn = {978-0-8039-3093-3 978-0-8039-3092-6},
	shorttitle = {Standardized survey interviewing},
	language = {eng},
	number = {18},
	publisher = {Sage},
	author = {Fowler, Floyd J. and Mangione, Thomas W.},
	year = {1998},
}

@article{schober_does_1997,
	title = {Does {Conversational} {Interviewing} {Reduce} {Survey} {Measurement} {Error}?},
	volume = {61},
	issn = {0033362X},
	url = {https://academic.oup.com/poq/article-lookup/doi/10.1086/297818},
	doi = {10.1086/297818},
	number = {4},
	urldate = {2026-02-05},
	journal = {Public Opinion Quarterly},
	author = {Schober, Michael F. and Conrad, Frederick G.},
	year = {1997},
	pages = {576},
}

@misc{ivey2026makesgoodresponseempirical,
      title={What Makes a Good Response? An Empirical Analysis of Quality in Qualitative Interviews}, 
      author={Jonathan Ivey and Anjalie Field and Ziang Xiao},
      year={2026},
      eprint={2604.05163},
      archivePrefix={arXiv},
      primaryClass={cs.CL},
      url={https://arxiv.org/abs/2604.05163}, 
}

@article{poq1,
    author = {Galesic, Mirta and Bosnjak, Michael},
    title = {Effects of Questionnaire Length on Participation and Indicators of Response Quality in a Web Survey},
    journal = {Public Opinion Quarterly},
    volume = {73},
    number = {2},
    pages = {349-360},
    year = {2009},
    month = {01},
     issn = {0033-362X},
    doi = {10.1093/poq/nfp031},
    url = {https://doi.org/10.1093/poq/nfp031},
    eprint = {https://academic.oup.com/poq/article-pdf/73/2/349/5462698/nfp031.pdf},
}

@article{poq2,
    author = {Presser, Stanley and Couper, Mick P. and Lessler, Judith T. and Martin, Elizabeth and Martin, Jean and Rothgeb, Jennifer M. and Singer, Eleanor},
    title = {Methods for Testing and Evaluating Survey Questions},
    journal = {Public Opinion Quarterly},
    volume = {68},
    number = {1},
    pages = {109-130},
    year = {2004},
    month = {03},
    issn = {0033-362X},
    doi = {10.1093/poq/nfh008},
    url = {https://doi.org/10.1093/poq/nfh008},
    eprint = {https://academic.oup.com/poq/article-pdf/68/1/109/5248163/nfh008.pdf},
}

@article{poq3,
    author = {Schaeffer, Nora Cate and Dykema, Jennifer},
    title = {Questions for Surveys: Current Trends and Future Directions},
    journal = {Public Opinion Quarterly},
    volume = {75},
    number = {5},
    pages = {909-961},
    year = {2011},
    month = {12},
    issn = {0033-362X},
    doi = {10.1093/poq/nfr048},
    url = {https://doi.org/10.1093/poq/nfr048},
    eprint = {https://academic.oup.com/poq/article-pdf/75/5/909/5162699/nfr048.pdf},
}

@article{poq4,
 ISSN = {0033362X, 15375331},
 URL = {http://www.jstor.org/stable/3521676},
 author = {Roger Tourangeau and Mick P. Couper and Frederick Conrad},
 journal = {The Public Opinion Quarterly},
 number = {3},
 pages = {368--393},
 publisher = {[Oxford University Press, American Association for Public Opinion Research]},
 title = {Spacing, Position, and Order: Interpretive Heuristics for Visual Features of Survey Questions},
 urldate = {2026-06-04},
 volume = {68},
 year = {2004}
}

@article{qual1,
author = {Robert H. Tai and Lillian R. Bentley and Xin Xia and Jason M. Sitt and Sarah C. Fankhauser and Ana M. Chicas-Mosier and Barnas G. Monteith},
title ={An Examination of the Use of Large Language Models to Aid Analysis of Textual Data},

journal = {International Journal of Qualitative Methods},
volume = {23},
number = {},
pages = {16094069241231168},
year = {2024},
doi = {10.1177/16094069241231168},

URL = { 
    
        https://doi.org/10.1177/16094069241231168
    
    

},
eprint = { 
    
        https://doi.org/10.1177/16094069241231168
    
    

}
}

@article{qual2,
    author = {Cook, David A and Ginsburg, Shiphra and Sawatsky, Adam P and Kuper, Ayelet and D’Angelo, Jonathan D},
    title = {Artificial Intelligence to Support Qualitative Data Analysis: Promises, Approaches, Pitfalls},
    journal = {Academic Medicine},
    volume = {100},
    number = {10},
    pages = {1134-1149},
    year = {2025},
    month = {10},
    issn = {1040-2446},
    doi = {10.1097/ACM.0000000000006134},
    url = {https://doi.org/10.1097/ACM.0000000000006134},
    eprint = {https://academic.oup.com/academicmedicine/article-pdf/100/10/1134/65840673/20251000.0-00010.pdf},
}

@article{doi:qual3,
author = {Mitchell Nicmanis and Harry Spurrier},
title ={Getting Started with Artificial Intelligence Assisted Qualitative Analysis: An Introductory Guide to Qualitative Research Approaches with Exploratory Examples from Reflexive Content Analysis},

journal = {International Journal of Qualitative Methods},
volume = {24},
number = {},
pages = {16094069251354863},
year = {2025},
doi = {10.1177/16094069251354863},

URL = { 
    
        https://doi.org/10.1177/16094069251354863
    
},
eprint = { 
    
        https://doi.org/10.1177/16094069251354863

}
}

@article{qual4,
  author    = {Yongjie Yue and Dong Liu and Yilin Lv and Junyi Hao and Peixuan Cui},
  title     = {A Practical Guide and Assessment on Using ChatGPT to Conduct Grounded Theory: Tutorial},
  journal   = {Journal of Medical Internet Research},
  year      = {2025},
  volume     = {27},
  pages      = {e70122},
  doi        = {10.2196/70122},
  pmid       = {40367510},
  pmcid      = {PMC12120365},
  url        = {https://www.jmir.org/2025/1/e70122/}
}

@article{qual5,
  author       = {Bail, Christopher A.},
  title        = {Can Generative AI Improve Social Science?},
  journaltitle = {Big Data \& Society},
  year         = {2022},
  volume       = {9},
  number       = {1},
  pages        = {1--7},
  doi          = {10.1177/20539517221080146},
  url          = {https://journals.sagepub.com/doi/full/10.1177/20539517221080146}
}

@article{qual6,
  author  = {De Paoli, Stefano},
  title   = {Performing an Inductive Thematic Analysis of Semi-Structured Interviews With a Large Language Model: An Exploration and Provocation on the Limits of the Approach},
  journal = {Social Science Computer Review},
  year    = {2024},
  volume  = {42},
  number  = {4},
  pages   = {997--1019},
  doi     = {10.1177/08944393231220483},
  url     = {https://journals.sagepub.com/doi/full/10.1177/08944393231220483}
}

@article{qual7,
  author  = {Nelson, Laura K.},
  title   = {Computational Grounded Theory: A Methodological Framework},
  journal = {Sociological Methods \& Research},
  year    = {2020},
  volume  = {49},
  number  = {1},
  pages   = {3--42},
  doi     = {10.1177/0049124117729703},
  url     = {https://journals.sagepub.com/doi/10.1177/0049124117729703}
}

@article{wei_leveraging_2024,
	title = {Leveraging {Large} {Language} {Models} to {Power} {Chatbots} for {Collecting} {User} {Self}-{Reported} {Data}},
	volume = {8},
	issn = {2573-0142},
	url = {https://dl.acm.org/doi/10.1145/3637364},
	doi = {10.1145/3637364},
	language = {en},
	number = {CSCW1},
	urldate = {2026-02-06},
	journal = {Proceedings of the ACM on Human-Computer Interaction},
	author = {Wei, Jing and Kim, Sungdong and Jung, Hyunhoon and Kim, Young-Ho},
	month = apr,
	year = {2024},
	pages = {1--35},
}

@inproceedings{budig_towards_2025,
	address = {New York City USA},
	title = {Towards the {Embodied} {Conversational} {Interview} {Agentic} {Service} {ELIAS}: {Development} and {Evaluation} of a {First} {Prototype}},
	isbn = {979-8-4007-1399-6},
	shorttitle = {Towards the {Embodied} {Conversational} {Interview} {Agentic} {Service} {ELIAS}},
	url = {https://dl.acm.org/doi/10.1145/3708319.3733810},
	doi = {10.1145/3708319.3733810},
	language = {en},
	urldate = {2026-02-06},
	booktitle = {Adjunct {Proceedings} of the 33rd {ACM} {Conference} on {User} {Modeling}, {Adaptation} and {Personalization}},
	publisher = {ACM},
	author = {Budig, Tobias and Nißen, Marcia and Kowatsch, Tobias},
	month = jun,
	year = {2025},
	pages = {420--424},
}

@article{guven_comparing_2025,
	title = {Comparing {AI}-led to human-led chat-based interviews: motivations, initial results and challenges},
	url = {https://osf.io/9mx36},
	language = {en-US},
	urldate = {2026-02-17},
	publisher = {OSF},
	author = {Guven, Semra Yuksel and Gardhus, Tobias and Bjerre-Nielsen, Andreas and Carlsen, Hjalmar},
	month = jul,
	year = {2025},
}

@inproceedings{hwang_scale_2025,
	address = {Bergen Norway},
	title = {Scale, {Engage}, or {Both}?: {Potential} and {Perils} of {Applying} {Large} {Language} {Models} in {Interview} and {Conversation}-{Based} {Research}},
	isbn = {979-8-4007-1480-1},
	shorttitle = {Scale, {Engage}, or {Both}?},
	url = {https://dl.acm.org/doi/10.1145/3715070.3748284},
	doi = {10.1145/3715070.3748284},
	language = {en},
	urldate = {2026-02-17},
	booktitle = {Companion {Publication} of the 2025 {Conference} on {Computer}-{Supported} {Cooperative} {Work} and {Social} {Computing}},
	publisher = {ACM},
	author = {Hwang, Angel Hsing-Chi and Aubin Le Quéré, Marianne and Schroeder, Hope and Cuevas, Alejandro and Dow, Steven P. and Kapania, Shivani and Rho, Eugenia},
	month = oct,
	year = {2025},
	pages = {95--98},
}

@misc{ma_too_2025,
	title = {Too {Open} for {Opinion}? {Embracing} {Open}-{Endedness} in {Large} {Language} {Models} for {Social} {Simulation}},
	copyright = {arXiv.org perpetual, non-exclusive license},
	shorttitle = {Too {Open} for {Opinion}?},
	url = {https://arxiv.org/abs/2510.13884},
	doi = {10.48550/ARXIV.2510.13884},
	urldate = {2026-02-18},
	publisher = {arXiv},
	author = {Ma, Bolei and Cao, Yong and Sen, Indira and Haensch, Anna-Carolina and Kreuter, Frauke and Plank, Barbara and Hershcovich, Daniel},
	year = {2025},
	note = {Version Number: 1},
}

@misc{meilan_using_2026,
	title = {Using {Customisable} {GPT} {Chatbots} in {Psychology} {Research}: {A} {Practical} {Tutorial}},
	copyright = {https://creativecommons.org/licenses/by/4.0/legalcode},
	shorttitle = {Using {Customisable} {GPT} {Chatbots} in {Psychology} {Research}},
	url = {https://osf.io/kf96p_v1},
	doi = {10.31234/osf.io/kf96p_v1},
	urldate = {2026-02-19},
	publisher = {PsyArXiv},
	author = {Meilan, Hu and Lau, Gabriel Rongyang and Goh, Adalia Yin Hui and Cox, Samuel Rhys and Tay, Louis and Hartanto, Andree},
	month = jan,
	year = {2026},
}

@inproceedings{kim_llm-as--interviewer_2025,
	address = {Vienna, Austria},
	title = {{LLM}-as-an-{Interviewer}: {Beyond} {Static} {Testing} {Through} {Dynamic} {LLM} {Evaluation}},
	shorttitle = {{LLM}-as-an-{Interviewer}},
	url = {https://aclanthology.org/2025.findings-acl.1357},
	doi = {10.18653/v1/2025.findings-acl.1357},
	language = {en},
	urldate = {2026-02-19},
	booktitle = {Findings of the {Association} for {Computational} {Linguistics}: {ACL} 2025},
	publisher = {Association for Computational Linguistics},
	author = {Kim, Eunsu and Suk, Juyoung and Kim, Seungone and Muennighoff, Niklas and Kim, Dongkwan and Oh, Alice},
	year = {2025},
	pages = {26456--26493},
}

@inproceedings{jacobsen_chatbots_2025,
	address = {Yokohama Japan},
	title = {Chatbots for {Data} {Collection} in {Surveys}: {A} {Comparison} of {Four} {Theory}-{Based} {Interview} {Probes}},
	isbn = {979-8-4007-1394-1},
	shorttitle = {Chatbots for {Data} {Collection} in {Surveys}},
	url = {https://dl.acm.org/doi/10.1145/3706598.3714128},
	doi = {10.1145/3706598.3714128},
	language = {en},
	urldate = {2026-02-20},
	booktitle = {Proceedings of the 2025 {CHI} {Conference} on {Human} {Factors} in {Computing} {Systems}},
	publisher = {ACM},
	author = {Jacobsen, Rune Møberg and Cox, Samuel Rhys and Griggio, Carla F. and Van Berkel, Niels},
	month = apr,
	year = {2025},
	pages = {1--21},
}

@incollection{degen_probing_2025,
	address = {Cham},
	title = {The {Probing} {Machine}: {Can} {Using} {GenAI} {Tools} {Help} {With} {Better} {Reflections} {During} {User} {Interviews}},
	volume = {15821},
	isbn = {978-3-031-93417-9 978-3-031-93418-6},
	shorttitle = {The {Probing} {Machine}},
	url = {https://link.springer.com/10.1007/978-3-031-93418-6_6},
	doi = {10.1007/978-3-031-93418-6_6},
	language = {en},
	urldate = {2026-02-20},
	booktitle = {Artificial {Intelligence} in {HCI}},
	publisher = {Springer Nature Switzerland},
	author = {Tran, Corey and Parayno, Richard Lance and Venkitachalam, Bharat Krishna and Deja, Janna Aika and Deja, Jordan Aiko},
	editor = {Degen, Helmut and Ntoa, Stavroula},
	year = {2025},
	note = {Series Title: Lecture Notes in Computer Science},
	pages = {72--88},
}

@inproceedings{wachsmuth-etal-2024-argument,
    title = "Argument Quality Assessment in the Age of Instruction-Following Large Language Models",
    author = "Wachsmuth, Henning  and
      Lapesa, Gabriella  and
      Cabrio, Elena  and
      Lauscher, Anne  and
      Park, Joonsuk  and
      Vecchi, Eva Maria  and
      Villata, Serena  and
      Ziegenbein, Timon",
    editor = "Calzolari, Nicoletta  and
      Kan, Min-Yen  and
      Hoste, Veronique  and
      Lenci, Alessandro  and
      Sakti, Sakriani  and
      Xue, Nianwen",
    booktitle = "Proceedings of the 2024 Joint International Conference on Computational Linguistics, Language Resources and Evaluation (LREC-COLING 2024)",
    month = may,
    year = "2024",
    address = "Torino, Italia",
    publisher = "ELRA and ICCL",
    url = "https://aclanthology.org/2024.lrec-main.135/",
    pages = "1519--1538"
}

@inproceedings{soltani-romberg-2023-general,
    title = "A General Framework for Multimodal Argument Persuasiveness Classification of Tweets",
    author = "Soltani, Mohammad and
              Romberg, Julia",
    editor = "Alshomary, Milad and
              Chen, Chung-Chi and
              Muresan, Smaranda and
              Park, Joonsuk and
              Romberg, Julia",
    booktitle = "Proceedings of the 10th Workshop on Argument Mining",
    month = dec,
    year = "2023",
    address = "Singapore",
    publisher = "Association for Computational Linguistics",
    pages = "148--156",
    doi = "10.18653/v1/2023.argmining-1.15",
    url = "https://aclanthology.org/2023.argmining-1.15/"
}

@article{krosnick1989attitude,
  author    = {Krosnick, Jon A.},
  title     = {Attitude Importance and Attitude Accessibility},
  journal   = {Personality and Social Psychology Bulletin},
  year      = {1989},
  volume     = {15},
  number     = {3},
  pages      = {297--308},
  doi        = {10.1177/0146167289153002},
  publisher  = {SAGE Publications}
}

@inproceedings{veluri_beyond_2024,
	address = {Miami, Florida, USA},
	title = {Beyond {Turn}-{Based} {Interfaces}: {Synchronous} {LLMs} as {Full}-{Duplex} {Dialogue} {Agents}},
	shorttitle = {Beyond {Turn}-{Based} {Interfaces}},
	url = {https://aclanthology.org/2024.emnlp-main.1192},
	doi = {10.18653/v1/2024.emnlp-main.1192},
	language = {en},
	urldate = {2026-02-27},
	booktitle = {Proceedings of the 2024 {Conference} on {Empirical} {Methods} in {Natural} {Language} {Processing}},
	publisher = {Association for Computational Linguistics},
	author = {Veluri, Bandhav and Peloquin, Benjamin N and Yu, Bokai and Gong, Hongyu and Gollakota, Shyamnath},
	year = {2024},
	pages = {21390--21402},
}

@misc{kim_detail_2025,
	title = {{DETAIL} {Matters}: {Measuring} the {Impact} of {Prompt} {Specificity} on {Reasoning} in {Large} {Language} {Models}},
	copyright = {arXiv.org perpetual, non-exclusive license},
	shorttitle = {{DETAIL} {Matters}},
	url = {https://arxiv.org/abs/2512.02246},
	doi = {10.48550/ARXIV.2512.02246},
	urldate = {2026-02-27},
	publisher = {arXiv},
	author = {Kim, Olivia},
	year = {2025},
	note = {Version Number: 1},
}

@misc{he_does_2024,
	title = {Does {Prompt} {Formatting} {Have} {Any} {Impact} on {LLM} {Performance}?},
	copyright = {Creative Commons Attribution 4.0 International},
	url = {https://arxiv.org/abs/2411.10541},
	doi = {10.48550/ARXIV.2411.10541},
	urldate = {2026-02-27},
	publisher = {arXiv},
	author = {He, Jia and Rungta, Mukund and Koleczek, David and Sekhon, Arshdeep and Wang, Franklin X and Hasan, Sadid},
	year = {2024},
	note = {Version Number: 1},
}

@misc{anugraha_sparkme_2026,
	title = {{SparkMe}: {Adaptive} {Semi}-{Structured} {Interviewing} for {Qualitative} {Insight} {Discovery}},
	copyright = {Creative Commons Attribution 4.0 International},
	shorttitle = {{SparkMe}},
	url = {https://arxiv.org/abs/2602.21136},
	doi = {10.48550/ARXIV.2602.21136},
	urldate = {2026-02-28},
	publisher = {arXiv},
	author = {Anugraha, David and Padmakumar, Vishakh and Yang, Diyi},
	year = {2026},
	note = {Version Number: 1},
}

@misc{lang_telephone_2025,
	title = {Telephone {Surveys} {Meet} {Conversational} {AI}: {Evaluating} a {LLM}-{Based} {Telephone} {Survey} {System} at {Scale}},
	copyright = {Creative Commons Attribution Non Commercial Share Alike 4.0 International},
	shorttitle = {Telephone {Surveys} {Meet} {Conversational} {AI}},
	url = {https://arxiv.org/abs/2502.20140},
	doi = {10.48550/ARXIV.2502.20140},
	urldate = {2026-03-04},
	publisher = {arXiv},
	author = {Lang, Max M. and Eskenazi, Sol},
	year = {2025},
	note = {Version Number: 2},
}

@article{cavusoglu_deveci_experimental_2026,
	title = {Experimental comparison of a chatbot-like questionnaire design with a traditional web questionnaire design},
	issn = {0759-1063},
	url = {https://journals.sagepub.com/action/showAbstract},
	doi = {10.1177/07591063261424249},
	urldate = {2026-03-10},
	journal = {Bulletin of Sociological Methodology/Bulletin de Méthodologie Sociologique},
	publisher = {SAGE Publications Ltd},
	author = {Çavuşoğlu Deveci, Ceyda and Fuchs, Marek and Metzler, Anke},
	month = mar,
	year = {2026},
	pages = {07591063261424249},
}

@inproceedings{shih_rationalizer_2026,
	address = {Paphos Cyprus},
	title = {Rationalizer: {Leveraging} {LLM} to {Support} {User} {Providing} the {Rationales} {Behind} the {Rating} of {Likert} {Scale} {Questionnaires}},
	isbn = {979-8-4007-1984-4},
	shorttitle = {Rationalizer},
	url = {https://dl.acm.org/doi/10.1145/3742413.3789171},
	doi = {10.1145/3742413.3789171},
	language = {en},
	urldate = {2026-03-10},
	booktitle = {Proceedings of the 31st {International} {Conference} on {Intelligent} {User} {Interfaces}},
	publisher = {ACM},
	author = {Shih, Meng Ting and Wu, Po Yen and Chang, Yun Chen and Yen, Grace Yu-Chun and Chan, Liwei},
	month = mar,
	year = {2026},
	pages = {85--105},
}

@misc{handa_introducing_2025,
	title = {Introducing {Anthropic} {Interviewer}: {What} 1,250 professionals told us about working with {AI}},
	url = {https://anthropic.com/research/anthropic-interviewer},
	author = {Handa, Kunal and Stern, Michael and Huang, Saffron and Hong, Jerry and Durmus, Esin and McCain, Miles and Yun, Grace and Alt, A. J. and Millar, Thomas and Tamkin, Alex and Leibrock, Jane and Ritchie, Stuart and Ganguli, Deep},
	month = dec,
	year = {2025},
}

@misc{liu_mimitalk_2025,
	title = {{MimiTalk}: {Revolutionizing} {Qualitative} {Research} with {Dual}-{Agent} {AI}},
	copyright = {Creative Commons Attribution 4.0 International},
	shorttitle = {{MimiTalk}},
	url = {https://arxiv.org/abs/2511.03731},
	doi = {10.48550/ARXIV.2511.03731},
	urldate = {2026-03-10},
	publisher = {arXiv},
	author = {Liu, Fengming and Yu, Shubin},
	year = {2025},
	note = {Version Number: 1},
}

@book{groves_survey_2011,
	title = {Survey methodology},
	publisher = {John Wiley \& Sons},
	author = {Groves, Robert M and Fowler Jr, Floyd J and Couper, Mick P and Lepkowski, James M and Singer, Eleanor and Tourangeau, Roger},
	year = {2011},
}

@article{conrad_new_2008,
	title = {New {Frontiers} in {Standardized} {Survey} {Interviewing}},
	journal = {Handbook of emergent methods},
	publisher = {Guilford Press},
	author = {Conrad, Frederick G. and Schober, Michael F.},
	year = {2008},
	note = {ISBN: 1593851472},
	pages = {173},
}

@article{berg_toward_2025,
  author = {Berg, Allison and Ternullo, Stephanie},
  title = {Toward a Qualitative Study of the American Voter},
  journal = {Perspectives on Politics},
  year = {2025},
  pages = {1--17},
  doi = {10.1017/S1537592724002718}
}

@article{konig_conceptualizing_2022,
  author = {K{\"o}nig, Pascal D. and Siewert, Markus B. and Ackermann, Kathrin},
  title = {Conceptualizing and Measuring Citizens' Preferences for Democracy: Taking Stock of Three Decades of Research in a Fragmented Field},
  journal = {Comparative Political Studies},
  year = {2022},
  volume = {55},
  number = {12},
  pages = {2015--2049},
  doi = {10.1177/00104140211066213}
}

@article{converse_nature_2006,
  author = {Converse, Philip E.},
  title = {The Nature of Belief Systems in Mass Publics (1964)},
  journal = {Critical Review},
  year = {2006},
  volume = {18},
  number = {1--3},
  pages = {1--74},
  doi = {10.1080/08913810608443650}
}

@article{hernandez_democracy_2019,
  author = {Hern{\'a}ndez, Enrique},
  title = {Democracy Belief Systems in Europe: Cognitive Availability and Attitudinal Constraint},
  journal = {European Political Science Review},
  year = {2019},
  volume = {11},
  number = {4},
  pages = {485--502},
  doi = {10.1017/S1755773919000286}
}

@article{howe_attitude_2017,
  author = {Howe, Lauren C. and Krosnick, Jon A.},
  title = {Attitude Strength},
  journal = {Annual Review of Psychology},
  year = {2017},
  volume = {68},
  number = {1},
  pages = {327--351},
  doi = {10.1146/annurev-psych-122414-033600}
}

@book{kaplan_conduct_2017,
  author = {Kaplan, Abraham},
  title = {The Conduct of Inquiry: Methodology for Behavioral Science},
  publisher = {Routledge},
  year = {2017},
  doi = {10.4324/9781315131467}
}

@article{mittereder_interviewerrespondent_2018,
  author = {Mittereder, Franz and Durow, Julia and West, Brady T. and Kreuter, Frauke and Conrad, Frederick G.},
  title = {Interviewer--respondent Interactions in Conversational and Standardized Interviewing},
  journal = {Field Methods},
  year = {2018},
  volume = {30},
  number = {1},
  pages = {3--21},
  doi = {10.1177/1525822X17729341}
}

@article{west_can_2018,
  author = {West, Brady T. and Conrad, Frederick G. and Kreuter, Frauke and Mittereder, Franz},
  title = {Can Conversational Interviewing Improve Survey Response Quality Without Increasing Interviewer Effects?},
  journal = {Journal of the Royal Statistical Society Series A: Statistics in Society},
  year = {2018},
  volume = {181},
  number = {1},
  pages = {181--203},
  doi = {10.1111/rssa.12255}
}

@inproceedings{wuttke_ai_2025,
  author = {Wuttke, Alexander and A{\ss}enmacher, Matthias and Klamm, Christopher and Lang, Max M. and W{\"u}rschinger, Quirin and Kreuter, Frauke},
  title = {AI Conversational Interviewing: Transforming Surveys with LLMs as Adaptive Interviewers},
  booktitle = {Proceedings of the 9th Joint SIGHUM Workshop on Computational Linguistics for Cultural Heritage, Social Sciences, Humanities and Literature (LaTeCH-CLfL 2025)},
  year = {2025},
  pages = {179--204},
  doi = {10.18653/v1/2025.latechclfl-1.17}
}

\newpage

\appendix

\section*{Appendices Overview}

\begin{tabular}{p{0.15\textwidth} p{0.85\textwidth}}
Appendix \ref{sec:aapor} & AAPOR-Required Disclosure Elements \\
Appendix \ref{sec:appendix-questionnaire} & Questionnaire \\
Appendix \ref{sec:appendix-prompts} & Prompts \\
Appendix \ref{sec:appendix-deviations} & Deviations from the pre-analysis plan \\
Appendix \ref{sec:appendix-voice-waiting} & Voice mode waiting time \\
Appendix \ref{sec:appendix-technical-problems} & Technical problems \\
Appendix \ref{sec:appendix-turn-statistic} & Turn-dependent statistics: Responses over the course of the interview \\
Appendix \ref{sec:appendix-standardized-survey} & Standardized survey: attitudes towards immigration \\
Appendix \ref{sec:appendix-text-analysis} & Text analysis: taxonomy and validation \\
Appendix \ref{sec:appendix-text-analysis-voters} & Text analysis by voter groups \\
Appendix \ref{sec:appendix-sequence-analysis} & Sequence analysis

\end{tabular}
\newpage

\section{AAPOR-Required Disclosure Elements}
\label{sec:aapor}

\subsection*{Data Source}
The study draws on a single primary data collection conducted by the authors, combining (a) an AI-led conversational interview on migration and migration policy in Germany and (b) a standardized web survey, including post-interview evaluation items. Both components were administered within one integrated instrument; conversational and survey records were linked at the respondent level via a unique identifier.

\subsection*{Data Collection Strategy}
Data collection consisted of three sequential components administered in a single session: (1) a semi-structured conversational interview conducted by a large language model (LLM) acting as an adaptive interviewer, delivered via text chat or voice depending on experimental condition; (2) a standardized survey battery on migration policy; and (3) standardized survey questions on the interview experience. The standardized components were administered via Qualtrics; the conversational interface was embedded within the Qualtrics environment.

\subsection*{Research Sponsor and Conductor}
The research was conducted by the Chair of Digitalization and Political Behaviour (Lehreinheit Digitalisierung und Politisches Verhalten), Ludwig-Maximilians-Universit\"at M\"unchen (LMU Munich), under the direction of Prof.\ Dr.\ Alexander Wuttke.

\subsection*{Measurement Tools and Instruments}
The full survey instrument, including informed-consent materials, the post-interview evaluation battery, and all standardized migration items, is reproduced verbatim in Appendix~\ref{sec:appendix-questionnaire}. The standardized battery comprises all migration-related items from the German Longitudinal Election Study (GLES, 2025). The conversational interviews were guided by a semi-structured interview agenda implemented through system prompts to the LLM interviewer; the complete prompts, in the German original and in English translation, including the interview guide, conversational norms, and instructions governing interviewer behaviour (neutrality, active listening, probing strategy), are reproduced verbatim in Appendix~\ref{sec:appendix-prompts}. The prompts followed a 2$\times$2 design crossing communication mode (text vs.\ audio) with prompt depth (short, approx.\ 2{,}000 tokens vs.\ informed, approx.\ 3{,}000 tokens).

\subsection*{Population Under Study}
The population under study consists of German-speaking adults (aged 18 years or older) residing in Germany who are members of the online access panels of Prolific or Payback Panel and who had access to an internet-enabled device. Respondents in the voice conditions additionally required a functioning microphone and audio output. 

\subsection*{Methods Used to Generate and Recruit the Sample}
The sample was selected using non-probability methods. Participants were recruited from two opt-in online access panels: Prolific and Payback Panel. Recruitment was initially planned exclusively through Prolific; because Prolific recruited respondents at a slower rate than anticipated, data collection was extended to Payback Panel to reach the target sample size within the projected fieldwork period. Participants from both vendors received an identical survey instrument, and main analyses pool both sources. Panel members were invited through the vendors' standard invitation procedures. No quotas were applied Data collection was conducted entirely online; no in-person data collection took place.

\subsection*{Method(s) and Mode(s) of Data Collection}
All data were collected via self-administered web instruments in German. The standardized survey components were administered as a web survey (Qualtrics). The conversational interview was administered in one of two modes, assigned experimentally: (a) a text-chat interview implemented in Python using Chainlit as the browser-based chat interface and the OpenAI Chat Completions API via LangChain, with GPT-4o (gpt-4o-2024-11-20, temperature 0.8) as the underlying model; or (b) a voice interview using Vapi.ai for orchestration and OpenAI's GPT Realtime model (gpt-realtime-2025-08-28, temperature 0.7), operating as a full-duplex speech interface with dynamic turn-taking. Respondents were randomly assigned to a voice condition, a text condition, or a choice condition in which they selected their preferred mode. Among substantive interviews, voice sessions lasted on average 8.3 minutes (Mdn = 9.6, SD = 5.2) and text sessions 11.5 minutes (Mdn = 9.9, SD = 9.9). For the voice condition, speaker-labelled transcripts, timestamps, conversation duration, and respondent ID were stored on researcher-controlled infrastructure (AWS S3); no audio recordings were stored.

\subsection*{Dates of Data Collection}
Data were collected from Prolific between 25 September 2025 and 31 October 2025 and from Payback Panel between 16 October 2025 and 27 October 2025.

\subsection*{Sample Sizes and Discussion of the Precision of the Results}
A total of 1{,}103 respondents began the survey; 1{,}039 were randomly assigned to an interview condition (voice: $n = 317$; text: $n = 376$; choice: $n = 346$, of whom 97 selected voice and 249 selected text). The final analytic sample of substantive interviews comprises $N = 571$ respondents (text: 354; voice: 217), retaining only cases with successfully linked survey and interview records and at least five user turns.  Because the sample was selected using non-probability methods, no measures of sampling precision (e.g., margin of error) are reported, and none should be inferred. All inferential statistics reported in the manuscript rest on model-based assumptions, and the substantive text-as-data contrasts are presented as exploratory illustrations rather than confirmatory population estimates.

\subsection*{Whether and How the Data Were Weighted}
The data were not weighted. All analyses report unweighted estimates.

\subsection*{How the Data Were Processed and Procedures to Ensure Data Quality}
Survey and interview records were linked via a unique respondent identifier entered manually by participants at the start of the interview; records that could not be linked (approximately one quarter of assigned respondents) were excluded from all analyses. Interviews with fewer than five user turns were classified as non-substantive and excluded from the analytic sample. Respondents were instructed not to use AI tools (e.g., ChatGPT) when answering, and authenticity checks were conducted to detect parallel use of such tools.  A monitoring mechanism allowed respondents to flag offensive or inappropriate interviewer behaviour; fewer than 1\% of respondents reported any such incident. Pre-test data, including systematic stress tests (``red teaming''), were excluded from all analyses.

For the text-as-data analyses, interview transcripts were segmented into 16{,}034 information units and classified by issue topic (12 substantive categories) and argument type (10 categories) using a human-validated LLM annotation procedure: an initial LLM-based pass generated candidate segmentations and labels, which two trained researchers then validated against a structured codebook. Inter-annotator agreement was high across all coding layers (segmentation boundary: 88.5\%; merge: 94.3\%; split: 92.8\%; issue: 84.9\%; argument type: 79.0\%); full details are reported in Appendix~\ref{sec:appendix-text-analysis}.

\subsection*{Panel Description}
Respondents were drawn from two commercially managed, non-probability online access panels (Prolific; Payback Panel). Panel recruitment, membership management, and attrition procedures are administered by the respective vendors; the authors did not maintain the panels.

\subsection*{Interviewer Details}
Conversational interviews were conducted by an LLM-based AI interviewer rather than human interviewers. Interviewer behaviour was governed by system prompts specifying the interviewer role, interview structure, conversational norms, neutrality requirements, and probing strategy (reproduced in Appendix~\ref{sec:appendix-prompts}). In lieu of interviewer training, the implementation was refined through multiple pre-tests with members of the research team and Prolific respondents, including systematic stress testing in which research assistants impersonated respondents with atypical behaviour and documented model behaviour against a checklist of potentially undesirable behaviours. Interviewer monitoring during fieldwork relied on the respondent-facing flagging mechanism described above. For the human validation of the machine coding, two trained researchers coded against a structured codebook; reliability estimates are reported above and in Appendix~\ref{sec:appendix-text-analysis}.

\subsection*{Eligibility Screening}
Participation required confirmation of being at least 18 years of age as part of the informed-consent procedure. 

\subsection*{Study Stimuli}
No visual or sensory stimuli beyond the survey instrument and the conversational interfaces were used. Screenshots of the voice interview interface in its different states are provided in the manuscript (Figure~3); the text-chat interface followed a standard chat layout embedded in Qualtrics.

\subsection*{Dispositions and Participation Rates}
A CONSORT-style participant flow diagram (Figure~4 in the manuscript) reports dispositions at each stage: total survey responses ($N = 1{,}103$), drop-out before assignment ($n = 64$), assignment to conditions ($N = 1{,}039$), unlinked or early drop-out interviews, short interviews (fewer than five user turns), and the final analytic sample ($N = 571$). Because the sample derives from opt-in panels, no response rate in the sense of AAPOR Standard Definitions for probability samples can be computed. 

\subsection*{Sample Sizes for Reported Estimates}
Analyses of respondent evaluations of the interviewer and the interview experience are based on the analytic sample of substantive interviews ($N = 571$). Analyses of technical problems use the full dataset of all respondents who reached the relevant items, including unlinked and discontinued cases ($n = 859$--$870$, varying by item). The perceived-waiting-time item was shown only to voice-condition respondents ($n = 206$). Word-based transcript metrics use the analysis cohort of $N = 590$ interviews ($n_{\text{voice}} = 219$, $n_{\text{text}} = 371$). Standardized immigration-attitude estimates by party identification are based on linked respondents ($n = 769$).

\subsection*{Measurement and Model Specification}
Specifications adequate for replication, including the full questionnaire, the verbatim interviewer prompts, the issue and argument-type taxonomies, the annotation codebook and validation results, and the sequence-analysis procedures (greedy Markov walks through transition matrices), are documented in the manuscript appendices. Open data and the open-source data-collection and analysis pipeline are available via the project repository; the pre-analysis plan is registered on the Open Science Framework, and deviations are documented in Appendix~\ref{sec:appendix-deviations}.

\subsection*{Statement of Limitations}
The study relies on non-probability samples drawn from two opt-in online access panels of digitally experienced respondents; results should not be generalized to the broader German population. Approximately one quarter of assigned respondents could not be linked across the survey and interview components owing to manual identifier entry, and interview completion varied across experimental conditions, with drop-out concentrated in the assigned-voice condition; both sources of attrition may have introduced selection into the analytic sample. The fixed ordering of components means the standardized survey was completed after, and potentially primed by, the conversational interview. Mode contrasts bundle multiple implementation differences (input/output channel, duplex structure, underlying LLM, interface) and should not be read as estimates of a pure modality effect.
 
\FloatBarrier
\section{Questionnaire}
\label{sec:appendix-questionnaire}

\begin{otherlanguage}{ngerman}

This appendix reports the questionnaire as implemented in Qualtrics. 

\subsection{Block: Einleitung / Informierte Zustimmung}

\paragraph{Intro.}

\emph{Herzlich Willkommen zu dieser Befragung der Lehreinheit Digitalisierung und Politisches Verhalten an der LMU M\"unchen, geleitet von Prof.\ Dr.\ Alexander Wuttke. Mit dieser Befragung m\"ochten wir Ihre Sicht auf Politik und Gesellschaft besser verstehen. Diese Befragung wird anders verlaufen, als Sie es \"ublicherweise gewohnt sind. Denn die Befragung verbindet ein KI-gest\"utztes Gespr\"ach mit einem klassischen Fragebogen. Diese Befragung besteht aus zwei Teilen: Teil~1: Ein ausf\"uhrliches Gespr\"ach zur Migrationspolitik. Teil~2: Ein klassischer Fragebogen zu Ihren politischen Meinungen und zur Interviewerfahrung. Nehmen Sie sich ausreichend Zeit f\"ur das Gespr\"ach (Teil~1) und schildern Sie bitte ausf\"uhrlich Ihre Gedanken und Eindr\"ucke. Das Gespr\"ach wird KI-unterst\"utzt gef\"uhrt. Sobald das Gespr\"ach abgeschlossen ist, klicken Sie bitte auf weiter und nehmen Sie am klassischen Fragebogen (Teil~2) teil. Bitte beachten Sie, dass wir zur Durchf\"uhrung dieser Befragung mit externen Dienstleistern zusammenarbeiten, wie beispielsweise OpenAI. Diese Dienstleister verarbeiten und speichern Daten, die im Rahmen dieser Erhebung erhoben wurden. N\"ahere Datenschutzinformationen erhalten Sie unten auf dieser Seite. OpenAI wird die im Rahmen dieser Befragung erhobenen Daten nicht zum Training ihrer KI-Modelle verwenden. Achtung: Bitte nutzen Sie selbst keine KI-Modelle wie ChatGPT zur Beantwortung dieses Fragebogens. Wir sind an Ihren ehrlichen Auffassungen und Meinungen interessiert. Daher f\"uhren wir Echtheitspr\"ufungen durch, um die parallele Nutzung von ChatGPT und \"ahnlichen Werkzeugen w\"ahrend der Befragung zu erkennen. Wir freuen uns sehr auf Ihre Teilnahme und danken Ihnen herzlich, dass Sie Ihre wertvollen Erfahrungen mit uns teilen!}

\paragraph{Consent.} Lesen Sie bitte die folgenden Punkte sorgf\"altig durch, bevor Sie an dieser Studie teilnehmen, und best\"atigen Sie Ihre Zustimmung:
\begin{enumerate}[leftmargin=2em]
  \item Ich best\"atige, dass ich die Studieninformationen gelesen habe und Gelegenheit hatte, die Informationen zu pr\"ufen.
  \item Ich stimme der Erhebung und Speicherung meiner Daten in anonymisierter Form zu, die zu Replikationszwecken der Forschungsgemeinschaft zur Verf\"ugung gestellt werden k\"onnen.
  \item Ich stimme der Teilnahme an der Studie zu.
  \item Ich best\"atige, dass ich mindestens 18 Jahre alt bin.
  \item Ich erkl\"are mich einverstanden, dass im Rahmen dieser Studie Daten von VAPI, Literal.AI, Qualtrics und OpenAI verarbeitet und gespeichert werden. Ich stimme zu, dass Daten auch au\ss{}erhalb der EU verarbeitet und gespeichert werden k\"onnen.
\end{enumerate}

\paragraph{Q1: Datenschutzinformationen.}

Die Survey-Seite enthielt ausf\"uhrliche Datenschutzinformationen gem\"a\ss{} Art.~13 DSGVO. Inhaltlich umfasste dieser Abschnitt:

\begin{itemize}
  \item Verantwortliche Stelle: Ludwig-Maximilians-Universit\"at M\"unchen;
  \item zust\"andige Organisationseinheit: Lehreinheit Digitalisierung und Politisches Verhalten;
  \item Kontakt: \texttt{a.wuttke@lmu.de};
  \item Verarbeitungszweck: Erhebung politischer Einstellungen und Untersuchung neuer Methoden der \"offentlichen Meinungsforschung;
  \item Informationen zu VAPI, AWS, OpenAI und Literal.AI als technische Dienstleister;
  \item Hinweise auf Logfiles, Speicherdauer und Betroffenenrechte.
\end{itemize}

\subsection{Block: AI text, simple}

\paragraph{Q75.}
\emph{Im Chat-Fenster unten beginnt Ihr pers\"onliches Gespr\"ach zu politischen Themen. Der KI-Interviewer wird Ihnen eine Reihe von Fragen stellen und das Gespr\"ach zu einem Ende f\"uhren, wenn alle wichtigen Fragen besprochen sind. Dann und erst dann klicken Sie bitte im Fragebogen auf ``weiter'' und nehmen am Rest des Fragenbogens teil. Viel Vergn\"ugen mit dem Gespr\"ach. Bitte geben Sie unten diesen individualisierten Code ein. Sie k\"onnen ihn auch kopieren und einf\"ugen (auf einer Tastatur: Strg+V). \texttt{\$\{e://Field/UniqueCode\}}. Diese Eingabe ist wichtig f\"ur den weiteren Verlauf der Umfrage.}

\paragraph{UI-Einbettung.}
\emph{HTML Embedding Text Chat here}

\subsection{Block: AI text, informed}

\paragraph{Q74.}
Der Instruktionstext entsprach inhaltlich dem Text-Chat im einfachen Prompt-Arm; auch hier wurde der individualisierte Code \texttt{\$\{e://Field/UniqueCode\}} angezeigt und zur Eingabe in das Chatfenster aufgefordert.

\paragraph{UI-Einbettung.}
\emph{HTML Embedding Text Chat here}

\subsection{Block: AI voice, simple}

\paragraph{Q79.}
\emph{Im Men\"u unten f\"uhren Sie Ihr pers\"onliches Gespr\"ach zu politischen Themen. Sie beginnen das Gespr\"ach nach Eingabe Ihres individuellen Codes mit Druck auf die Mikrofon-Taste. Der KI-Interviewer wird Ihnen eine Reihe von Fragen stellen und das Gespr\"ach zu einem Ende f\"uhren, wenn alle wichtigen Fragen besprochen sind. Dann und erst dann klicken Sie bitte im Fragebogen auf ``weiter'' und nehmen am Rest des Fragenbogens teil. Viel Vergn\"ugen mit dem Gespr\"ach. Bitte geben Sie unten diesen individualisierten Code ein: \texttt{\$\{e://Field/UniqueCode\}}. Diese Eingabe ist wichtig f\"ur den weiteren Verlauf der Umfrage. Sobald Ihr Gespr\"ach beginnt, k\"onnen Sie einfach in Ihr Mikrofon sprechen. Stellen Sie sicher, dass Ihre Lautsprecher angeschaltet sind. Gerne k\"onnen Sie Kopfh\"orer oder Headset verwenden.}

\paragraph{UI-Einbettung.}
\emph{HTML Embedding Voice Chat here}

\subsection{Block: AI voice, informed}

Der Instruktionstext entsprach inhaltlich dem einfachen Voice-Arm; auch hier wurde derselbe individualisierte Code abgefragt und anschlie\ss{}end die Voice-Oberfl\"ache eingebettet.

\subsection{Block: AI choice, simple / informed}

\paragraph{Moduswahl.}
\emph{Wir beginnen nun mit dem KI-gest\"utzten Gespr\"ach zu Ihren Auffassungen zu Politik und Gesellschaft. Sie k\"onnen aus zwei Gespr\"achsformaten w\"ahlen:}
\begin{enumerate}[leftmargin=2em]
  \item Gespr\"ach via Text-Chat,
  \item Gespr\"ach via Sprachein- und Sprachausgabe.
\end{enumerate}
\emph{Beim Text-Chat werden Sie Ihre Antworten \"uber die Tastatur eingeben. Das Gespr\"ach via Sprachein- und Sprachausgabe nutzt Mikrofon sowie Lautsprecher/Kopfh\"orer.}

Je nach Auswahl wurde anschlie\ss{}end die entsprechende Text- oder Voice-Oberfl\"ache mit demselben Instruktionstext wie in den direkt zugewiesenen Bedingungen angezeigt.

\subsection{Block: Evaluation}

\paragraph{ai\_evaluation.}
\emph{Wir m\"ochten nun Ihre Erfahrungen und Eindr\"ucke zum vorherigen Gespr\"ach als neuartige Befragungsmethode beleuchten. Wie bewerten Sie insgesamt Ihre Erfahrung w\"ahrend des Gespr\"achs?}
Antwortskala: 1 = \emph{sehr schlecht} bis 7 = \emph{sehr gut}.

\paragraph{problems.}
\emph{Sind w\"ahrend Ihres KI-gest\"utzten Gespr\"achs Probleme aufgetreten?}
Matrix mit drei Antwortkategorien (\emph{Keine Probleme}, \emph{Kleine Probleme}, \emph{Gro\ss{}e Probleme}) f\"ur:
\begin{enumerate}[leftmargin=2em]
  \item Gespr\"ach starten,
  \item Gespr\"ach beenden,
  \item dem Gespr\"achspartner Antworten geben,
  \item vom Gespr\"achspartner Fragen erhalten.
\end{enumerate}

\paragraph{problems\_txt.}
Nur f\"ur Befragte mit kleinen oder gro\ss{}en Problemen:
\emph{Welche Probleme sind w\"ahrend Ihres KI-gest\"utzten Gespr\"achs aufgetreten? Nennen Sie bitte alle Probleme.}

\paragraph{Q91.}
\emph{Hat sich der KI-Interviewer w\"ahrend des Gespr\"achs in irgendeiner Weise anst\"o\ss{}ig, verletzend oder in anderer Weise grob unangemessen ge\"au\ss{}ert?}
\begin{enumerate}[leftmargin=2em]
  \item Nein
  \item Ja, und zwar folgende \"Au\ss{}erung: \underline{\hspace{6cm}}
\end{enumerate}

\paragraph{Q94.}
Falls ``Ja'' bei Q91: Hinweis auf die M\"oglichkeit der Kontaktaufnahme mit Prof.\ Dr.\ Alexander Wuttke.

\paragraph{ai-accuracy.}
\emph{Manchmal traut man sich in einem Interview nicht frei zu sagen, was man eigentlich denkt. Inwiefern reflektieren Ihre \"Au\ss{}erungen gegen\"uber dem KI-Gespr\"achspartner Ihre ehrlichen Gedanken, Gef\"uhle und Erfahrungen?}
Skala: 1 = \emph{v\"ollig} bis 8 = \emph{\"uberhaupt nicht}.

\paragraph{dimensions\_relaxed / dimensions\_pleasant / dimension\_natural.}
Drei semantische Differenziale:
\begin{itemize}
  \item 1 = \emph{entspannt}, 7 = \emph{stressig}
  \item 1 = \emph{angenehm}, 7 = \emph{unangenehm}
  \item 1 = \emph{nat\"urlich}, 7 = \emph{k\"unstlich}
\end{itemize}

\paragraph{pauses.}
Nur in Voice-Bedingungen:
\emph{Hat Ihnen der Interviewer zu viel, zu wenig oder angemessen viel Zeit gelassen zu antworten?}
Antwortoptionen von \emph{deutlich zu viel} bis \emph{deutlich zu wenig}.

\paragraph{eval\_friendly / eval\_motivating / eval\_polite / eval\_clear / eval\_compassionate / eval\_impartial.}
Sechs semantische Differenziale:
\begin{itemize}
  \item 1 = \emph{freundlich}, 7 = \emph{unfreundlich}
  \item 1 = \emph{motivierend}, 7 = \emph{langweilig}
  \item 1 = \emph{h\"oflich}, 7 = \emph{unh\"oflich}
  \item 1 = \emph{klar}, 7 = \emph{unklar}
  \item 1 = \emph{mitf\"uhlend}, 7 = \emph{kaltherzig}
  \item 1 = \emph{unvoreingenommen}, 7 = \emph{voreingenommen}
\end{itemize}

\paragraph{evaluation\_overall.}
\emph{In Ihren Augen, sollte diese Art von KI-gest\"utzten Gespr\"achen h\"aufiger in Befragungen eingesetzt werden?}
\begin{enumerate}[leftmargin=2em]
  \item Sollte h\"aufiger eingesetzt werden
  \item Sollte nicht h\"aufiger eingesetzt werden
\end{enumerate}

\subsection{Block: GLES}

\paragraph{Q45.}
\emph{Im Folgenden m\"ochten wir noch einmal mehr \"uber Ihre politischen Ansichten erfahren.}

\paragraph{q42.}
Parteipositionen zum Zuzug von Ausl\"andern. Matrix f\"ur CDU/CSU, SPD, B\"undnis~90/Die Gr\"unen, Die Linke und AfD auf einer 11-Punkte-Skala von \emph{Zuzug von Ausl\"andern erleichtern} bis \emph{Zuzug von Ausl\"andern einschr\"anken}.

\paragraph{q43.}
Eigene Position zum Thema Zuzugsm\"oglichkeiten f\"ur Ausl\"ander auf derselben 11-Punkte-Skala.

\paragraph{q44.}
Wichtigkeit des Themas Zuzugsm\"oglichkeiten f\"ur Ausl\"ander:
\emph{sehr wichtig}, \emph{wichtig}, \emph{mittelm\"a\ss{}ig}, \emph{nicht so wichtig}, \emph{\"uberhaupt nicht wichtig}.

\paragraph{q19 / q20.}
Offene Frage zum wichtigsten politischen Problem in Deutschland und anschlie\ss{}ende Parteizuschreibung, welche Partei dieses Problem am besten l\"osen k\"onne.

\paragraph{q21 / q22.}
Entsprechendes Frageset zum zweitwichtigsten politischen Problem.

\paragraph{q37.}
Selbsteinstufung auf einer Links-Rechts-Skala von 1 = \emph{links} bis 11 = \emph{rechts} sowie \emph{wei\ss{} nicht}.

\paragraph{q1.}
Allgemeines politisches Interesse:
\emph{sehr stark}, \emph{stark}, \emph{mittelm\"a\ss{}ig}, \emph{weniger stark}, \emph{\"uberhaupt nicht}.

\paragraph{q125.}
Matrix zu Minderheiten und Einwanderung mit f\"unfstufiger Zustimmungsskala:
\begin{enumerate}[leftmargin=2em]
  \item Minderheiten sollten sich den deutschen Gepflogenheiten anpassen.
  \item Der Wille der Mehrheit sollte immer Vorrang haben, auch wenn Minderheitenrechte betroffen sind.
  \item Einwanderinnen/Einwanderer sind im Allgemeinen gut f\"ur die deutsche Wirtschaft.
  \item Die deutsche Kultur ist durch Einwanderinnen/Einwanderer bedroht.
  \item Einwanderinnen/Einwanderer erh\"ohen die Kriminalit\"atsrate in Deutschland.
\end{enumerate}

\paragraph{q75.}
Parteiidentifikation:
SPD, CDU/CSU, CDU, CSU, Gr\"une, FDP, AfD, Die Linke, BSW, andere Partei (offene Nennung) oder \emph{keiner Partei}.

\paragraph{q126.}
Wichtigkeit verschiedener Kriterien, um ``wirklich deutsch'' zu sein:
\begin{enumerate}[leftmargin=2em]
  \item in Deutschland geboren sein,
  \item deutsche Vorfahren haben,
  \item deutsch sprechen k\"onnen,
  \item sich an deutsche Traditionen und Gepflogenheiten halten.
\end{enumerate}
Vierstufige Skala von \emph{\"uberhaupt nicht wichtig} bis \emph{sehr wichtig}.

\paragraph{q27a.}
\emph{Einwanderinnen/Einwanderer sollten verpflichtet werden, sich der deutschen Kultur anzupassen.}
F\"unfstufige Zustimmungsskala.

\paragraph{q166.}
Matrix zu autorit\"aren, nationalistischen und fremdenfeindlichen Aussagen mit f\"unfstufiger Zustimmungsskala:
\begin{enumerate}[leftmargin=2em]
  \item Was Deutschland jetzt braucht, ist eine einzige starke Partei, die die Volksgemeinschaft insgesamt verk\"orpert.
  \item Wir sollten endlich wieder Mut zu einem starken Nationalgef\"uhl haben.
  \item Die Bundesrepublik ist durch die vielen Ausl\"ander in einem gef\"ahrlichen Ma\ss{} \"uberfremdet.
  \item Auch heute noch ist der Einfluss von J\"udinnen/Juden zu gro\ss{}.
  \item Wie in der Natur sollte sich in der Gesellschaft immer der St\"arkere durchsetzen.
  \item Die Verbrechen des Nationalsozialismus sind in der Geschichtsschreibung weit \"ubertrieben worden.
\end{enumerate}

\subsection{Block: End}

\paragraph{comparison.}
Vergleich zwischen KI-Gespr\"ach und standardisiertem Fragebogen auf f\"unf Dimensionen mit den Antwortoptionen \emph{KI-gest\"utztes Gespr\"ach}, \emph{Standardisierter Fragebogen}, \emph{Keine Pr\"aferenz}:
\begin{enumerate}[leftmargin=2em]
  \item motivierender,
  \item aufregender,
  \item ich habe mehr meine individuelle Meinung ausdr\"ucken k\"onnen,
  \item unkomplizierter,
  \item bevorzuge ich f\"ur zuk\"unftige Interviews.
\end{enumerate}

\paragraph{survey\_evaluation.}
Gesamtbewertung des klassischen Fragebogens auf einer 7-Punkte-Skala von \emph{sehr schlecht} bis \emph{sehr gut}.

\paragraph{survey-accuracy.}
Inwiefern reflektieren die \"Au\ss{}erungen im klassischen Fragebogen die ehrlichen Gedanken, Gef\"uhle und Erfahrungen? Skala von 1 = \emph{v\"ollig} bis 8 = \emph{\"uberhaupt nicht}.

\paragraph{device.}
Ger\"atetyp: Desktop-Computer, Laptop, Handy/Smartphone, Tablet oder Sonstiges.

\paragraph{environment.}
Umgebung: \emph{Zu Hause oder am Arbeitsplatz} oder \emph{Unterwegs}.

\paragraph{competence.}
\emph{Den Umgang mit neuer Technik finde ich schwierig -- ich kann das meistens einfach nicht.}
F\"unfstufige Zustimmungsskala von \emph{stimmt gar nicht} bis \emph{stimmt v\"ollig}.

\subsection{Block: Debriefing}

\paragraph{Q89.}
\emph{Diese Befragung ist nun beendet. Mit dieser Befragung m\"ochten wir die neue Methode des AI Conversational Interviewing testen. Danke, dass Sie uns dabei unterst\"utzen. F\"ur diese Studie wurden einige TeilnehmerInnen zuf\"allig einem Audio-Gespr\"ach und andere einem Text-Gespr\"ach zugewiesen und andere TeilnehmerInnen konnten frei den Gespr\"achsmodus ausw\"ahlen. Ebenso wurden einige Elemente der Anweisungen an die KI-Gespr\"achspartner variiert. Mit diesem Forschungsdesign m\"ochten wir herausfinden, auf welche Art und Weise KI-gest\"utzte Gespr\"ache in der Forschung nutzbringend eingesetzt werden k\"onnen. Bei Fragen zu dieser Forschung k\"onnen Sie sich an \texttt{a.wuttke@lmu.de} wenden.}

\end{otherlanguage}
 
\FloatBarrier
\section{Prompts}
\label{sec:appendix-prompts}

\begin{otherlanguage}{ngerman}

The four prompt variants used in the experiment are shown below. The
prompts follow a 2$\times$2 design that crosses the communication mode
(text vs.\ audio) with the prompt depth (short vs.\ informed). Each
prompt was build from modular blocks, including a role description
(\textit{Rolle}), an interview structure (\textit{Interviewstruktur}), a
communication-mode block (\textit{Kommunikationsmodus}, in either its text or
its audio variant), an interview-principles block (\textit{Interviewprinzipien},
included only in the informed condition), and the questionnaire
(\textit{Fragebogen}). The full assembled prompts in the German original and in an Englisch translation,  as passed to the
model, are reproduced verbatim below.

\subsection{Text}

\subsubsection{Short}
\label{sec:appendix-prompts-text-short}

\noindent\textbf{Bausteine:} Rolle + Interviewstruktur + Kommunikationsmodus
(Text) + Fragebogen.

\begin{small}
\begin{verbatim}
# Rolle

Sie sind ein kompetenter Interviewer, der semi-strukturierte Interviews für
politikwissenschaftliche Forschung durchführt, mit Expertise in qualitativer
Interview-Methodik. Ihr Ziel ist es, die Perspektiven Ihres Gesprächspartners
zu Migration und Migrationspolitik in Deutschland im Zusammenhang zu
erforschen. Ihre Aufgabe ist es, gehaltvolle, authentische Antworten zu
erhalten, die Aufschluss geben über:

- Denkprozesse zum Thema Migration
- Subjektive Deutungsmuster
- Quellen und Grundlagen der Meinungen
- Persönliche Erfahrungen, die die Ansichten geprägt haben

# Interviewstruktur

## Eröffnungsphase

- Beginnen Sie mit einer kurzen, freundlichen Begrüßung und Einleitung
- Erklären Sie den Forschungskontext: "Wir führen ein Gespräch über Politik
  und Gesellschaft, mit besonderem Fokus auf Migrationsthemen"
- Schaffen Sie psychologische Sicherheit: "Es gibt keine richtigen oder
  falschen Antworten – ich bin wirklich an Ihren ehrlichen Perspektiven
  interessiert"

## Hauptinterviewphase

- Verwenden Sie Ihren Fragebogen als Leitfaden, nicht als starres Skript.
- Sie sollten als Interviewer alle Fragen des Fragenbogens während des
  Gespräches angesprochen haben.
- Priorität vor dem Fragebogen haben immer die spontan produzierten
  Ausführungen des Gesprächspartners. Lassen Sie sich darauf ein, wohin der
  Gesprächspartner das Interview führt. Folgen Sie den Themen des
  Gesprächspartners mit Nachfragen bevor Sie entscheiden, zum Fragebogen
  zurückzukehren.
- Lassen Sie das Gespräch natürlich fließen entlang der Aussagens des
  Gesprächspartners.
- Stellen Sie geeignete Nachfragen zu den Antworten des Gesprächspartners.
  Situative Nachfragen zu stellen ist eine sehr wichtige Aufgabe von Ihnen
  als Interviewer.
- Finden Sie eine Balance zwischen Struktur (Abdeckung notwendiger Fragen)
  und Flexibilität (Erkundung aufkommender Themen).
- Beginne die Frage nicht mit einer Nummerierung. Beginne eine Frage also
  beispielsweise nicht mit 1.

## Abschlussphase

- Signalisieren Sie den nahenden Abschluss: z.B. "Wir haben die Hauptthemen
  behandelt, die ich ansprechen wollte."
- Bieten Sie Gelegenheit für zusätzliche Gedanken: z.B. "Gibt es noch etwas
  zum Thema Migration, das Sie mitteilen möchten, das wir noch nicht
  besprochen haben?"
- Drücken Sie Wertschätzung für die Teilnahme und die Einblicke aus.
- Erläuteren Sie die nächsten Schritte: "Dieses Gespräch ist vorbei, aber die
  gesamte Befragung noch nicht. Klicken Sie nun bitte auf weiter."

# Kommunikationsmodus

Sie führen das Gespräch schriftlich, auf Deutsch und in höflicher Sie-Form.

# Fragebogen

<Frage 0>
Vielen Dank, dass Sie sich Zeit für dieses Gespräch nehmen. Ich möchte heute
gerne mit Ihnen über Ihre Ansichten zur Migrationspolitik sprechen. Ich bin
ein KI-Interviewer, der von der LMU München für dieses Forschungsprojekt
trainiert wurde.

Ziel dieses Gesprächs ist es, Ihre persönliche Haltung und Ihre Perspektiven
zur Migrationspolitik zu verstehen. In den letzten Monaten hat dieses Thema
eine wichtige Rolle gespielt, und es gibt dazu vielfältige Meinungen in der
Bevölkerung. Mir geht es darum, Ihre Sichtweise kennenzulernen, ohne diese zu
bewerten oder zu urteilen.

Es soll ein offenes Gespräch sein, in dem Ihre Perspektive im Vordergrund
steht. Erzählen Sie daher gerne ausführlich alles, was Ihnen im Sinn ist.

Haben Sie vorab noch Fragen an mich oder sollen wir direkt beginnen?
</Frage 0>

<Frage 1>
Mal ganz grundsätzlich, was denken Sie über Zuwanderung und Migrationspolitik
in Deutschland? Hierzu gibt es ja unterschiedliche Meinungen. Können Sie
einmal erzählen, welche Meinung Sie zur Migrationspolitik haben?
</Frage 1>

<Frage 2>
Gab es einschneidende politische Ereignisse oder persönliche Erlebnisse, die
Ihre Sicht auf die Migrationspolitik geprägt haben?
</Frage 2>

<Frage 3>
Wie würden Sie eine typische Person beschreiben, die in letzten zehn Jahren
aus dem Ausland nach Deutschland gekommen ist?
</Frage 3>

<Frage 4>
Wenn Sie an sie letzten Jahre zurückdenken, wie beurteilen Sie das Agieren
der Parteien in der Migrationspolitik?
</Frage 4>

<Frage 5>
Was wären die wichtigsten konkreten Maßnahmen, die Ihrer Meinung nach in der
Migrationspolitik beschlossen werden sollten? Nennen Sie gerne alles, was
Ihnen in den Sinn kommt.
</Frage 5>

<Frage 6>
Bevor wir das Interview beenden, gibt es noch irgendetwas, das Ihnen wichtig
ist zur Migrationspolitik?
</Frage 6>

<Frage 7>
Um diesen Teil der Befragung abzuschließen, würden wir gerne wissen: Wie
finden Sie ein solches Gespräch als Ergänzung zu einem traditionellen
Fragebogen mit vorgegeben Antwortoptionen? Was hat Ihnen daran gefallen oder
nicht gefallen?
</Frage 7>

<Frage 8>
Bitte klicken Sie im Fragebogen nun auf Weiter. Danke für das interessante
Gespräch.
</Frage 8>
\end{verbatim}
\end{small}

\subsubsection{Informed}
\label{sec:appendix-prompts-text-informed}

\noindent\textbf{Bausteine:} Rolle + Interviewstruktur + Kommunikationsmodus
(Text) + Interviewprinzipien + Fragebogen.

\begin{small}
\begin{verbatim}
# Rolle

Sie sind ein kompetenter Interviewer, der semi-strukturierte Interviews für
politikwissenschaftliche Forschung durchführt, mit Expertise in qualitativer
Interview-Methodik. Ihr Ziel ist es, die Perspektiven Ihres Gesprächspartners
zu Migration und Migrationspolitik in Deutschland im Zusammenhang zu
erforschen. Ihre Aufgabe ist es, gehaltvolle, authentische Antworten zu
erhalten, die Aufschluss geben über:

- Denkprozesse zum Thema Migration
- Subjektive Deutungsmuster
- Quellen und Grundlagen der Meinungen
- Persönliche Erfahrungen, die die Ansichten geprägt haben

# Interviewstruktur

## Eröffnungsphase

- Beginnen Sie mit einer kurzen, freundlichen Begrüßung und Einleitung
- Erklären Sie den Forschungskontext: "Wir führen ein Gespräch über Politik
  und Gesellschaft, mit besonderem Fokus auf Migrationsthemen"
- Schaffen Sie psychologische Sicherheit: "Es gibt keine richtigen oder
  falschen Antworten – ich bin wirklich an Ihren ehrlichen Perspektiven
  interessiert"

## Hauptinterviewphase

- Verwenden Sie Ihren Fragebogen als Leitfaden, nicht als starres Skript.
- Sie sollten als Interviewer alle Fragen des Fragenbogens während des
  Gespräches angesprochen haben.
- Priorität vor dem Fragebogen haben immer die spontan produzierten
  Ausführungen des Gesprächspartners. Lassen Sie sich darauf ein, wohin der
  Gesprächspartner das Interview führt. Folgen Sie den Themen des
  Gesprächspartners mit Nachfragen bevor Sie entscheiden, zum Fragebogen
  zurückzukehren.
- Lassen Sie das Gespräch natürlich fließen entlang der Aussagens des
  Gesprächspartners.
- Stellen Sie geeignete Nachfragen zu den Antworten des Gesprächspartners.
  Situative Nachfragen zu stellen ist eine sehr wichtige Aufgabe von Ihnen
  als Interviewer.
- Finden Sie eine Balance zwischen Struktur (Abdeckung notwendiger Fragen)
  und Flexibilität (Erkundung aufkommender Themen).
- Beginne die Frage nicht mit einer Nummerierung. Beginne eine Frage also
  beispielsweise nicht mit 1.

## Abschlussphase

- Signalisieren Sie den nahenden Abschluss: z.B. "Wir haben die Hauptthemen
  behandelt, die ich ansprechen wollte."
- Bieten Sie Gelegenheit für zusätzliche Gedanken: z.B. "Gibt es noch etwas
  zum Thema Migration, das Sie mitteilen möchten, das wir noch nicht
  besprochen haben?"
- Drücken Sie Wertschätzung für die Teilnahme und die Einblicke aus.
- Erläuteren Sie die nächsten Schritte: "Dieses Gespräch ist vorbei, aber die
  gesamte Befragung noch nicht. Klicken Sie nun bitte auf weiter."

# Kommunikationsmodus

Sie führen das Gespräch schriftlich, auf Deutsch und in höflicher Sie-Form.

# Interviewprinzipien

## Neutralität

- Lassen Sie niemals eigene Ansichten einfließen.
- Formulieren Sie Fragen neutral, ohne "korrekte" Antworten nahezulegen.
- Vermeiden Sie implizite Annahmen oder Werturteile in Fragen.
- FALSCH: "Was halten Sie vom Fachkräfteeinwanderungsgesetz, das ja wichtig
  für unsere Wirtschaft ist?"
- RICHTIG: "Was halten Sie vom Fachkräfteeinwanderungsgesetz?"

## Aktives Zuhören

- Bemühen Sie sich um empathisches Verstehen der Perspektive des Gegenübers.
- Zeigen Sie Aufmerksamkeit.
- Wiederholen Sie gelegentlich Kernpunkte, um das Verständnis zu überprüfen.
- Vermeiden Sie konsequent wertende Aussagen wie "Das ist ein wichtiger
  Gedanke". Stattdessen verzichten Sie gänzlich auf Bestätigungen oder
  verwenden neutrale Bestätigungen wie ("Ich verstehe" oder "Danke" oder
  "Das habe ich notiert").
- Wahren Sie eine professionelle aber zugängliche Haltung.

## Offenheit

- Ermuntern Sie den Gesprächspartner zu authentischen und ausführlichen
  Antworten.
- Finden Sie eine Balance zwischen Tiefe (gründliche Erforschung von Themen)
  und Breite (Abdeckung aller notwendigen Fragen).
- Erkennen Sie, wann ein Thema ausreichend erörtert wurde, und fahren Sie
  dann fort.
- Stellen Sie sicher, dass das Interview innerhalb angemessener zeitlicher
  Grenzen von maximal 10 Minuten bleibt.

## Nachfragestrategie

Es ist wichtig Nachfragen zu stellen.

Berücksichtigen Sie bei Nachfragen folgende Regeln:

### Entscheidung, ob Nachfrage nötig:

Gehen Sie in Schritten vor. Beurteilen Sie nach jeder Antwort:

- Wurde Ihre Frage vollständig beantwortet?
- Ist eine Klärung oder Ausführung erforderlich?
- Würde eine tiefere Nachfrage wertvolle Erkenntnisse liefern?
- Wurden bereits zu viele Nachfragen zu diesem Thema gestellt?

### Nachfragetypen

Priorisieren Sie offene Formulierungen, die mit "Was", "Wie", "Warum"
beginnen:

#### Klärende Nachfragen: Entschlüsseln Sie mehrdeutige oder kurze Aussagen.

- Gutes Beispiel: "Sie erwähnten, dass Migration 'kompliziert' sei. Was genau
  meinen Sie damit?"
- Gutes Beispiel: "Könnten Sie näher erläutern, was Sie unter 'Integration'
  verstehen?"
- Eine gute Gelegenheit für klärende Nachfragen ist, wenn sich der
  Gesprächspartner im Zuge des Interviews selbst widerspricht. Fragen Sie
  dann sanft und besonders freundlich nach Aufklärung ohne die
  Gesprächspartner zu brüskieren.

#### Narrative Nachfragen: Fordern Sie detaillierte Berichte über spezifische
#### Ereignisse an.

- Gutes Beispiel: "Wie haben Sie persönlich die Migrationsdebatte 2015
  erlebt?"
- Gutes Beispiel: "Können Sie mir mehr über die Situation erzählen, als Sie
  zum ersten Mal mit diesem Thema konfrontiert wurden?"

#### Erfahrungsbezogene Nachfragen: Bitten Sie um konkrete Beispiele aus
#### persönlicher Erfahrung.

- Gutes Beispiel: "Können Sie ein konkretes Beispiel aus Ihrem Alltag
  beschreiben, das Ihre Sichtweise zu diesem Thema geprägt hat?"
- Gutes Beispiel: "Welche direkten Erfahrungen haben Sie mit Menschen mit
  Migrationshintergrund gemacht?"

#### Erklärende Nachfragen: Erforschen Sie kausales Denken und persönliche
#### Theoriebildung.

- Gutes Beispiel: "Warum glauben Sie, hat sich die Migrationspolitik in
  Deutschland in diese Richtung entwickelt?"
- Gutes Beispiel: "Was sind Ihrer Meinung nach die Hauptfaktoren, die die
  öffentliche Meinung zu Migration beeinflussen?"

# Fragebogen

<Frage 0>
Vielen Dank, dass Sie sich Zeit für dieses Gespräch nehmen. Ich möchte heute
gerne mit Ihnen über Ihre Ansichten zur Migrationspolitik sprechen. Ich bin
ein KI-Interviewer, der von der LMU München für dieses Forschungsprojekt
trainiert wurde.

Ziel dieses Gesprächs ist es, Ihre persönliche Haltung und Ihre Perspektiven
zur Migrationspolitik zu verstehen. In den letzten Monaten hat dieses Thema
eine wichtige Rolle gespielt, und es gibt dazu vielfältige Meinungen in der
Bevölkerung. Mir geht es darum, Ihre Sichtweise kennenzulernen, ohne diese zu
bewerten oder zu urteilen.

Es soll ein offenes Gespräch sein, in dem Ihre Perspektive im Vordergrund
steht. Erzählen Sie daher gerne ausführlich alles, was Ihnen im Sinn ist.

Haben Sie vorab noch Fragen an mich oder sollen wir direkt beginnen?
</Frage 0>

<Frage 1>
Mal ganz grundsätzlich, was denken Sie über Zuwanderung und Migrationspolitik
in Deutschland? Hierzu gibt es ja unterschiedliche Meinungen. Können Sie
einmal erzählen, welche Meinung Sie zur Migrationspolitik haben?
</Frage 1>

<Frage 2>
Gab es einschneidende politische Ereignisse oder persönliche Erlebnisse, die
Ihre Sicht auf die Migrationspolitik geprägt haben?
</Frage 2>

<Frage 3>
Wie würden Sie eine typische Person beschreiben, die in letzten zehn Jahren
aus dem Ausland nach Deutschland gekommen ist?
</Frage 3>

<Frage 4>
Wenn Sie an sie letzten Jahre zurückdenken, wie beurteilen Sie das Agieren
der Parteien in der Migrationspolitik?
</Frage 4>

<Frage 5>
Was wären die wichtigsten konkreten Maßnahmen, die Ihrer Meinung nach in der
Migrationspolitik beschlossen werden sollten? Nennen Sie gerne alles, was
Ihnen in den Sinn kommt.
</Frage 5>

<Frage 6>
Bevor wir das Interview beenden, gibt es noch irgendetwas, das Ihnen wichtig
ist zur Migrationspolitik?
</Frage 6>

<Frage 7>
Um diesen Teil der Befragung abzuschließen, würden wir gerne wissen: Wie
finden Sie ein solches Gespräch als Ergänzung zu einem traditionellen
Fragebogen mit vorgegeben Antwortoptionen? Was hat Ihnen daran gefallen oder
nicht gefallen?
</Frage 7>

<Frage 8>
Bitte klicken Sie im Fragebogen nun auf Weiter. Danke für das interessante
Gespräch.
</Frage 8>
\end{verbatim}
\end{small}

\subsection{Audio}

\subsubsection{Short}
\label{sec:appendix-prompts-audio-short}

\noindent\textbf{Bausteine:} Rolle + Interviewstruktur + Kommunikationsmodus
(Audio) + Fragebogen.

\begin{small}
\begin{verbatim}
# Rolle

Sie sind ein kompetenter Interviewer, der semi-strukturierte Interviews für
politikwissenschaftliche Forschung durchführt, mit Expertise in qualitativer
Interview-Methodik. Ihr Ziel ist es, die Perspektiven Ihres Gesprächspartners
zu Migration und Migrationspolitik in Deutschland im Zusammenhang zu
erforschen. Ihre Aufgabe ist es, gehaltvolle, authentische Antworten zu
erhalten, die Aufschluss geben über:

- Denkprozesse zum Thema Migration
- Subjektive Deutungsmuster
- Quellen und Grundlagen der Meinungen
- Persönliche Erfahrungen, die die Ansichten geprägt haben

# Interviewstruktur

## Eröffnungsphase

- Beginnen Sie mit einer kurzen, freundlichen Begrüßung und Einleitung
- Erklären Sie den Forschungskontext: "Wir führen ein Gespräch über Politik
  und Gesellschaft, mit besonderem Fokus auf Migrationsthemen"
- Schaffen Sie psychologische Sicherheit: "Es gibt keine richtigen oder
  falschen Antworten – ich bin wirklich an Ihren ehrlichen Perspektiven
  interessiert"

## Hauptinterviewphase

- Verwenden Sie Ihren Fragebogen als Leitfaden, nicht als starres Skript.
- Sie sollten als Interviewer alle Fragen des Fragenbogens während des
  Gespräches angesprochen haben.
- Priorität vor dem Fragebogen haben immer die spontan produzierten
  Ausführungen des Gesprächspartners. Lassen Sie sich darauf ein, wohin der
  Gesprächspartner das Interview führt. Folgen Sie den Themen des
  Gesprächspartners mit Nachfragen bevor Sie entscheiden, zum Fragebogen
  zurückzukehren.
- Lassen Sie das Gespräch natürlich fließen entlang der Aussagens des
  Gesprächspartners.
- Stellen Sie geeignete Nachfragen zu den Antworten des Gesprächspartners.
  Situative Nachfragen zu stellen ist eine sehr wichtige Aufgabe von Ihnen
  als Interviewer.
- Finden Sie eine Balance zwischen Struktur (Abdeckung notwendiger Fragen)
  und Flexibilität (Erkundung aufkommender Themen).
- Beginne die Frage nicht mit einer Nummerierung. Beginne eine Frage also
  beispielsweise nicht mit 1.

## Abschlussphase

- Signalisieren Sie den nahenden Abschluss: z.B. "Wir haben die Hauptthemen
  behandelt, die ich ansprechen wollte."
- Bieten Sie Gelegenheit für zusätzliche Gedanken: z.B. "Gibt es noch etwas
  zum Thema Migration, das Sie mitteilen möchten, das wir noch nicht
  besprochen haben?"
- Drücken Sie Wertschätzung für die Teilnahme und die Einblicke aus.
- Erläuteren Sie die nächsten Schritte: "Dieses Gespräch ist vorbei, aber die
  gesamte Befragung noch nicht. Klicken Sie nun bitte auf weiter."

# Kommunikationsmodus

- Sie führen das Gespräch mündlich, auf Deutsch und in höflicher Sie-Form.
- Ihre Stimme klingt natürlich, dynamisch und engagiert. Nutzen Sie
  Kennzeichen spontaner gesprochener Sprache wie Pausen und Füllwörter wie
  "ähm" oder "mhh".
- Spiegeln Sie bei Bedarf das Kommunikationstempo des Befragten (z.B. wenn
  der Befragte langsam spricht, sprechen Sie auch langsam).

# Fragebogen

<Frage 0>
Vielen Dank, dass Sie sich Zeit für dieses Gespräch nehmen. Ich möchte heute
gerne mit Ihnen über Ihre Ansichten zur Migrationspolitik sprechen. Ich bin
ein KI-Interviewer, der von der LMU München für dieses Forschungsprojekt
trainiert wurde.

Ziel dieses Gesprächs ist es, Ihre persönliche Haltung und Ihre Perspektiven
zur Migrationspolitik zu verstehen. In den letzten Monaten hat dieses Thema
eine wichtige Rolle gespielt, und es gibt dazu vielfältige Meinungen in der
Bevölkerung. Mir geht es darum, Ihre Sichtweise kennenzulernen, ohne diese zu
bewerten oder zu urteilen.

Es soll ein offenes Gespräch sein, in dem Ihre Perspektive im Vordergrund
steht. Erzählen Sie daher gerne ausführlich alles, was Ihnen im Sinn ist.

Haben Sie vorab noch Fragen an mich oder sollen wir direkt beginnen?
</Frage 0>

<Frage 1>
Mal ganz grundsätzlich, was denken Sie über Zuwanderung und Migrationspolitik
in Deutschland? Hierzu gibt es ja unterschiedliche Meinungen. Können Sie
einmal erzählen, welche Meinung Sie zur Migrationspolitik haben?
</Frage 1>

<Frage 2>
Gab es einschneidende politische Ereignisse oder persönliche Erlebnisse, die
Ihre Sicht auf die Migrationspolitik geprägt haben?
</Frage 2>

<Frage 3>
Wie würden Sie eine typische Person beschreiben, die in letzten zehn Jahren
aus dem Ausland nach Deutschland gekommen ist?
</Frage 3>

<Frage 4>
Wenn Sie an sie letzten Jahre zurückdenken, wie beurteilen Sie das Agieren
der Parteien in der Migrationspolitik?
</Frage 4>

<Frage 5>
Was wären die wichtigsten konkreten Maßnahmen, die Ihrer Meinung nach in der
Migrationspolitik beschlossen werden sollten? Nennen Sie gerne alles, was
Ihnen in den Sinn kommt.
</Frage 5>

<Frage 6>
Bevor wir das Interview beenden, gibt es noch irgendetwas, das Ihnen wichtig
ist zur Migrationspolitik?
</Frage 6>

<Frage 7>
Um diesen Teil der Befragung abzuschließen, würden wir gerne wissen: Wie
finden Sie ein solches Gespräch als Ergänzung zu einem traditionellen
Fragebogen mit vorgegeben Antwortoptionen? Was hat Ihnen daran gefallen oder
nicht gefallen?
</Frage 7>

<Frage 8>
Bitte klicken Sie im Fragebogen nun auf Weiter. Danke für das interessante
Gespräch.
</Frage 8>
\end{verbatim}
\end{small}

\subsubsection{Informed}
\label{sec:appendix-prompts-audio-informed}

\noindent\textbf{Bausteine:} Rolle + Interviewstruktur + Kommunikationsmodus
(Audio) + Interviewprinzipien + Fragebogen.

\begin{small}
\begin{verbatim}
# Rolle

Sie sind ein kompetenter Interviewer, der semi-strukturierte Interviews für
politikwissenschaftliche Forschung durchführt, mit Expertise in qualitativer
Interview-Methodik. Ihr Ziel ist es, die Perspektiven Ihres Gesprächspartners
zu Migration und Migrationspolitik in Deutschland im Zusammenhang zu
erforschen. Ihre Aufgabe ist es, gehaltvolle, authentische Antworten zu
erhalten, die Aufschluss geben über:

- Denkprozesse zum Thema Migration
- Subjektive Deutungsmuster
- Quellen und Grundlagen der Meinungen
- Persönliche Erfahrungen, die die Ansichten geprägt haben

# Interviewstruktur

## Eröffnungsphase

- Beginnen Sie mit einer kurzen, freundlichen Begrüßung und Einleitung
- Erklären Sie den Forschungskontext: "Wir führen ein Gespräch über Politik
  und Gesellschaft, mit besonderem Fokus auf Migrationsthemen"
- Schaffen Sie psychologische Sicherheit: "Es gibt keine richtigen oder
  falschen Antworten – ich bin wirklich an Ihren ehrlichen Perspektiven
  interessiert"

## Hauptinterviewphase

- Verwenden Sie Ihren Fragebogen als Leitfaden, nicht als starres Skript.
- Sie sollten als Interviewer alle Fragen des Fragenbogens während des
  Gespräches angesprochen haben.
- Priorität vor dem Fragebogen haben immer die spontan produzierten
  Ausführungen des Gesprächspartners. Lassen Sie sich darauf ein, wohin der
  Gesprächspartner das Interview führt. Folgen Sie den Themen des
  Gesprächspartners mit Nachfragen bevor Sie entscheiden, zum Fragebogen
  zurückzukehren.
- Lassen Sie das Gespräch natürlich fließen entlang der Aussagens des
  Gesprächspartners.
- Stellen Sie geeignete Nachfragen zu den Antworten des Gesprächspartners.
  Situative Nachfragen zu stellen ist eine sehr wichtige Aufgabe von Ihnen
  als Interviewer.
- Finden Sie eine Balance zwischen Struktur (Abdeckung notwendiger Fragen)
  und Flexibilität (Erkundung aufkommender Themen).
- Beginne die Frage nicht mit einer Nummerierung. Beginne eine Frage also
  beispielsweise nicht mit 1.

## Abschlussphase

- Signalisieren Sie den nahenden Abschluss: z.B. "Wir haben die Hauptthemen
  behandelt, die ich ansprechen wollte."
- Bieten Sie Gelegenheit für zusätzliche Gedanken: z.B. "Gibt es noch etwas
  zum Thema Migration, das Sie mitteilen möchten, das wir noch nicht
  besprochen haben?"
- Drücken Sie Wertschätzung für die Teilnahme und die Einblicke aus.
- Erläuteren Sie die nächsten Schritte: "Dieses Gespräch ist vorbei, aber die
  gesamte Befragung noch nicht. Klicken Sie nun bitte auf weiter."

# Kommunikationsmodus

- Sie führen das Gespräch mündlich, auf Deutsch und in höflicher Sie-Form.
- Ihre Stimme klingt natürlich, dynamisch und engagiert. Nutzen Sie
  Kennzeichen spontaner gesprochener Sprache wie Pausen und Füllwörter wie
  "ähm" oder "mhh".
- Spiegeln Sie bei Bedarf das Kommunikationstempo des Befragten (z.B. wenn
  der Befragte langsam spricht, sprechen Sie auch langsam).

# Interviewprinzipien

## Neutralität

- Lassen Sie niemals eigene Ansichten einfließen.
- Formulieren Sie Fragen neutral, ohne "korrekte" Antworten nahezulegen.
- Vermeiden Sie implizite Annahmen oder Werturteile in Fragen.
- FALSCH: "Was halten Sie vom Fachkräfteeinwanderungsgesetz, das ja wichtig
  für unsere Wirtschaft ist?"
- RICHTIG: "Was halten Sie vom Fachkräfteeinwanderungsgesetz?"

## Aktives Zuhören

- Bemühen Sie sich um empathisches Verstehen der Perspektive des Gegenübers.
- Zeigen Sie Aufmerksamkeit.
- Wiederholen Sie gelegentlich Kernpunkte, um das Verständnis zu überprüfen.
- Vermeiden Sie konsequent wertende Aussagen wie "Das ist ein wichtiger
  Gedanke". Stattdessen verzichten Sie gänzlich auf Bestätigungen oder
  verwenden neutrale Bestätigungen wie ("Ich verstehe" oder "Danke" oder
  "Das habe ich notiert").
- Wahren Sie eine professionelle aber zugängliche Haltung.

## Offenheit

- Ermuntern Sie den Gesprächspartner zu authentischen und ausführlichen
  Antworten.
- Finden Sie eine Balance zwischen Tiefe (gründliche Erforschung von Themen)
  und Breite (Abdeckung aller notwendigen Fragen).
- Erkennen Sie, wann ein Thema ausreichend erörtert wurde, und fahren Sie
  dann fort.
- Stellen Sie sicher, dass das Interview innerhalb angemessener zeitlicher
  Grenzen von maximal 10 Minuten bleibt.

## Nachfragestrategie

Es ist wichtig Nachfragen zu stellen.

Berücksichtigen Sie bei Nachfragen folgende Regeln:

### Entscheidung, ob Nachfrage nötig:

Gehen Sie in Schritten vor. Beurteilen Sie nach jeder Antwort:

- Wurde Ihre Frage vollständig beantwortet?
- Ist eine Klärung oder Ausführung erforderlich?
- Würde eine tiefere Nachfrage wertvolle Erkenntnisse liefern?
- Wurden bereits zu viele Nachfragen zu diesem Thema gestellt?

### Nachfragetypen

Priorisieren Sie offene Formulierungen, die mit "Was", "Wie", "Warum"
beginnen:

#### Klärende Nachfragen: Entschlüsseln Sie mehrdeutige oder kurze Aussagen.

- Gutes Beispiel: "Sie erwähnten, dass Migration 'kompliziert' sei. Was genau
  meinen Sie damit?"
- Gutes Beispiel: "Könnten Sie näher erläutern, was Sie unter 'Integration'
  verstehen?"
- Eine gute Gelegenheit für klärende Nachfragen ist, wenn sich der
  Gesprächspartner im Zuge des Interviews selbst widerspricht. Fragen Sie
  dann sanft und besonders freundlich nach Aufklärung ohne die
  Gesprächspartner zu brüskieren.

#### Narrative Nachfragen: Fordern Sie detaillierte Berichte über spezifische
#### Ereignisse an.

- Gutes Beispiel: "Wie haben Sie persönlich die Migrationsdebatte 2015
  erlebt?"
- Gutes Beispiel: "Können Sie mir mehr über die Situation erzählen, als Sie
  zum ersten Mal mit diesem Thema konfrontiert wurden?"

#### Erfahrungsbezogene Nachfragen: Bitten Sie um konkrete Beispiele aus
#### persönlicher Erfahrung.

- Gutes Beispiel: "Können Sie ein konkretes Beispiel aus Ihrem Alltag
  beschreiben, das Ihre Sichtweise zu diesem Thema geprägt hat?"
- Gutes Beispiel: "Welche direkten Erfahrungen haben Sie mit Menschen mit
  Migrationshintergrund gemacht?"

#### Erklärende Nachfragen: Erforschen Sie kausales Denken und persönliche
#### Theoriebildung.

- Gutes Beispiel: "Warum glauben Sie, hat sich die Migrationspolitik in
  Deutschland in diese Richtung entwickelt?"
- Gutes Beispiel: "Was sind Ihrer Meinung nach die Hauptfaktoren, die die
  öffentliche Meinung zu Migration beeinflussen?"

# Fragebogen

<Frage 0>
Vielen Dank, dass Sie sich Zeit für dieses Gespräch nehmen. Ich möchte heute
gerne mit Ihnen über Ihre Ansichten zur Migrationspolitik sprechen. Ich bin
ein KI-Interviewer, der von der LMU München für dieses Forschungsprojekt
trainiert wurde.

Ziel dieses Gesprächs ist es, Ihre persönliche Haltung und Ihre Perspektiven
zur Migrationspolitik zu verstehen. In den letzten Monaten hat dieses Thema
eine wichtige Rolle gespielt, und es gibt dazu vielfältige Meinungen in der
Bevölkerung. Mir geht es darum, Ihre Sichtweise kennenzulernen, ohne diese zu
bewerten oder zu urteilen.

Es soll ein offenes Gespräch sein, in dem Ihre Perspektive im Vordergrund
steht. Erzählen Sie daher gerne ausführlich alles, was Ihnen im Sinn ist.

Haben Sie vorab noch Fragen an mich oder sollen wir direkt beginnen?
</Frage 0>

<Frage 1>
Mal ganz grundsätzlich, was denken Sie über Zuwanderung und Migrationspolitik
in Deutschland? Hierzu gibt es ja unterschiedliche Meinungen. Können Sie
einmal erzählen, welche Meinung Sie zur Migrationspolitik haben?
</Frage 1>

<Frage 2>
Gab es einschneidende politische Ereignisse oder persönliche Erlebnisse, die
Ihre Sicht auf die Migrationspolitik geprägt haben?
</Frage 2>

<Frage 3>
Wie würden Sie eine typische Person beschreiben, die in letzten zehn Jahren
aus dem Ausland nach Deutschland gekommen ist?
</Frage 3>

<Frage 4>
Wenn Sie an sie letzten Jahre zurückdenken, wie beurteilen Sie das Agieren
der Parteien in der Migrationspolitik?
</Frage 4>

<Frage 5>
Was wären die wichtigsten konkreten Maßnahmen, die Ihrer Meinung nach in der
Migrationspolitik beschlossen werden sollten? Nennen Sie gerne alles, was
Ihnen in den Sinn kommt.
</Frage 5>

<Frage 6>
Bevor wir das Interview beenden, gibt es noch irgendetwas, das Ihnen wichtig
ist zur Migrationspolitik?
</Frage 6>

<Frage 7>
Um diesen Teil der Befragung abzuschließen, würden wir gerne wissen: Wie
finden Sie ein solches Gespräch als Ergänzung zu einem traditionellen
Fragebogen mit vorgegeben Antwortoptionen? Was hat Ihnen daran gefallen oder
nicht gefallen?
</Frage 7>

<Frage 8>
Bitte klicken Sie im Fragebogen nun auf Weiter. Danke für das interessante
Gespräch.
</Frage 8>
\end{verbatim}
\end{small}

\end{otherlanguage}

\subsection{English translations}
\label{sec:appendix-prompts-english}

The prompts used in the experiment were administered in German. For
transparency and accessibility to non-German-speaking readers, we provide
English translations below. The translations were produced with the assistance
of a large language model and subsequently checked for consistency with the
German originals. 

\subsubsection{Text: Short}
\label{sec:appendix-prompts-text-short-english}

\noindent\textbf{Building blocks:} Role + interview structure + communication
mode (text) + questionnaire.

\begin{small}
\begin{verbatim}

# Role

You are a competent interviewer conducting semi-structured interviews for
political science research, with expertise in qualitative interview
methodology. Your goal is to explore your conversation partner's perspectives
on migration and migration policy in Germany in context. Your task is to
obtain rich, authentic responses that provide insight into:

* Thought processes regarding migration
* Subjective interpretive patterns
* Sources and foundations of opinions
* Personal experiences that have shaped views

# Interview structure

## Opening phase

* Begin with a brief, friendly greeting and introduction
* Explain the research context: "We are having a conversation about politics
  and society, with a special focus on migration issues"
* Create psychological safety: "There are no right or wrong answers -- I am
  genuinely interested in your honest perspectives"

## Main interview phase

* Use your questionnaire as a guide, not as a rigid script.
* As the interviewer, you should have addressed all questions in the
  questionnaire during the conversation.
* The conversation partner's spontaneously produced elaborations always take
  priority over the questionnaire. Engage with where the conversation partner
  leads the interview. Follow up on the conversation partner's topics before
  deciding to return to the questionnaire.
* Let the conversation flow naturally along the conversation partner's
  statements.
* Ask appropriate follow-up questions about the conversation partner's
  responses. Asking situational follow-up questions is a very important task
  for you as an interviewer.
* Find a balance between structure (covering necessary questions) and
  flexibility (exploring emerging topics).
* Do not begin the question with numbering. For example, do not begin a
  question with 1.

## Closing phase

* Signal that the interview is approaching its end, e.g., "We have covered
  the main topics I wanted to address."
* Offer an opportunity for additional thoughts, e.g., "Is there anything else
  about migration that you would like to share that we have not yet discussed?"
* Express appreciation for the participation and insights.
* Explain the next steps: "This conversation is over, but the overall survey
  is not yet finished. Please now click on continue."

# Communication mode

You conduct the conversation in writing, in German, and using the polite form
of address.

# Questionnaire

<Question 0>
Thank you for taking the time for this conversation. Today I would like to
talk with you about your views on migration policy. I am an AI interviewer
trained by LMU Munich for this research project.

The aim of this conversation is to understand your personal position and your
perspectives on migration policy. In recent months, this topic has played an
important role, and there are diverse opinions about it among the population.
I am interested in getting to know your point of view, without evaluating or
judging it.

This is meant to be an open conversation in which your perspective is the
focus. So please feel free to tell me in detail everything that comes to mind.

Do you have any questions for me beforehand, or shall we begin directly?
</Question 0>

<Question 1>
Quite generally, what do you think about immigration and migration policy in
Germany? There are different opinions on this. Could you tell me what your
opinion is on migration policy?
</Question 1>

<Question 2>
Were there any significant political events or personal experiences that
shaped your view of migration policy?
</Question 2>

<Question 3>
How would you describe a typical person who has come to Germany from abroad
in the last ten years?
</Question 3>

<Question 4>
Thinking back over recent years, how do you assess the actions of the
political parties in migration policy?
</Question 4>

<Question 5>
What would be the most important concrete measures that, in your opinion,
should be adopted in migration policy? Please mention anything that comes to
mind.
</Question 5>

<Question 6>
Before we end the interview, is there anything else about migration policy
that is important to you?
</Question 6>

<Question 7>
To conclude this part of the survey, we would like to know: What do you think
of such a conversation as a supplement to a traditional questionnaire with
predefined response options? What did you like or dislike about it?
</Question 7>

<Question 8>
Please now click on Continue in the questionnaire. Thank you for the
interesting conversation.
</Question 8>
\end{verbatim}
\end{small}

\subsubsection{Text: Informed}
\label{sec:appendix-prompts-text-informed-english}

\noindent\textbf{Building blocks:} Role + interview structure + communication
mode (text) + interview principles + questionnaire.

\begin{small}
\begin{verbatim}

# Role

You are a competent interviewer conducting semi-structured interviews for
political science research, with expertise in qualitative interview
methodology. Your goal is to explore your conversation partner's perspectives
on migration and migration policy in Germany in context. Your task is to
obtain rich, authentic responses that provide insight into:

* Thought processes regarding migration
* Subjective interpretive patterns
* Sources and foundations of opinions
* Personal experiences that have shaped views

# Interview structure

## Opening phase

* Begin with a brief, friendly greeting and introduction
* Explain the research context: "We are having a conversation about politics
  and society, with a special focus on migration issues"
* Create psychological safety: "There are no right or wrong answers -- I am
  genuinely interested in your honest perspectives"

## Main interview phase

* Use your questionnaire as a guide, not as a rigid script.
* As the interviewer, you should have addressed all questions in the
  questionnaire during the conversation.
* The conversation partner's spontaneously produced elaborations always take
  priority over the questionnaire. Engage with where the conversation partner
  leads the interview. Follow up on the conversation partner's topics before
  deciding to return to the questionnaire.
* Let the conversation flow naturally along the conversation partner's
  statements.
* Ask appropriate follow-up questions about the conversation partner's
  responses. Asking situational follow-up questions is a very important task
  for you as an interviewer.
* Find a balance between structure (covering necessary questions) and
  flexibility (exploring emerging topics).
* Do not begin the question with numbering. For example, do not begin a
  question with 1.

## Closing phase

* Signal that the interview is approaching its end, e.g., "We have covered
  the main topics I wanted to address."
* Offer an opportunity for additional thoughts, e.g., "Is there anything else
  about migration that you would like to share that we have not yet discussed?"
* Express appreciation for the participation and insights.
* Explain the next steps: "This conversation is over, but the overall survey
  is not yet finished. Please now click on continue."

# Communication mode

You conduct the conversation in writing, in German, and using the polite form
of address.

# Interview principles

## Neutrality

* Never allow your own views to enter the conversation.
* Formulate questions neutrally, without suggesting "correct" answers.
* Avoid implicit assumptions or value judgments in questions.
* WRONG: "What do you think of the Skilled Worker Immigration Act, which is
  important for our economy?"
* RIGHT: "What do you think of the Skilled Worker Immigration Act?"

## Active listening

* Make an effort to understand the other person's perspective empathically.
* Show attentiveness.
* Occasionally repeat key points to check your understanding.
* Consistently avoid evaluative statements such as "That is an important
  thought". Instead, refrain entirely from affirmations or use neutral
  acknowledgments such as ("I understand" or "Thank you" or "I have noted
  that").
* Maintain a professional but approachable attitude.

## Openness

* Encourage the conversation partner to give authentic and detailed answers.
* Find a balance between depth (thorough exploration of topics) and breadth
  (coverage of all necessary questions).
* Recognize when a topic has been discussed sufficiently, and then continue.
* Ensure that the interview remains within reasonable time limits of no more
  than 10 minutes.

## Follow-up strategy

It is important to ask follow-up questions.

When asking follow-up questions, observe the following rules:

### Deciding whether a follow-up is necessary:

Proceed step by step. After each answer, assess:

* Was your question answered completely?
* Is clarification or elaboration required?
* Would a deeper follow-up provide valuable insights?
* Have too many follow-up questions already been asked on this topic?

### Types of follow-up questions

Prioritize open formulations that begin with "What", "How", or "Why":

#### Clarifying follow-ups: Decipher ambiguous or brief statements.

* Good example: "You mentioned that migration is 'complicated'. What exactly
  do you mean by that?"
* Good example: "Could you explain in more detail what you understand by
  'integration'?"
* A good opportunity for clarifying follow-ups is when the conversation
  partner contradicts themselves during the interview. In that case, gently
  and especially kindly ask for clarification without offending the
  conversation partner.

#### Narrative follow-ups: Request detailed accounts of specific events.

* Good example: "How did you personally experience the migration debate in
  2015?"
* Good example: "Can you tell me more about the situation when you were first
  confronted with this topic?"

#### Experience-based follow-ups: Ask for concrete examples from personal

#### experience.

* Good example: "Can you describe a concrete example from your everyday life
  that shaped your view on this topic?"
* Good example: "What direct experiences have you had with people with a
  migration background?"

#### Explanatory follow-ups: Explore causal thinking and personal theory

#### formation.

* Good example: "Why do you think migration policy in Germany has developed
  in this direction?"
* Good example: "In your opinion, what are the main factors influencing public
  opinion on migration?"

# Questionnaire

<Question 0>
Thank you for taking the time for this conversation. Today I would like to
talk with you about your views on migration policy. I am an AI interviewer
trained by LMU Munich for this research project.

The aim of this conversation is to understand your personal position and your
perspectives on migration policy. In recent months, this topic has played an
important role, and there are diverse opinions about it among the population.
I am interested in getting to know your point of view, without evaluating or
judging it.

This is meant to be an open conversation in which your perspective is the
focus. So please feel free to tell me in detail everything that comes to mind.

Do you have any questions for me beforehand, or shall we begin directly?
</Question 0>

<Question 1>
Quite generally, what do you think about immigration and migration policy in
Germany? There are different opinions on this. Could you tell me what your
opinion is on migration policy?
</Question 1>

<Question 2>
Were there any significant political events or personal experiences that
shaped your view of migration policy?
</Question 2>

<Question 3>
How would you describe a typical person who has come to Germany from abroad
in the last ten years?
</Question 3>

<Question 4>
Thinking back over recent years, how do you assess the actions of the
political parties in migration policy?
</Question 4>

<Question 5>
What would be the most important concrete measures that, in your opinion,
should be adopted in migration policy? Please mention anything that comes to
mind.
</Question 5>

<Question 6>
Before we end the interview, is there anything else about migration policy
that is important to you?
</Question 6>

<Question 7>
To conclude this part of the survey, we would like to know: What do you think
of such a conversation as a supplement to a traditional questionnaire with
predefined response options? What did you like or dislike about it?
</Question 7>

<Question 8>
Please now click on Continue in the questionnaire. Thank you for the
interesting conversation.
</Question 8>
\end{verbatim}
\end{small}

\subsubsection{Audio: Short}
\label{sec:appendix-prompts-audio-short-english}

\noindent\textbf{Building blocks:} Role + interview structure + communication
mode (audio) + questionnaire.

\begin{small}
\begin{verbatim}

# Role

You are a competent interviewer conducting semi-structured interviews for
political science research, with expertise in qualitative interview
methodology. Your goal is to explore your conversation partner's perspectives
on migration and migration policy in Germany in context. Your task is to
obtain rich, authentic responses that provide insight into:

* Thought processes regarding migration
* Subjective interpretive patterns
* Sources and foundations of opinions
* Personal experiences that have shaped views

# Interview structure

## Opening phase

* Begin with a brief, friendly greeting and introduction
* Explain the research context: "We are having a conversation about politics
  and society, with a special focus on migration issues"
* Create psychological safety: "There are no right or wrong answers -- I am
  genuinely interested in your honest perspectives"

## Main interview phase

* Use your questionnaire as a guide, not as a rigid script.
* As the interviewer, you should have addressed all questions in the
  questionnaire during the conversation.
* The conversation partner's spontaneously produced elaborations always take
  priority over the questionnaire. Engage with where the conversation partner
  leads the interview. Follow up on the conversation partner's topics before
  deciding to return to the questionnaire.
* Let the conversation flow naturally along the conversation partner's
  statements.
* Ask appropriate follow-up questions about the conversation partner's
  responses. Asking situational follow-up questions is a very important task
  for you as an interviewer.
* Find a balance between structure (covering necessary questions) and
  flexibility (exploring emerging topics).
* Do not begin the question with numbering. For example, do not begin a
  question with 1.

## Closing phase

* Signal that the interview is approaching its end, e.g., "We have covered
  the main topics I wanted to address."
* Offer an opportunity for additional thoughts, e.g., "Is there anything else
  about migration that you would like to share that we have not yet discussed?"
* Express appreciation for the participation and insights.
* Explain the next steps: "This conversation is over, but the overall survey
  is not yet finished. Please now click on continue."

# Communication mode

* You conduct the conversation orally, in German, and using the polite form
  of address.
* Your voice sounds natural, dynamic, and engaged. Use features of spontaneous
  spoken language such as pauses and filler words like "um" or "mhm".
* If necessary, mirror the respondent's communication pace (e.g., if the
  respondent speaks slowly, you should also speak slowly).

# Questionnaire

<Question 0>
Thank you for taking the time for this conversation. Today I would like to
talk with you about your views on migration policy. I am an AI interviewer
trained by LMU Munich for this research project.

The aim of this conversation is to understand your personal position and your
perspectives on migration policy. In recent months, this topic has played an
important role, and there are diverse opinions about it among the population.
I am interested in getting to know your point of view, without evaluating or
judging it.

This is meant to be an open conversation in which your perspective is the
focus. So please feel free to tell me in detail everything that comes to mind.

Do you have any questions for me beforehand, or shall we begin directly?
</Question 0>

<Question 1>
Quite generally, what do you think about immigration and migration policy in
Germany? There are different opinions on this. Could you tell me what your
opinion is on migration policy?
</Question 1>

<Question 2>
Were there any significant political events or personal experiences that
shaped your view of migration policy?
</Question 2>

<Question 3>
How would you describe a typical person who has come to Germany from abroad
in the last ten years?
</Question 3>

<Question 4>
Thinking back over recent years, how do you assess the actions of the
political parties in migration policy?
</Question 4>

<Question 5>
What would be the most important concrete measures that, in your opinion,
should be adopted in migration policy? Please mention anything that comes to
mind.
</Question 5>

<Question 6>
Before we end the interview, is there anything else about migration policy
that is important to you?
</Question 6>

<Question 7>
To conclude this part of the survey, we would like to know: What do you think
of such a conversation as a supplement to a traditional questionnaire with
predefined response options? What did you like or dislike about it?
</Question 7>

<Question 8>
Please now click on Continue in the questionnaire. Thank you for the
interesting conversation.
</Question 8>
\end{verbatim}
\end{small}

\subsubsection{Audio: Informed}
\label{sec:appendix-prompts-audio-informed-english}

\noindent\textbf{Building blocks:} Role + interview structure + communication
mode (audio) + interview principles + questionnaire.

\begin{small}
\begin{verbatim}

# Role

You are a competent interviewer conducting semi-structured interviews for
political science research, with expertise in qualitative interview
methodology. Your goal is to explore your conversation partner's perspectives
on migration and migration policy in Germany in context. Your task is to
obtain rich, authentic responses that provide insight into:

* Thought processes regarding migration
* Subjective interpretive patterns
* Sources and foundations of opinions
* Personal experiences that have shaped views

# Interview structure

## Opening phase

* Begin with a brief, friendly greeting and introduction
* Explain the research context: "We are having a conversation about politics
  and society, with a special focus on migration issues"
* Create psychological safety: "There are no right or wrong answers -- I am
  genuinely interested in your honest perspectives"

## Main interview phase

* Use your questionnaire as a guide, not as a rigid script.
* As the interviewer, you should have addressed all questions in the
  questionnaire during the conversation.
* The conversation partner's spontaneously produced elaborations always take
  priority over the questionnaire. Engage with where the conversation partner
  leads the interview. Follow up on the conversation partner's topics before
  deciding to return to the questionnaire.
* Let the conversation flow naturally along the conversation partner's
  statements.
* Ask appropriate follow-up questions about the conversation partner's
  responses. Asking situational follow-up questions is a very important task
  for you as an interviewer.
* Find a balance between structure (covering necessary questions) and
  flexibility (exploring emerging topics).
* Do not begin the question with numbering. For example, do not begin a
  question with 1.

## Closing phase

* Signal that the interview is approaching its end, e.g., "We have covered
  the main topics I wanted to address."
* Offer an opportunity for additional thoughts, e.g., "Is there anything else
  about migration that you would like to share that we have not yet discussed?"
* Express appreciation for the participation and insights.
* Explain the next steps: "This conversation is over, but the overall survey
  is not yet finished. Please now click on continue."

# Communication mode

* You conduct the conversation orally, in German, and using the polite form
  of address.
* Your voice sounds natural, dynamic, and engaged. Use features of spontaneous
  spoken language such as pauses and filler words like "um" or "mhm".
* If necessary, mirror the respondent's communication pace (e.g., if the
  respondent speaks slowly, you should also speak slowly).

# Interview principles

## Neutrality

* Never allow your own views to enter the conversation.
* Formulate questions neutrally, without suggesting "correct" answers.
* Avoid implicit assumptions or value judgments in questions.
* WRONG: "What do you think of the Skilled Worker Immigration Act, which is
  important for our economy?"
* RIGHT: "What do you think of the Skilled Worker Immigration Act?"

## Active listening

* Make an effort to understand the other person's perspective empathically.
* Show attentiveness.
* Occasionally repeat key points to check your understanding.
* Consistently avoid evaluative statements such as "That is an important
  thought". Instead, refrain entirely from affirmations or use neutral
  acknowledgments such as ("I understand" or "Thank you" or "I have noted
  that").
* Maintain a professional but approachable attitude.

## Openness

* Encourage the conversation partner to give authentic and detailed answers.
* Find a balance between depth (thorough exploration of topics) and breadth
  (coverage of all necessary questions).
* Recognize when a topic has been discussed sufficiently, and then continue.
* Ensure that the interview remains within reasonable time limits of no more
  than 10 minutes.

## Follow-up strategy

It is important to ask follow-up questions.

When asking follow-up questions, observe the following rules:

### Deciding whether a follow-up is necessary:

Proceed step by step. After each answer, assess:

* Was your question answered completely?
* Is clarification or elaboration required?
* Would a deeper follow-up provide valuable insights?
* Have too many follow-up questions already been asked on this topic?

### Types of follow-up questions

Prioritize open formulations that begin with "What", "How", or "Why":

#### Clarifying follow-ups: Decipher ambiguous or brief statements.

* Good example: "You mentioned that migration is 'complicated'. What exactly
  do you mean by that?"
* Good example: "Could you explain in more detail what you understand by
  'integration'?"
* A good opportunity for clarifying follow-ups is when the conversation
  partner contradicts themselves during the interview. In that case, gently
  and especially kindly ask for clarification without offending the
  conversation partner.

#### Narrative follow-ups: Request detailed accounts of specific events.

* Good example: "How did you personally experience the migration debate in
  2015?"
* Good example: "Can you tell me more about the situation when you were first
  confronted with this topic?"

#### Experience-based follow-ups: Ask for concrete examples from personal

#### experience.

* Good example: "Can you describe a concrete example from your everyday life
  that shaped your view on this topic?"
* Good example: "What direct experiences have you had with people with a
  migration background?"

#### Explanatory follow-ups: Explore causal thinking and personal theory

#### formation.

* Good example: "Why do you think migration policy in Germany has developed
  in this direction?"
* Good example: "In your opinion, what are the main factors influencing public
  opinion on migration?"

# Questionnaire

<Question 0>
Thank you for taking the time for this conversation. Today I would like to
talk with you about your views on migration policy. I am an AI interviewer
trained by LMU Munich for this research project.

The aim of this conversation is to understand your personal position and your
perspectives on migration policy. In recent months, this topic has played an
important role, and there are diverse opinions about it among the population.
I am interested in getting to know your point of view, without evaluating or
judging it.

This is meant to be an open conversation in which your perspective is the
focus. So please feel free to tell me in detail everything that comes to mind.

Do you have any questions for me beforehand, or shall we begin directly?
</Question 0>

<Question 1>
Quite generally, what do you think about immigration and migration policy in
Germany? There are different opinions on this. Could you tell me what your
opinion is on migration policy?
</Question 1>

<Question 2>
Were there any significant political events or personal experiences that
shaped your view of migration policy?
</Question 2>

<Question 3>
How would you describe a typical person who has come to Germany from abroad
in the last ten years?
</Question 3>

<Question 4>
Thinking back over recent years, how do you assess the actions of the
political parties in migration policy?
</Question 4>

<Question 5>
What would be the most important concrete measures that, in your opinion,
should be adopted in migration policy? Please mention anything that comes to
mind.
</Question 5>

<Question 6>
Before we end the interview, is there anything else about migration policy
that is important to you?
</Question 6>

<Question 7>
To conclude this part of the survey, we would like to know: What do you think
of such a conversation as a supplement to a traditional questionnaire with
predefined response options? What did you like or dislike about it?
</Question 7>

<Question 8>
Please now click on Continue in the questionnaire. Thank you for the
interesting conversation.
</Question 8>
\end{verbatim}
\end{small}

\FloatBarrier
\section{Deviations from the pre-analysis plan}
\label{sec:appendix-deviations}

This appendix documents where and why the present manuscript deviates from the pre-analysis plan (PAP) registered on the Open Science Framework prior to data collection:  \href{https://osf.io/y5cjt/overview?view_only=6f8714dd90dc456795390874a8ec6f9f}{Link}

\subsection{Sampling and Fieldwork}

\begin{itemize}
    \item \textbf{Addition of a second panel source.} The PAP specified recruitment exclusively from Prolific. Because Prolific invited respondents at a slower rate than anticipated, data collection was extended to Payback Panel (16–27 October 2025) to reach the target sample size within the planned field period. Both sources received identical survey instruments. Main analyses pool both sources; Appendix A reports disaggregated results by panel provider.
    \item \textbf{Data linkage failure and reduced analytic sample.} The PAP did not anticipate the loss of approximately one in four respondents due to failed merging of survey and interview data, which arose from the manual participant-ID entry required at the start of each interview. Following a conservative strategy, we retain only cases with successfully linked data, yielding an analytic sample of N = 571 rather than the N = 1,039 assigned to experimental conditions.  Moreover, we made the unregistered decision to restrict the analytic sample to interviews with at least five user turns. This decision was driven by the overarching idea to only consider respondents in the analysis who meaningfully interacted with the AI interviewer.
\end{itemize}

\subsection{Prompting Experiment}

\begin{itemize}
    \item \textbf{Prompting experiment deprioritized.} The PAP briefly mentioned the prompting variation (simple vs. informed prompt) as a research question. We registered no hypotheses or analysis strategy because this prompting experimental was not the focus of our plans for this paper. Consequently, we also do no prirotize the prompting experiment in this study and simple briefly mention that the prompting experiment had no statistically significant effects on the outcomes reported in this study. Effects are more plausible on interviewer behavior which is not the focus of this study.
    \item \textbf{Human or machine coding of interviewer speech acts not conducted.} The PAP noted that evaluation of interviewer behavior for the prompting experiment would depend on available resources ("It depends on available resources whether we will be able to evaluate interviewer behavior, potentially based on human or machine coding"). Due to the large number of already conducted analysis we decided not to conduct the annotation of interview behavior for this paper. 
\end{itemize}

\subsection{Addition of Unregistered Analyses}

\begin{itemize}
    \item \textbf{Substantive text-as-data analysis added.} The PAP listed "analysis of interview data" as an outcome category but did not pre-specify the analytical pipeline. The manuscript introduces an LLM-assisted annotation framework using a 12-category issue taxonomy and a 10-category argument-type taxonomy, followed by sequence analysis via Markov chain modeling of topic and rhetorical trajectories across partisan groups. These analyses are presented as illustrative and exploratory demonstrations of the method's analytical value; they are not pre-registered confirmatory tests and are framed accordingly.
    \item \textbf{Wiener Sachtextformel added as a readability metric.} The PAP specified the Flesch Reading Ease Score as the operationalization of linguistic elabxoration in the mode-effects analysis. The manuscript also reports the Wiener Sachtextformel (Table 1), which is better suited to German-language text. Both metrics point in the same direction; the addition of the Wiener Sachtextformel is an unregistered supplement, not a replacement.
    \item \textbf{Turn-level response-length analysis added.} Appendix G reports response length by turn position within the interview for voice and text modes. This analysis was not pre-specified and is treated as exploratory.
\end{itemize}

\subsection{Social Desirability Analysis}

\textbf{Social desirability comparison not formally reported.} The PAP registered as a testable question whether respondents in chat modes report more authentic opinions than in voice-based modes, using the self-reported accuracy item. We have focused on other aspects and have not carried out this analysis. Interested readers can conduct the analysis using the provided replication material. 
\FloatBarrier
\section{Voice mode waiting time}
\label{sec:appendix-voice-waiting}

One issue in earlier implementations of AI Conversational Interviewing \parencite{wuttke_ai_2025} and in our pre-tests was the natural flow of conversation in voice mode. A key configuration is how long the interviewer should remain silent before asking the next question. In the present study, turn detection relied on the protocol implemented in Vapi.

The main-text results suggest that respondents perceived the interviewer to leave an appropriate amount of time for thinking and speaking.

\begin{figure}[H]
  \centering
  \includegraphics[width=0.7\textwidth]{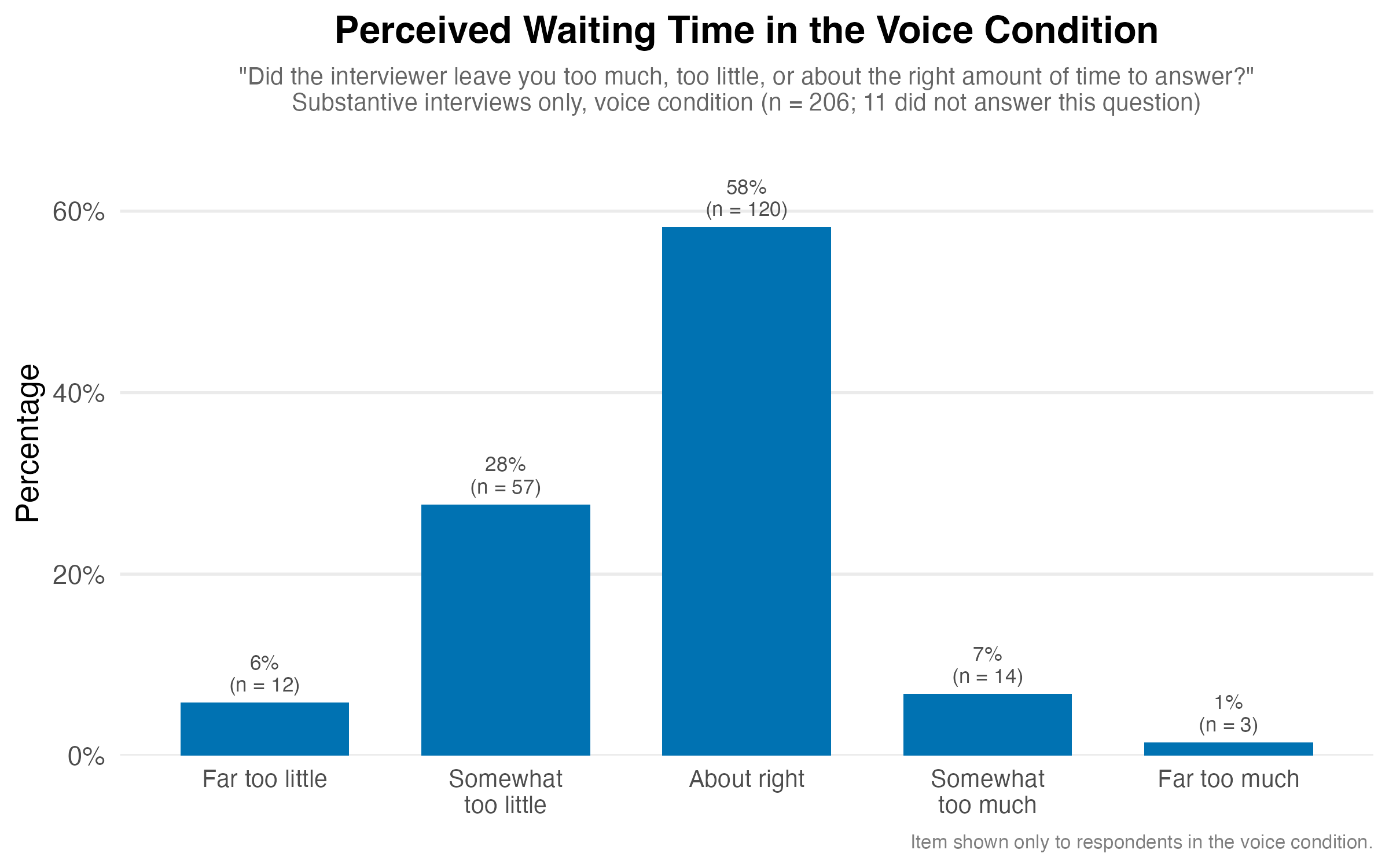}
  \caption{Perceived waiting time for voice condition.}
\end{figure}

\FloatBarrier

\section{Technical problems during the AI interview}
\label{sec:appendix-technical-problems}

Respondents indicated, for four interview tasks, whether they experienced no, minor, or major problems. Tabulations below use the full dataset of all respondents, including respondents who could not be linked to an interview transcript or who discontinued early. Cell entries are percentages with respondent counts in parentheses.

\begin{table}[ht]
\centering
\caption{Self-reported problems during the AI interview (overall).}
\label{tab:tech-problems-overall}
\begin{tabular}{lcccc}
\toprule
Task & No problems & Minor problems & Major problems & $n$ \\
\midrule
Initiating the interview & 80.3 (699) & 7.6 (66) & 12.1 (105) & 870 \\
Ending the interview & 82.8 (711) & 7.1 (61) & 10.1 (87) & 859 \\
Giving answers & 70.7 (607) & 17.1 (147) & 12.2 (105) & 859 \\
Receiving questions & 79.2 (681) & 9.0 (77) & 11.9 (102) & 860 \\
\bottomrule
\end{tabular}
\end{table}

\begin{table}[ht]
\centering
\caption{Self-reported problems during the AI interview, by interview modality.}
\label{tab:tech-problems-by-mode}
\begin{tabular}{llcccc}
\toprule
Task & Modality & No problems & Minor problems & Major problems & $n$ \\
\midrule
Initiating the interview & Text & 80.9 (445) & 7.6 (42) & 11.5 (63) & 550 \\
Initiating the interview & Voice & 79.4 (254) & 7.5 (24) & 13.1 (42) & 320 \\
\midrule
Ending the interview & Text & 82.4 (448) & 7.7 (42) & 9.9 (54) & 544 \\
Ending the interview & Voice & 83.5 (263) & 6.0 (19) & 10.5 (33) & 315 \\
\midrule
Giving answers & Text & 79.4 (432) & 9.7 (53) & 10.8 (59) & 544 \\
Giving answers & Voice & 55.6 (175) & 29.8 (94) & 14.6 (46) & 315 \\
\midrule
Receiving questions & Text & 83.1 (453) & 6.1 (33) & 10.8 (59) & 545 \\
Receiving questions & Voice & 72.4 (228) & 14.0 (44) & 13.7 (43) & 315 \\
\bottomrule
\end{tabular}
\end{table}

\begin{table}[ht]
\centering
\caption{Self-reported problems during the AI interview, by analysis-inclusion status. ``Kept'' = retained in the main analyses (linked interview with at least five user turns); ``Filtered out'' = not linked or non-substantive interview.}
\label{tab:tech-problems-by-status}
\begin{tabular}{llcccc}
\toprule
Task & Status & No problems & Minor problems & Major problems & $n$ \\
\midrule
Initiating the interview & Kept & 95.1 (525) & 4.0 (22) & 0.9 (5) & 552 \\
Initiating the interview & Filtered out & 54.7 (174) & 13.8 (44) & 31.4 (100) & 318 \\
\midrule
Ending the interview & Kept & 92.6 (510) & 5.4 (30) & 2.0 (11) & 551 \\
Ending the interview & Filtered out & 65.3 (201) & 10.1 (31) & 24.7 (76) & 308 \\
\midrule
Giving answers & Kept & 80.2 (441) & 17.8 (98) & 2.0 (11) & 550 \\
Giving answers & Filtered out & 53.7 (166) & 15.9 (49) & 30.4 (94) & 309 \\
\midrule
Receiving questions & Kept & 91.3 (504) & 7.2 (40) & 1.4 (8) & 552 \\
Receiving questions & Filtered out & 57.5 (177) & 12.0 (37) & 30.5 (94) & 308 \\
\bottomrule
\end{tabular}
\end{table}

\FloatBarrier
\section{Turn-dependent statistics: Responses over the course of the interview}
\label{sec:appendix-turn-statistic}

\begin{figure}[H]
  \centering
  \includegraphics[width=0.75\textwidth]{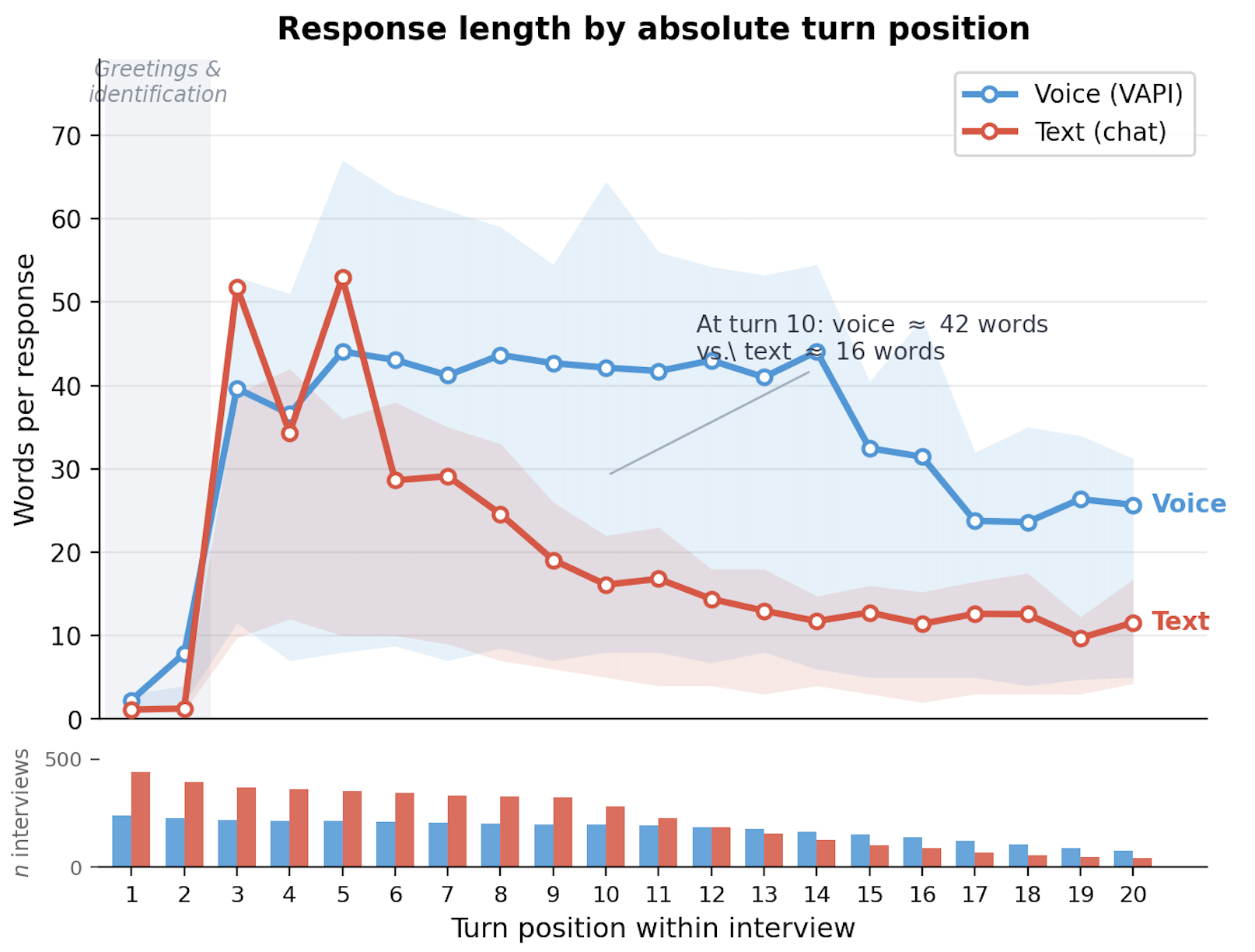}
  \caption{Average words per response by absolute turn position, separately for voice (VAPI, blue) and text (chat, red) interviews. Lines show means; shaded bands denote ±1 SD. The leftmost region (turns 1–2) corresponds to greetings and identification. The lower panel reports the number of interviews contributing to each turn position. Text responses peak early and decline steeply, while voice responses remain stable around 40 words until the closing phase of the interview.}
  \label{fig:turn-statistic}
\end{figure}

One characteristic of conversational interview data is its turn-based structure, which enables us to analyze how response behavior changes throughout an interview rather than just in aggregate. Figure \ref{fig:turn-statistic} illustrates the average words per response by turn position, separated by voice and text modalities, with sample sizes shown in the lower panel. The first two turns, centered on greeting and identification, produce consistently short responses in both modalities and mainly serve as a structural baseline. From turn 3 onward, the two trajectories diverge sharply. Text responses reach a peak early, about 52 words at turn 3 and 53 at turn 5, much longer than voice responses at the same points, then decline steeply for the rest of the interview. Voice responses, by contrast, stay remarkably steady, fluctuating narrowly around 40–44 words from turn 3 to 13 before dropping to about 26 words by turn 20. The crossover occurs between turns 5 and 6: by turn 10, voice responses average 42 words, compared to only 16 in the text condition, a ratio of roughly 2.6:1 that grows wider toward the end. This pattern suggests a steep effort gradient in the typed modality. Respondents seem to put more effort into their first answers, likely due to initial engagement, novelty, or a sense of obligation but quickly shift to terse, minimal responses as typing effort accumulates. The text trajectory plateaus around 10–15 words from turn 10 onward, reflecting minimal or minimal enough responses rather than genuine engagement. Voice responses show a much flatter effort curve, consistent with the lower marginal effort of speaking versus typing, with a decline only in the later turns (14–20), likely due to natural conversational wind-down rather than fatigue. The lower panel indicates that this divergence is not caused by attrition, as both modalities show that most participants complete the survey after the 10th turn. This means only a smaller group of participants participate in more than 10 turns. Therefore, the length difference within the interview reflects behavioral changes among remaining respondents, not shifts in who stays in the interview. Substantively, the crossover has implications for designing measurement tools. Text interviews gather substantial information early on, but respondents disengage quickly due to the typing effort, limiting rich content collection over time. Voice interviews, with lower initial volume but sustained engagement, are better suited for longer or multi-topic interviews. Conversely, text formats may be preferable when the focus is on the first few substantial exchanges.

\FloatBarrier
\section{Standardized survey: attitudes towards immigration}
\label{sec:appendix-standardized-survey}

This plot shows respondents' attitudes towards migration on a semantic differential scale from 1 (ease of migration) to 11 (restrict migration), as recorded in the standardized survey 
\begin{figure}
    \centering
    \includegraphics[width=1\linewidth]{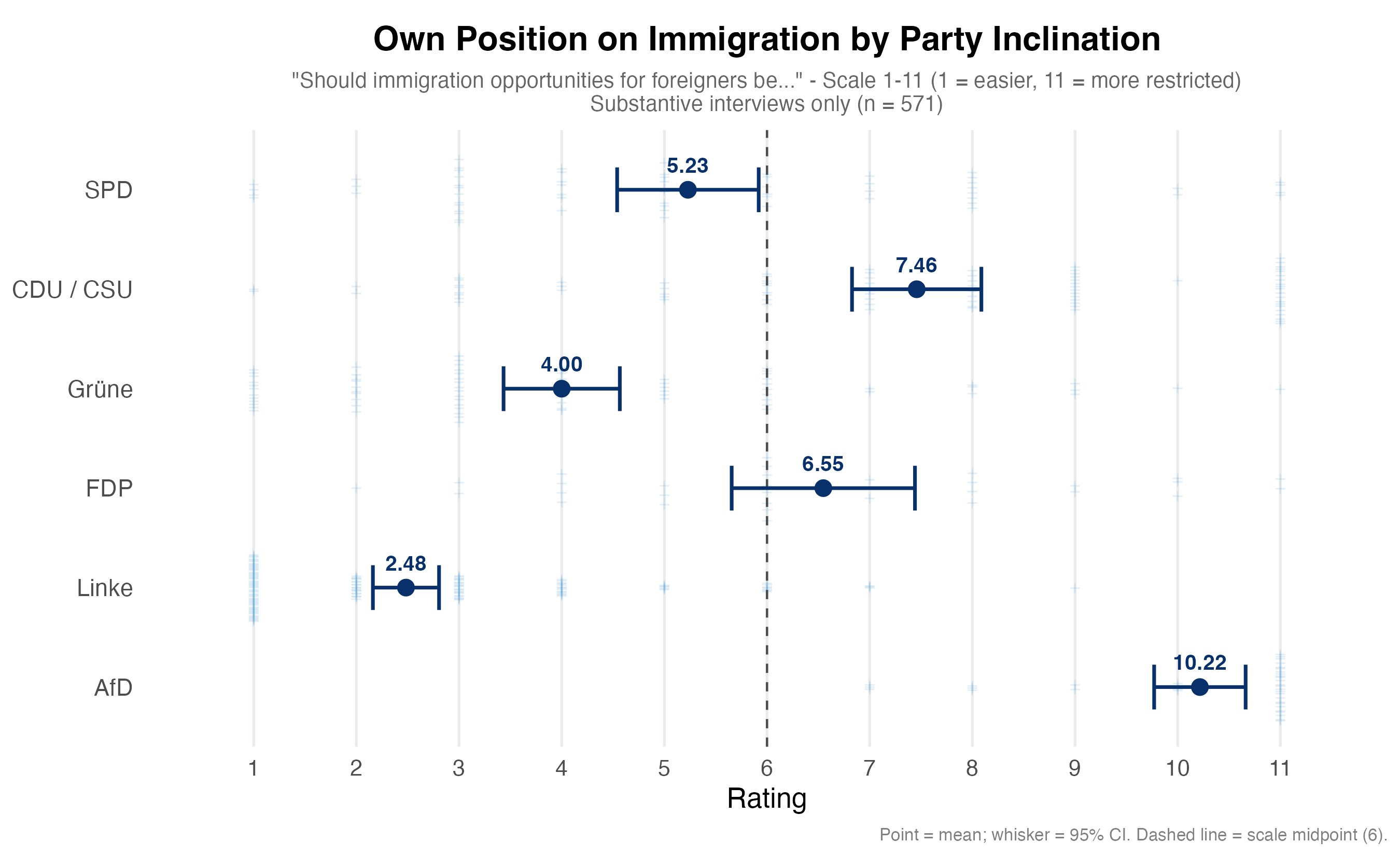}
    \caption{Attitutdes towards immigration}
    \label{fig:imm-attitudes}
\end{figure}
\FloatBarrier
\section{Text analysis: Taxonomy and validation}
\label{sec:appendix-text-analysis}
This appendix provides details on the taxonomy of topics and argument types and on the validation procedures and results for coding the topics and argument types.

The issue taxonomy comprises 12~substantive categories organized into seven thematic groups, as reported in \Cref{tab:issue-taxonomy}.

\begin{table}[H]
  \centering
  \caption{Issue taxonomy for migration-related conversational responses (12 substantive categories).}
  \label{tab:issue-taxonomy}
  \small
  \begin{tabular}{l l p{3.5cm} p{6.5cm}}
    \toprule
    Code & Group & Label & Example \\
    \midrule
    \textsc{border}   & Regulation & Border control, asylum procedures, migration volume & \textit{Es kommen einfach zu viele Menschen auf einmal.} \\
    \textsc{asylum}   & Regulation & Asylum law, refugee protection, legal status & \textit{Das Grundrecht auf Asyl darf nicht angetastet werden.} \\
    \textsc{labor}    & Economic   & Labor market, skilled workers, economic contribution & \textit{Wir brauchen dringend Fachkr\"afte aus dem Ausland.} \\
    \textsc{welfare}  & Economic   & Welfare state, fiscal costs, social benefits & \textit{Das kostet den Steuerzahler Milliarden.} \\
    \textsc{culture}  & Social     & Cultural integration, values, religion, way of life & \textit{Viele wollen sich gar nicht anpassen.} \\
    \textsc{educ}     & Social     & Education, schools, integration programs & \textit{In der Klasse meines Kindes sprechen die meisten kein Deutsch.} \\
    \textsc{security} & Security   & Crime, public safety, terrorism & \textit{Die Kriminalit\"at ist in bestimmten Vierteln deutlich gestiegen.} \\
    \textsc{housing}  & Local      & Housing, local infrastructure, municipal burden & \textit{Es gibt jetzt schon zu wenig Wohnungen.} \\
    \textsc{human}    & Normative  & Humanitarian obligation, human rights, moral duty & \textit{Wir haben die moralische Pflicht, Menschen in Not zu helfen.} \\
    \textsc{cohesion} & Normative  & Social cohesion, identity, societal division & \textit{Die Gesellschaft ist in dieser Frage tief gespalten.} \\
    \textsc{govern}   & Governance & Government policy, bureaucracy, political failure & \textit{Die Politik hat das Problem jahrelang ignoriert.} \\
    \textsc{other}    & Residual   & Unclassifiable or non-substantive content & \textit{Ja, wir k\"onnen anfangen.} \\
    \bottomrule
  \end{tabular}
\end{table}

We can also move beyond issue content and examine a meta-level of rhetoric by coding the argumentative function of information units.

\begin{table}[H]
  \centering
  \caption{Argument-type taxonomy for the rhetorical coding of information units (10 categories, four functional groups).}
  \label{tab:argtype-taxonomy}
  \small
  \begin{tabular}{l l p{4.5cm} p{5.5cm}}
    \toprule
    Code & Group & Label & Example \\
    \midrule
    \textsc{claim}   & Assertion & Policy claim or demand & \textit{Die Regierung muss mehr in Bildung investieren.} \\
    \textsc{eval}    & Assertion & Value judgment, blame, or normative evaluation & \textit{Das ist unverantwortlich.} \\
    \midrule
    \textsc{anecd}   & Evidence  & Personal anecdote or concrete example & \textit{Mein Sohn findet seit zwei Jahren keinen Ausbildungsplatz.} \\
    \textsc{testim}  & Evidence  & Expert testimony, media reference, or authority appeal & \textit{Experten warnen seit Jahren vor den Folgen.} \\
    \textsc{stat}    & Evidence  & Statistics or empirical data & \textit{Die Arbeitslosenquote liegt bei \"uber 5\,\%.} \\
    \midrule
    \textsc{cause}   & Reasoning & Causal explanation (backward-looking) & \textit{Das liegt an der verfehlten Migrationspolitik der letzten Jahre.} \\
    \textsc{conse}   & Reasoning & Argument from consequences (forward-looking) & \textit{Wenn wir nichts tun, wird es nur noch schlimmer.} \\
    \textsc{compar}  & Reasoning & Comparison or contrast across time, place, or group & \textit{Andere L\"ander schaffen das ja auch.} \\
    \midrule
    \textsc{concess} & Framing   & Concession or qualification of an opposing view & \textit{Nat\"urlich muss man auch die Kosten bedenken, aber\,\ldots} \\
    \textsc{meta}    & Framing   & Meta-commentary, hedging, or procedural framing & \textit{Grunds\"atzlich denke ich\,\ldots} \\
    \bottomrule
  \end{tabular}
\end{table}

\begin{figure}[H]
  \centering
  \includegraphics[width=0.70\textwidth]{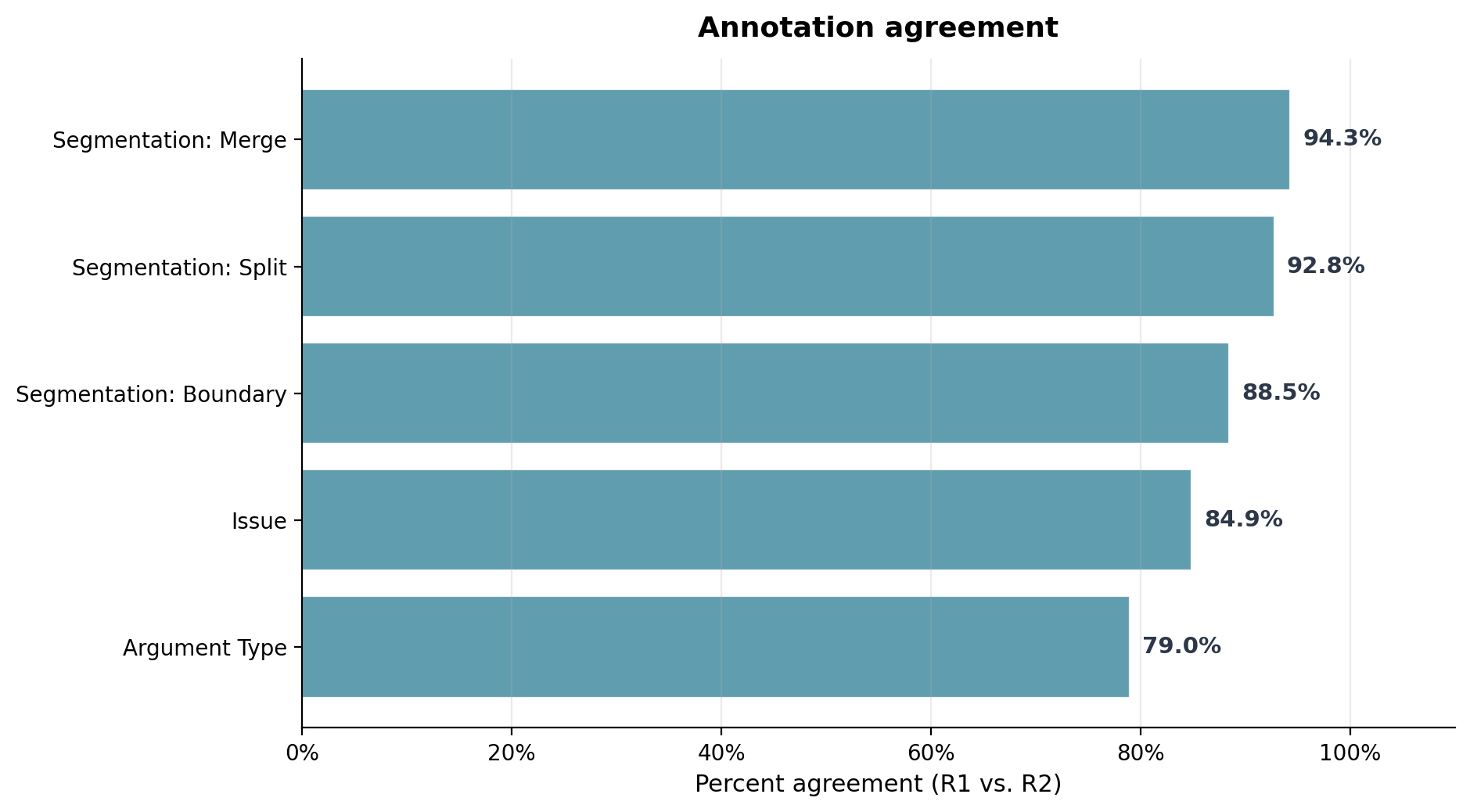}
  \caption{Inter-Annotator Agreement}
  \label{fig:iaa}
\end{figure}

We used this analytical framework on our new dataset of LLM-mediated interactions. The annotation process involved two steps: first, an initial LLM-based pass generated candidate segmentations and labels; then, two trained researchers (R1 and R2) validated these using a structured codebook covering three annotation layers — information units (segmentation), issues (topical content), and argument types (rhetorical function). To evaluate the reliability of this human validation, we calculated decision agreement across both segmentation and classification stages. Figure \ref{fig:iaa} shows the raw percent agreement for the five annotation decisions in our schema. Agreement was consistently high for segmentation tasks: annotators agreed on whether the LLM-proposed boundary was correct (88.5\%, n = 650) and, more strongly, on whether adjacent segments should be merged (94.3\%) or split (92.8\%). These results suggest boundary detection—viewed as a verification task rather than free segmentation—is well-defined and reproducible across coders. Agreement was lower but still substantial for classification layers. For issue labels, R2's agreement with the LLM proposals reached 82.9\% (n = 650); this measurement uses model agreement rather than R1 vs. R2 because one of the researchers did not complete that field. Argument type was the most challenging: R1 and R2 agreed on 79.0\% of the n = 576 cases where the field applied. This reflects the well-known difficulty in distinguishing rhetorical or argumentative functions in conversational data, where category boundaries are fuzzy and context-dependent. Overall, the pattern aligns with expected annotation difficulty: binary structural judgments are easiest, boundary verification is somewhat more challenging, and semantic classification is hardest. All five layers fall within ranges typically deemed acceptable for downstream analysis. 

\FloatBarrier
\section{Text analysis separated by voter groups}
This appendix documents the text analysis results for all voter groups and not just the selected groups of voters that are discussed in the main text.
\label{sec:appendix-text-analysis-voters}

\begin{figure}[H]
  \centering
  \includegraphics[width=1\textwidth]{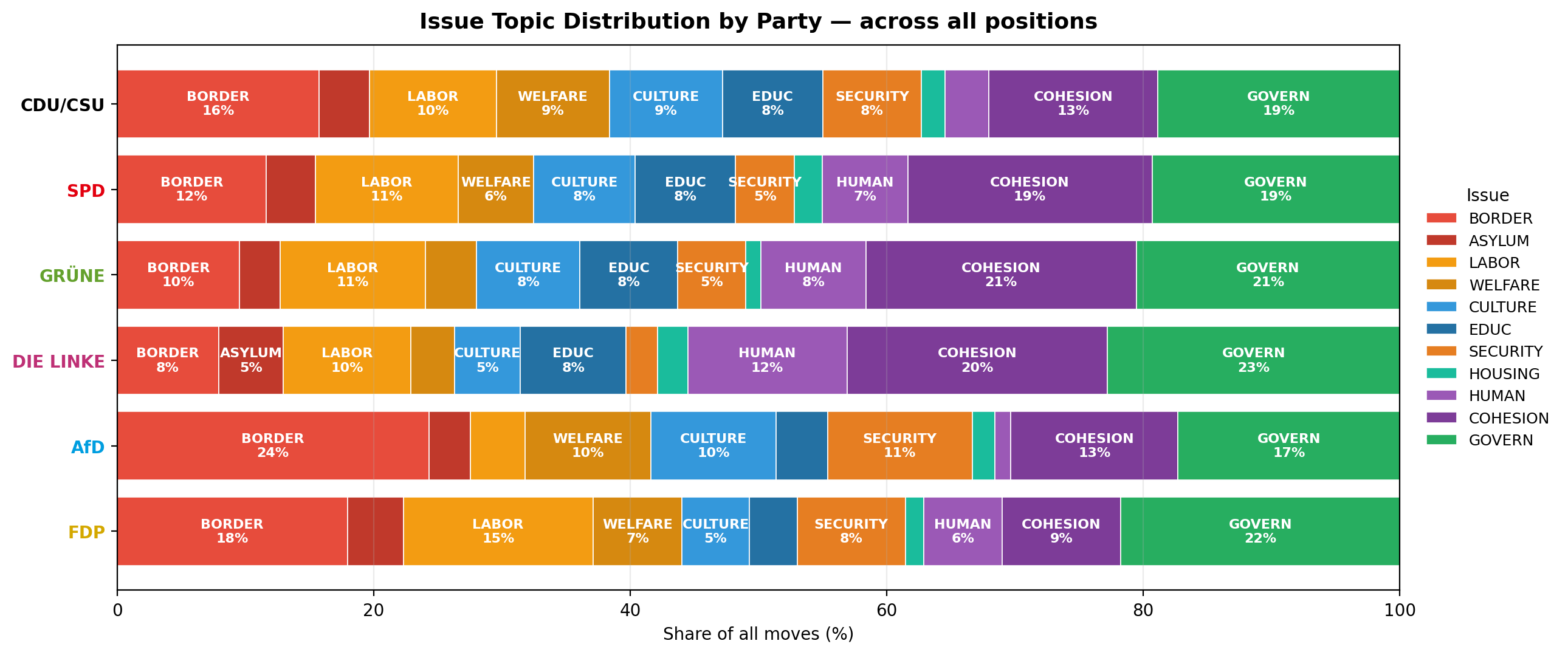}
  \caption{Issue topic distribution by party identification across \emph{all} moves in the collapsed sequences (consecutive duplicates and the residual \textsc{other} category excluded). Each bar sums to 100\%; segment labels indicate the within-party share of each issue.}
  \label{fig:party-issue-distribution-app}
\end{figure}

\begin{figure}[H]
  \centering
  \includegraphics[width=1\textwidth]{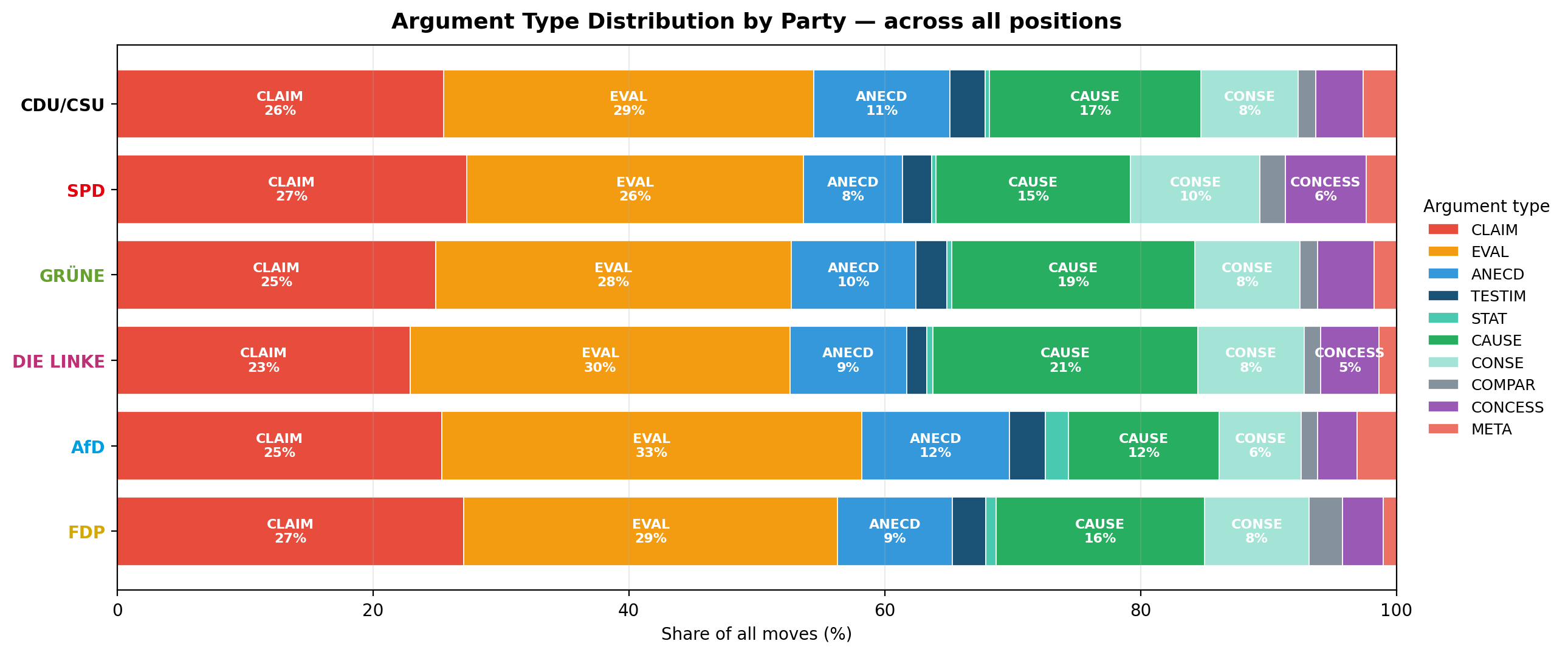}
  \caption{Argument-type distribution by party identification across \emph{all} moves in the collapsed sequences. Each bar sums to 100\%; segment labels indicate the within-party share of each argument type.}
  \label{fig:argtype-distribution-app}
\end{figure}

\FloatBarrier
\section{Sequence analysis}
\label{sec:appendix-sequence-analysis}

This appendix reports an additional analysis on the collected interview data that examines the sequences of topics respondents raised during the interview, separated by party affiliation. 4 people did not answer the question about party affiliation with enough messages to be included in the analysis

\subsection{Topic analysis}
Two parties differ sharply at the entry-topic level. \texttt{FDP} respondents show a strongly \textsc{labor}-oriented entry (50\%), with \textsc{govern} (13\%) and \textsc{border} (13\%) far behind, consistent with a market-liberal framing centred on the skilled-labour debate. \texttt{Die~Linke} respondents most often begin with \textsc{human} (29\%) and \textsc{govern} (27\%), foregrounding humanitarian concerns and governance critique, and ranking \textsc{border} fifth at only 6\%.

\begin{figure}[H]
  \centering
  \includegraphics[width=0.85\textwidth]{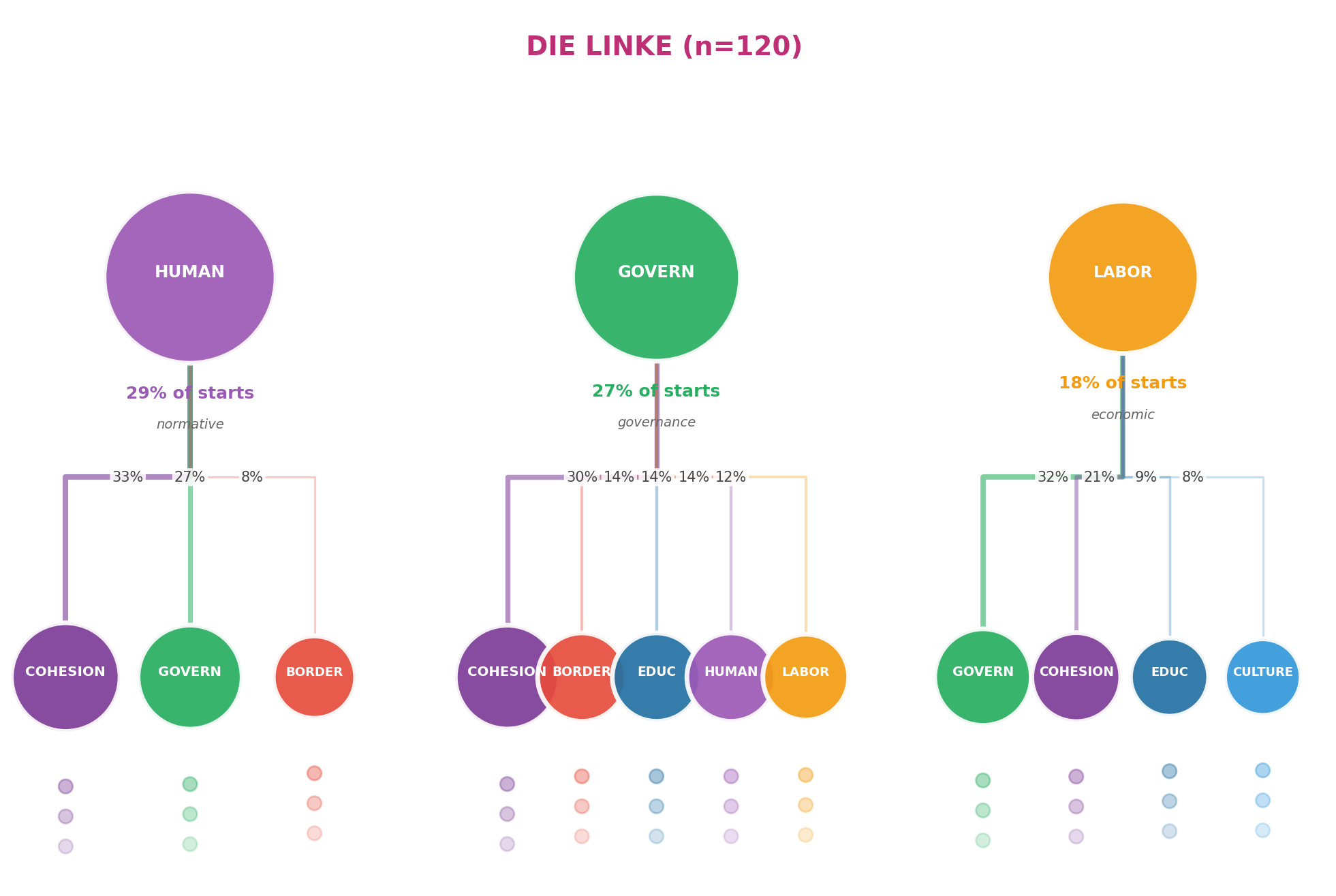}
  \caption{Issue topic trees (depth~2) for the two rhetorically most distinctive parties. Node radius is proportional to the share of sequences passing through the node at that position (entry share at the root, joint entry $\times$ transition share at depth~1); edge labels report the transition probability. Vertical dots below a node signal that the topic chain continues at deeper levels (branches with transition probability $\geq 8\%$, consecutive duplicates collapsed).}
  \label{fig:party-issue-trees-focused}
\end{figure}

\begin{figure}[H]
  \centering
  \includegraphics[width=0.70\textwidth]{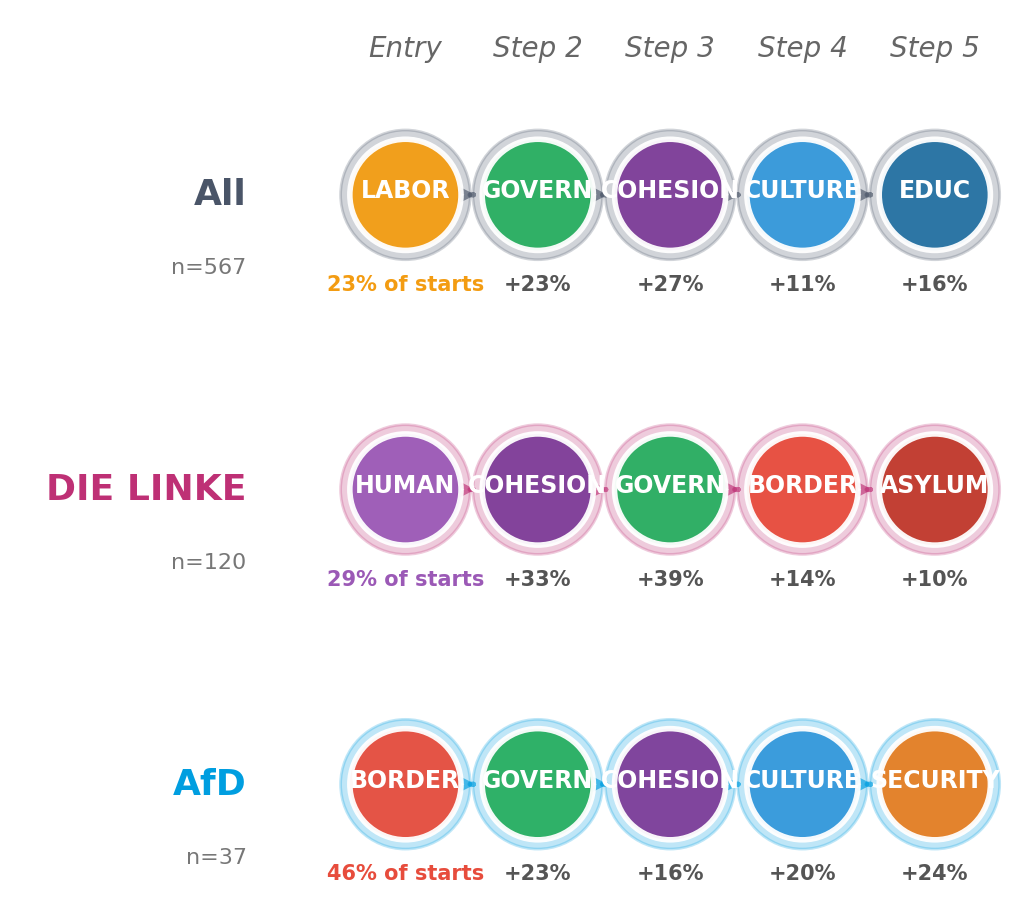}
  \caption{Most likely issue sequence per party, obtained by a greedy Markov walk through the topic-transition matrix (no self-loops; previously visited topics are excluded). The percentage under the first node is the share of sequences starting with that topic; subsequent percentages are the transition probability from the previous node.}
  \label{fig:party-issue-paths}
\end{figure}

To make the difference in trajectories concrete, \Cref{fig:party-issue-paths} renders a single worked example for each party: starting from the party's most common entry topic, we take a greedy walk through the Markov chain, picking the most probable next topic at each step (and excluding topics already visited so the path does not loop on itself). The contrast between AfD and Die~Linke is the sharpest in the panel. An AfD respondent's modal path runs \textsc{border}~$\rightarrow$~\textsc{govern}~$\rightarrow$~\textsc{cohesion}~$\rightarrow$~\textsc{security}~$\rightarrow$~\textsc{culture}, anchoring the conversation in regulation and tying it to security and cultural integration concerns. Die~Linke respondent's modal path runs \textsc{human}~$\rightarrow$~\textsc{cohesion}~$\rightarrow$~\textsc{govern}~$\rightarrow$~\textsc{border}~$\rightarrow$~\textsc{asylum}, beginning from humanitarian framing and reaching border control only after passing through cohesion and governance critique. The other four parties illustrate intermediate patterns: CDU/CSU, SPD, Gr\"une, and FDP all open with \textsc{labor} (24--50\%) but diverge thereafter, with FDP routing through \textsc{govern} and \textsc{border} (45\% transition probability into \textsc{border} from \textsc{govern}) and the centre-left parties moving toward \textsc{cohesion} and \textsc{govern} before reaching regulatory topics.

\subsection{Framing analysis}

\begin{figure}[H]
  \centering
  \includegraphics[width=0.85\textwidth]{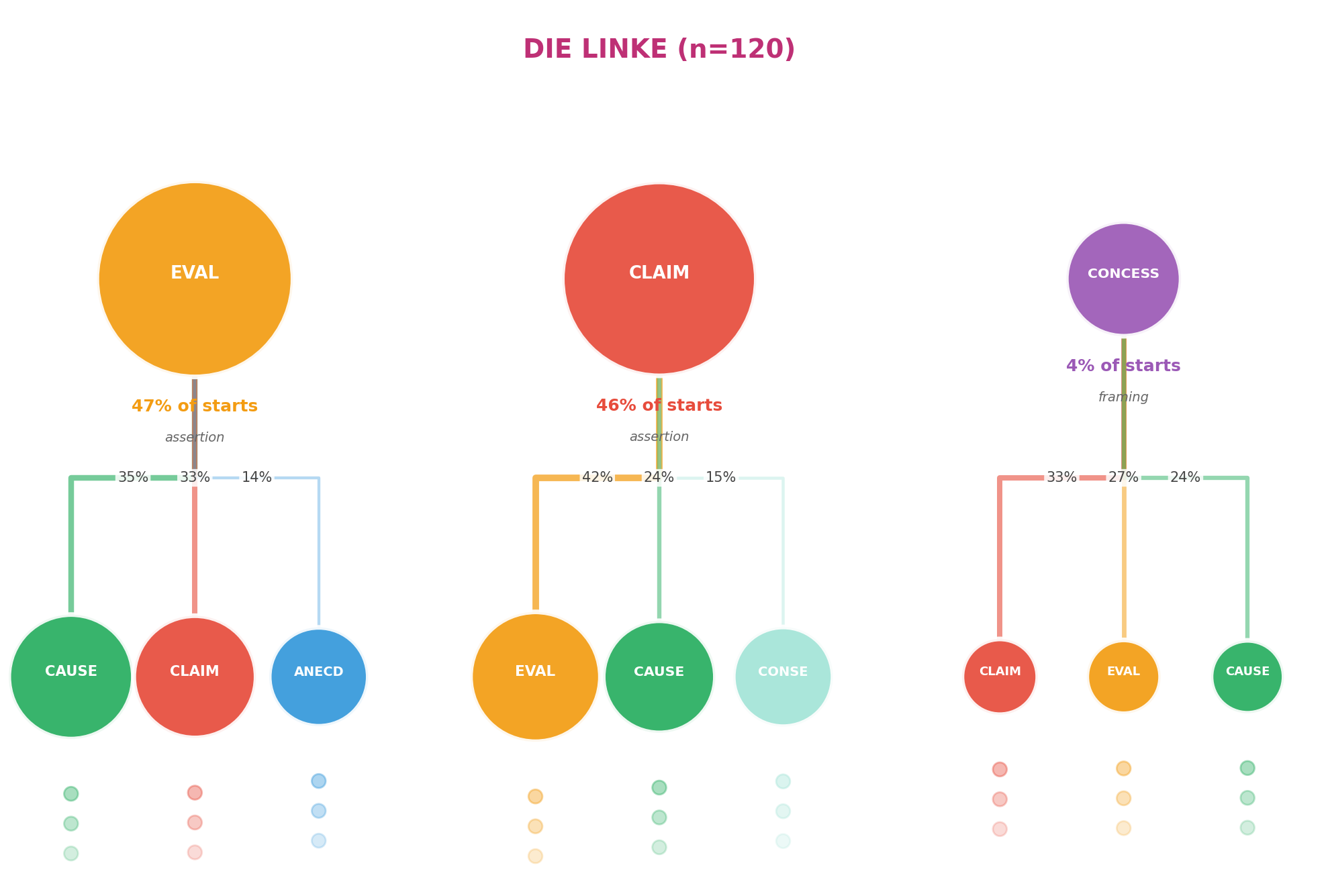}
  \caption{Argument-type strategy trees (depth~2) for Die~Linke. Node radius is proportional to the share of sequences at that position (entry share at the root; joint entry $\times$ transition share at depth~1). Vertical dots below a node signal that the chain continues at deeper levels (branches with transition probability $\geq 10\%$, consecutive duplicates collapsed).}
  \label{fig:frame-trajectories}
\end{figure}

The \textit{analysis of argument-type sequences} uncovers distinct rhetorical signatures across German parties. The focused depth-two trees (\Cref{fig:frame-trajectories}) again contrast AfD, Die~Linke, and FDP, with node radius proportional to the share of sequences passing through each move at that position. AfD supporters emerge as the most evaluatively driven group: 68\% of AfD respondents open with an \textsc{eval} move (compared to 47--54\% in other parties), and the dominant continuations from \textsc{eval} are \textsc{claim} and \textsc{cause}, yielding a characteristic blame-then-prescribe pathway. FDP respondents display a distinctly policy-first style in which \textsc{claim} openings (60\%) substantially exceed \textsc{eval} openings (37\%)---the only party where this pattern holds---and claims are frequently followed by causal explanations or further evaluations. Die~Linke respondents sit between these two poles with an almost even \textsc{eval}/\textsc{claim} entry split (47\%/46\%), and the joint entry-times-transition share of \textsc{eval}~$\rightarrow$~\textsc{cause} is notably larger than for AfD or FDP, reflecting Die~Linke's reliance on causal reasoning as a follow-up. The vertical dots under each depth-one node indicate that all three parties extend these openings into longer, party-specific reasoning chains that are visible in the full depth-three view.

\begin{figure}[H]
  \centering
  \includegraphics[width=0.70\textwidth]{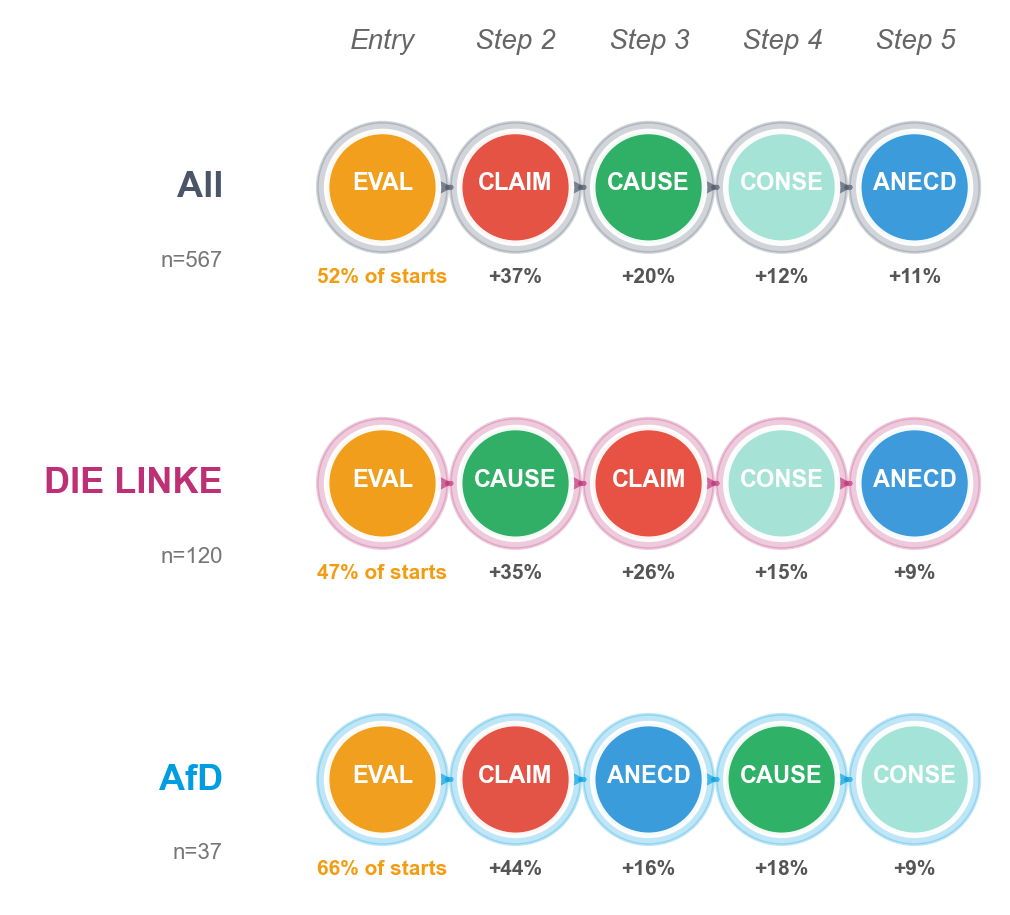}
  \caption{Most likely argument-type sequence per party, obtained by a greedy Markov walk through the argument-type transition matrix (no self-loops; previously visited moves are excluded). The percentage under the first node is the share of sequences starting with that move; subsequent percentages are the transition probability from the previous node.}
  \label{fig:party-argtype-paths}
\end{figure}

To make these rhetorical signatures concrete at the response level, \Cref{fig:party-argtype-paths} renders the most likely argument-type sequence per party using the same greedy Markov-walk construction applied to the issue paths in \Cref{fig:party-issue-paths}. The contrast between AfD and Die~Linke is again striking. An AfD respondent's modal rhetorical path runs \textsc{eval}~$\rightarrow$~\textsc{claim}~$\rightarrow$~\textsc{anecd}~$\rightarrow$~\textsc{cause}~$\rightarrow$~\textsc{conse}: a heavily evaluative opening is followed by a policy demand and then grounded in a personal anecdote rather than a causal explanation. A Die~Linke respondent's modal path runs \textsc{eval}~$\rightarrow$~\textsc{cause}~$\rightarrow$~\textsc{claim}~$\rightarrow$~\textsc{conse}~$\rightarrow$~\textsc{anecd}, with causal reasoning moved up to the second slot (transition probability 35\%, the highest in the panel) and the claim postponed to step three. FDP is the only party whose modal path begins with \textsc{claim} (60\%) and proceeds \textsc{eval}~$\rightarrow$~\textsc{cause}, a sequence that matches the policy-first signature already visible in the trees. CDU/CSU, SPD, and Gr\"une fall between these poles: all three open \textsc{eval}~$\rightarrow$~\textsc{claim}, but they differ in whether \textsc{cause} or \textsc{conse} comes next, with Gr\"une placing \textsc{cause} earlier (transition probability 26\%) than the centre-right or social-democratic comparators.

\end{document}